\documentclass[authoryear,iop]{emulateapj}

\usepackage[]{graphicx}
\usepackage{epsfig}
\usepackage{hyperref}
\usepackage{natbib}
\usepackage[titletoc]{appendix}
\usepackage{multirow}

\newcommand{\beq}{\begin{equation}}
\newcommand{\eeq}{\end{equation}}
\newcommand{\nbeq}{\begin{equation*}}
\newcommand{\neeq}{\end{equation*}}

\shortauthors{Potter et al. 2014}

\begin{document}


\title{Multi-dimensional simulations of the expanding supernova remnant of SN 1987A}



\author{T.M. Potter\altaffilmark{1,2}, L.Staveley-Smith\altaffilmark{1,3}, B. Reville \altaffilmark{4}, C.-Y. Ng \altaffilmark{5}, G. V. Bicknell \altaffilmark{6}, R. S. Sutherland \altaffilmark{6}, A. Y. Wagner \altaffilmark{7}} 
\email{tobympotter@gmail.com}


\altaffiltext{1}{International Centre for Radio Astronomy Research (ICRAR) M468, The University of Western Australia, 35 Stirling Highway, Crawley, WA 6009, Australia}
\altaffiltext{2}{School of Earth and Environment, M004, The University of Western Australia, 35 Stirling Highway, Crawley, WA 6009, Australia}
\altaffiltext{3}{ARC Centre of Excellence for All-sky Astrophysics (CAASTRO)}
\altaffiltext{4}{Centre for Plasma Physics, Queen's University Belfast, University Road, Belfast BT7 1NN, United Kingdom}
\altaffiltext{5}{Department of Physics, The University of Hong Kong, Pokfulam Road, Hong Kong}
\altaffiltext{6}{Research School of Astronomy and Astrophysics, Australian National University, Canberra, ACT 0200, Australia.}
\altaffiltext{7}{Center for Computational Sciences, Tsukuba University, Tsukuba, Ibaraki, 305-8577, Japan}



\begin{abstract}

The expanding remnant from SN 1987A is an excellent laboratory for investigating the physics of supernovae explosions. There are still a large number of outstanding questions, such the reason for the asymmetric radio morphology, the structure of the pre-supernova environment, and the efficiency of particle acceleration at the supernova shock. We explore these questions using three-dimensional simulations of the expanding remnant between days 820 and 10,000 after the supernova. We combine a hydrodynamical simulation with semi-analytic treatments of diffusive shock acceleration and magnetic field amplification to derive radio emission as part of an inverse problem.  Simulations show that an asymmetric explosion, combined with magnetic field amplification at the expanding shock, is able to replicate the persistent one-sided radio morphology of the remnant. We use an asymmetric Truelove \& McKee progenitor with an envelope mass of $10 M_{\sun}$ and an energy of $1.5 \times 10^{44} J$.  A termination shock in the progenitor's stellar wind at a distance of $0\farcs43-0\farcs51$ provides a good fit to the turn on of radio emission around day 1200. For the H\textsc{ii} region, a minimum distance of $0\farcs63\pm0\farcs01$ and maximum particle number density of $(7.11\pm1.78) \times 10^7$ m$^{-3}$ 
produces a good fit to the evolving average radius and velocity of the expanding shocks from day 2000 to day 7000 after explosion. The model predicts a noticeable reduction, and possibly a temporary reversal, in the asymmetric radio morphology of the remnant after day 7000, when the forward shock left the eastern lobe of the equatorial ring.

\end{abstract}




\maketitle

\section{Introduction}

Supernovae play an important role in the evolution of the Universe: providing a source of heavy elements, driving winds to regulate star formation, producing cosmic rays, magnetic fields, neutron stars and black holes. As the brightest supernova since 1604, the type II-P supernova SN 1987A has been the most well studied supernova in history. It was the only supernova to be associated with a neutrino detection \citep{Hirata:1987p4127,Bionta:1987p13260, Aglietta:1987p18205, Alexeyev:1988p18204}, and one of the few supernovae whose progenitor star was observed prior to explosion. \citet{Rousseau:1978p18031} and \citet{Walborn:1989p19593}, classified the progenitor star Sk $69^{\circ}202$ as a B3 I blue supergiant (BSG). The BSG had an estimated surface temperature of 16,000 K; a mass of $19 \pm 3$ $M_{\sun}$; an envelope mass of $5-10$ $M_{\sun}$ \citep{Woosley:1988p18039}; and an estimated wind velocity and mass loss rate of $450 \ \mathrm{km} \ \mathrm{s}^{-1}$ and $7.5 \times 10^{-8}$ $M_{\sun}$  $\mathrm{yr}^{-1}$ \citep{Chevalier:1995p4450}. This was surprising, as the expected progenitors of core-collapse supernovae were red supergiants (RSG's).
Soon after core collapse, a UV flash from shock breakout ionised the material surrounding the BSG and revealed a central equatorial ring, accompanied above and below by two fainter rings \citep{Gouiffes:1989p17578, Plait:1995p25}. Subsequent analysis of the echoes from the UV flash \citet{Sugerman:2005p11362} showed that, at an assumed distance of $50$ kpc, the equatorial ring is the waist of a much larger peanut shaped structure which extends around $2\times 10^{17}$ m ($6.1$ pc) in the direction normal to the plane of the central ring and  $1\times10^{17}$ m ($3.4$ pc) in the plane of the ring. \citep{Sugerman:2005p11362} estimated a total mass of 1.7 $M_\sun$ for the nebula. The amount of material in the circumstellar environment suggests the progenitor previously went through a phase of high mass loss as a RSG before transforming into a BSG prior to explosion. The cause of the transformation is uncertain. Possible explanations for the transformation involve a binary merger \citep{Podsiadlowski:1989p3992}, or low metallicity in the progenitor \citep{ Woosley:1987p22419}. General consensus is that the transformation took place approximately 20,000 years prior to the explosion, and the fast wind from the BSG interacted with the relic RSG wind to form the hourglass and rings \citep{Crotts:1991p17975,Blondin:1993p14977,Crotts:2000p17977, Podsiadlowski:2007p18006}. This interaction has been successfully modelled with hydrodynamical \citep{Blondin:1993p14977}, magnetohydrodynamical \citep{Tanaka:2002p19} and smoothed particle hydrodynamics simulations \citep{Podsiadlowski:2007p18006}.


 \subsection{Radio observations}

Over the last 25 years the interaction of the expanding supernova remnant has been monitored at wavelengths spanning the electromagnetic spectrum. Observations at radio frequencies ranging from 843 MHz to 92 GHz have traced the evolution of flux density, spectral index, and radio morphology of the remnant, from a few days after explosion to the present day \citep{Turtle:1987p2190,StaveleySmith:1992p680,Ball:1992p17710,Gaensler:1997p7998,Ball:2001p3450,Manchester:2005p7378,Gaensler:2007p42,StaveleySmith:2007p2,Ng:2008,Potter:2009p16343,Zanardo:2010p17425,Ng:2011p18225,Lakicevic:2012p22004}.
 Approximately four days after core collapse, radio emission peaked around 150 mJy at 1 GHz as the supernova transitioned from optically thick to optically thin regimes \citep{Turtle:1987p2190}. Over the next few months the emission faded to undetectable levels  as an expanding set of forward and reverse shocks propagated through a rarefied BSG wind. About 1200 days after core collapse, the shocks crashed into the termination shock of the pre-supernova BSG wind. Radio emission from this interaction  became visible, and the remnant was re-detected at 843 MHz by the Molonglo Observatory Synthesis Telescope (MOST) \citep{Ball:2001p3450} and at 1-8 GHz by the Australia Telescope Compact Array (ATCA) \citet{StaveleySmith:1992p680}. The flux density increased rapidly following re-detection as a new set of shocks began propagating away from the BSG wind boundary. Since day 2500, flux density has been growing exponentially at all frequencies \citep{Ball:2001p3450,Manchester:2002, StaveleySmith:2007p2, Ng:2008, Zanardo:2010p17425} as the shocks encounter relics from the RSG wind in the equatorial plane, and a hot BSG wind beyond the termination region at high latitudes. When the radio emission returned, the spectral index $\alpha (F(\nu) \propto \nu^{-\alpha})$ was between $0.8$ and $0.7$. Around day 2200 the spectrum had become its softest, with $\alpha$ around $1.05$. Since day 2500 the spectral index has been hardening linearly as $\alpha(t)=0.825 - 0.018 \times (t-5000)/365$, where $t$ is expressed in days \citep{Zanardo:2010p17425}. Presently, (day 9200), the spectral index has returned to a value between $0.7$ and $0.8$. 

 Since the return of radio emission, the morphology has been consistently measured as a double lobed ring. Interestingly, measurements report that the brightness of the eastern lobe has been consistently 30\% higher than the western lobe \citep{Ng:2008,Potter:2009p16343} for at least 7000 days following the explosion. Beyond that there is observational evidence that the asymmetry is beginning to decline \citep{Ng:2013p25184}. Exactly how the persistent asymmetry is generated is a puzzle. \citet{Gaensler:1997p7998} canvassed three possible explanations including the effect of a central pulsar, an asymmetric circumstellar environment, or an asymmetric explosion. After ruling out the effect of a central pulsar, they concluded that an asymmetric explosion is a likely cause of the radio asymmetry. In addition, there is 
strong evidence that the expansion of the remnant is asymmetric. Early radio images of the remnant made between days 2000-3000 (1992-1995) indicate that the eastern lobe was around $0\farcs1$ further from the measured position of the progenitor than the western lobe \citep{Reynolds:1995p3810,Gaensler:1997p7998}. Observations made at 18, 36 and 44 GHz  \citep{Manchester:2005p7378, Potter:2009p16343, Zanardo:2013p25168} between 2003 and 2011 (days 6000 to 8700) have shown that the eastern lobe is expanding with an average velocity of $6100 \pm 200 $ km s$^{-1}$; around three times faster than the $1900 \pm 400$ km s$^{-1}$ obtained for the western lobe. 
 
 In \citet{Ng:2008}, the topology of the radio emitting shell from SN 1987A was modelled using a shell of finite width and truncated to lie within a half-opening angle of the equatorial plane. They  projected the truncated shells to the $u-v$ plane, and used least squares optimisation to find shells that fitted $u-v$ data from the 8GHz ATCA monitoring observations. As a result, we have estimates of the radius, opening angle, and thickness of the expanding shell of emitting material. The estimated shell radius from the models indicate that the emitting region had a minimum average expansion of 30,000 km s$^{-1}$ from 1987 to 1992 \citep{Gaensler:1997p7998,Ng:2008}. After encountering the relic RSG material inside the ring \citep{Chevalier:1995p4450} the average speed of the supernova shocks slowed to around $4000 \pm 400$ km s$^{-1}$ and remained at that rate of expansion until day 7000 when the emitting region appears to become more ringlike \citep{Ng:2013p25184}.
 

\subsection{Theoretical models of radio emission from the expanding shocks}

 Radio emission from SN 1987A is thought to arise from relativistic electrons accelerated at the supernova shock front. Diffusive shock acceleration (DSA) \citep{Krymskii:1977p17010,Axford:1977p16686,Blandford:1978p17016, Bell:1978p16876} is believed to be the main source of relativistic electrons at such shocks \citep{Melrose:2009p11879}. It produces a non-thermal population of energetic electrons, whose isotropic phase-space distribution in momentum $f(p)$, has a power-law form $f(p)\propto p^{-b}$. For a strong shock with a compression ratio of $\zeta=4$, and a ratio of specific heats of $\gamma=5/3$, diffusive shock acceleration predicts the index on the distribution is $b=\frac{3\zeta}{\zeta - 1}$. 
 
 Early models of the radio emission from SN 1987A were constructed by calculating radio emission using power-law distributions evolving in an expanding shell of hot gas \citet{Turtle:1987p2190,Storey:1987p2168}. The underlying hydrodynamics of the shock were greatly simplified by the assumption that the shell of hot, radio-emitting gas underwent self-similar expansion.  \citet{Turtle:1987p2190} obtained $b \approx 5$ by fitting the analytic shell model of \citet{Chevalier:1982p686} to 843 GHz emission from the first 12 days after core collapse.  \citet{Storey:1987p2168} obtained $b$ in the range $3.79-5.33$ by fitting an expanding shell model to the same data. Their model included synchrotron self-absorption and free-free absorption.  \citet{Wamann:1991p17724} also proposed a model for the early rise and fall in radio emission in which shock-accelerated electrons ``surf'' outwards from the shock along a pre-existing spiral magnetic field line. 
  \citet{Ball:1992p51} and \citet{Kirk:1994p50} developed a time-dependent, two-zone model to evolve the radio-emitting electrons in the adiabatically expanding downstream. They were able to fit the $4.3$ GHz and $843$ MHz radio observations to 1800 days after core collapse by assuming the shock encounters clumps of material and deducing that $b\approx 4.8$. They postulated that the softening of the electron spectrum was due to cosmic ray feedback on the shock. 
  
  The later models of \citet{Duffy:1995p18382} and \citet{Berezhko:2000p56}, included cosmic ray feedback. The resulting weakening of the shock, as it decelerated in the cosmic-ray pressure gradient, modified the compression ratio at the density discontinuity to around 2.7 and softened the electron momentum spectrum index $b$ from $4$ to $4.8$. This provided a physical motivation for the index observed in SN 1987A.

Models of radio emission for SN 1987A up to this point used a pre-existing magnetic field that was compressed by the shock. In recent years it has been shown that plasma-instabilities excited by cosmic rays can amplify a background magnetic field $B$ \citep{Bell:2004p12496} by up to a factor of $\approx 45$ \citep{Riquelme:2009p25347,Riquelme:2010p25349}. The efficiency of magnetic field amplification is an active topic being studied with simulations. From \citet{Bell:2004p12496}, the resulting dependence of magnetic field $B$ on shock velocity $v_s$ scales as $B \propto v_s^{3/2}$. Subsequent models of radio emission from SN 1987A  included prescriptions for magnetic field amplification \citep{Berezhko:2006p55,Berezhko:2011p19115}. These models produce a strong downstream magnetic field of $2 \times 10^{-6}$ T, in contrast to previous models with an assumed magnetic field of $10^{-7}-10^{-8}$ T \citep{,Duffy:1995p18382, Berezhko:2000p56}. The strong dependence of magnetic field upon shock velocity raises the interesting possibility that synchrotron emission is strongly dependent on the shock velocity and thus the asymmetry of the radio remnant may be a byproduct of an asymmetric explosion. The fraction of electrons injected into the shock, $\chi_{el}$, is a product of the microphysics of the shock and is not well understood. Kinetic plasma simulations have made progress in this direction in recent years \citep[e.g.]{mcclementsetal,matsumotoetal,caprioli:2014}. The total synchrotron luminosity from supernovae is dependent upon both the strength of the magnetic field and $\chi_{el}$. With a pre-existing magnetic field model \citet{Berezhko:2000p56} found $\chi_{el}=(1-4)\times10^{-2}$ to be a good fit to radio observations. This was later modified to $\chi_{e}=6 \times 10^{-6}$ in \citet{Berezhko:2011p19115}, after allowing for additional non-linear magnetic field amplification. 

\subsection{The case for a multi-dimensional simulation}

A feature of previous models of radio emission from SN 1987A is their incorporation of varying degrees of spherical-symmetry. Such models cannot account for the interaction of the shock with the ring, nor can they replicate the evolving asymmetrical radio morphology of the remnant. The need for multi-dimensional simulations of SNR 1987A has been made clear e.g \citet{dwarkadas:2007}. Until recently, modelling SNR 1987A and other supernova remnants in multiple dimensions has been regarded e.g \citet{Dewey:2012p25187}, as highly complex and computationally challenging, thus limiting the potential for iterative exploration in  parameter space. However, recent advances in computing have reduced model realisation times, enabling more possibilities for model exploration in higher dimensions.  

With an aim to address the above questions and challenges we present results from a new three-dimensional simulation of the interaction of the shock from SN 1987A with its pre-supernova environment. This work is motivated by the need to overcome some of the limitations with previous one-dimensional models, such as the inability to adequately model the evolving radio morphology of the remnant. In a fully three dimensional simulation we can: (1) Test a hypothesis that magnetic field amplification in combination with an asymmetric explosion is the cause of the observed persistent asymmetry in the radio morphology ; (2) gain insight into the 3D structure of the pre-supernova material;  (3) obtain an estimate of the injection efficiency of shock acceleration by direct comparison with observations; (4) and make a prediction on how the remnant might evolve if the model is accurate.

The quality and relative abundance of observational monitoring data makes SN 1987A an ideal candidate for an inverse modelling problem. 





\section{Simulation technique} \label{sim_tech}

Non-thermal emission from the remnant is computed in two stages. First we use a hydrodynamics code to simulate the expansion of the supernova shock into a model environment. For this we use FLASH \citep{Fryxell:2000p299} to propagate the fluid according to Eulerian conservation equations of inviscid ideal gas hydrodynamics in a Cartesian grid. The equations solved for in FLASH are as follows:

\begin{eqnarray}
\frac{\partial \rho}{\partial t}+\nabla \cdot \rho \textbf{v} & = & 0, \\ 
\frac{\partial \rho \textbf v}{\partial t} + \nabla \cdot (\rho \textbf{v} \textbf{v}) + \nabla P & = & 0, \\
\frac{\partial \rho E}{\partial t}+\nabla \cdot [(\rho E+P) \textbf v] & = & 0 .
\end{eqnarray}

The density and pressure is $\rho$ and $P$, and $\textbf{v}$ represents the three components of fluid velocity. The specific energy $E=\epsilon+\frac{1}{2} \textbf{v}^2$ is the sum of specific internal ($\epsilon$) and specific kinetic $ \left ( \frac{1}{2}\textbf{v}^2 \right )$  energies. We assume an ideal monatomic plasma with $\gamma=5/3$ as the ratio of specific heats, and use the  
 ideal gas equation of state $P=(\gamma-1)\rho \epsilon$ to close the set of equations. Radiative cooling is implemented using a temperature-dependent cooling function (discussed in Section \ref{radiative_cooling}). The second stage involves post-processing the hydrodynamics output. We locate both forward and reverse shocks and apply semi-analytic models of diffusive shock acceleration to fix the momentum distribution $f(p)$ at locations in the grid immediately downstream from the shocks. The magnetic field is assumed to be amplified at the shock through cosmic-ray current driven instabilities \citep{Bell:2004p12496} and is evolved adiabatically downstream of the shock. The development of both the momentum distribution and magnetic field intensity for shocked voxels is implemented using a simple up-winded advection scheme, with source terms where necessary (see sections \ref{sec:advb} \& \ref{sec:advf}). Synchrotron emission and absorption is calculated from the resulting momentum distribution and magnetic field using standard analytic expressions. Comparisons of the simulated radius, flux density and morphology with observations are used to fit parameters in the initial environment.  

\subsection{Initial environment} \label{orientation}

The initial conditions for the pre-supernova environment are designed to be physically motivated as much as possible,  while fitting the monitoring observations. We use estimates from the literature to build an approximate model and  refine it using inverse modelling by comparison with observations. In regions of the model where we do not have good data we use results from a pre-supernova environment formation simulation (EFS, see Appendix \ref{remnant_formation} for details). We report the results of our best 3D supernova model, whose parameters have been tuned to fit radio observations such as the expanding radius, flux density and radio morphology. We do not claim that the residual of the model has been minimised, rather we report on a model that fits the majority of radio observations and can be used to make meaningful deductions about the real supernova. 

\subsubsection{Computational domain}

The scope of our simulation is to model the interaction of the supernova shock with the inner hourglass and equatorial ring for a simulated period of 10,000 days after explosion. In order to simplify radiative transfer calculations we used a Cartesian grid aligned with the plane of the sky and centred on the progenitor. The environment was then inclined within this grid in order to match the orientation of the equatorial ring. The positive X axis corresponds to West on the plane of the sky, the positive Y axis is North, and the positive Z axis points to Earth. The grid is a cube with length 256 cells ($3.36\times 10^{16}$ m) on a side. This corresponds to an angular separation of $4\farcs5$ at the assumed distance of 50 kpc, and is enough to encapsulate most of the innermost hourglass and expanding supernova shocks over a simulated period of 10,000 days. The somewhat low resolution model was chosen to permit reasonably fast and flexible model realisation times of around 10 hours for the complete inverse problem.

\begin{figure}[h]
\centering
\epsfig{file=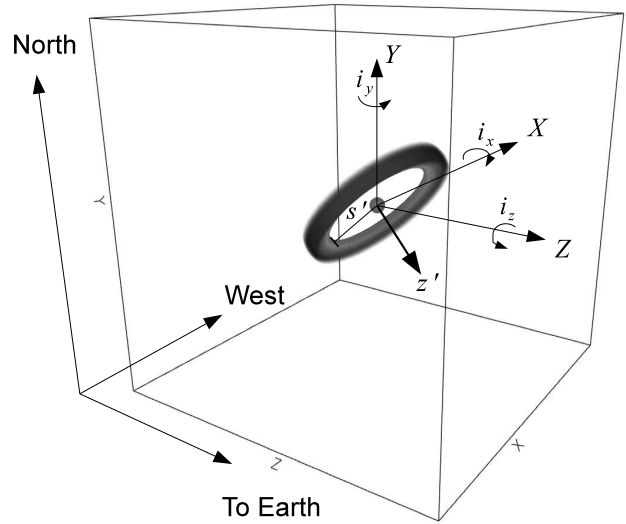,width=8.5cm} \caption{Cartoon of the equatorial ring showing the inclination of the environment at angles $i_x=41^\circ$, $i_y=-8^\circ$, $i_z=-9^\circ$. The rotated cylindrical coordinate system radial coordinate $s^{\prime}$ and vertical coordinate $z^{\prime}$ has $z^{\prime}$ parallel to the plane normal. }  \label{orientation_diagram}
\end{figure}

The equatorial ring and hourglass were inclined within the grid using a series of counter-clockwise rotations when looking down the axis toward the origin. For example a positive Z axis rotation is counter-clockwise when looking toward the origin from Earth. In Figure \ref{orientation_diagram} is a cartoon of the inclined environment \citet{Sugerman:2005p11362} found a best fit inclination of the equatorial ring and hourglass at $i_x=41^\circ$, $i_y=-8^\circ$, $i_z=-9^\circ$. In practice we use a series of successive X, Y, and Z rotations to achieve the observed inclination. The required rotations are $(x_{\mathrm{rot}}=41^\circ,y_{\mathrm{rot}}=-5^\circ, z_{\mathrm{rot}}=-3^\circ)$. Within this inclined environment we use the Cartesian coordinates $(x^{\prime},y^{\prime},z^{\prime})$ and cylindrical coordinates $(s^{\prime}, \phi^{\prime}, z^{\prime})$ centred on the progenitor. If the environment were not inclined, the positive  $z^{\prime}$ axis would point to Earth and the angle $\phi$ would be measured as a counterclockwise rotation from the X axis. We also use $r$, the radial distance from the progenitor. 

\subsubsection{Model features}

Within our domain, the main components of the pre-supernova environment surrounding SN 1987A are as follows. Outwards from the progenitor a supersonic, low density wind extends to a termination shock located at a radius approximately $3.5 \times 10^{15}$ m $(0\farcs47)$. Exterior to the termination shock lies a bipolar bubble of higher density hot, shocked BSG gas. Based on the environment formation simulations in Appendix \ref{remnant_formation} we found that the hot BSG wind re-accelerated to form another shock at a Mach disk at a radius of $1.3 \times 10^{16}$ m ($1\farcs8$). The expanding bubble is the driver that shapes the hourglass and rings. The material at the edge of the bubble is referred to as the H\textsc{ii} region \citep{Chevalier:1995p4450}. The equatorial ring lies within the H\textsc{ii} region at a distance of $(6.4 \pm 0.8) \times 10^{15} $ m or $(0\farcs86 \pm 0\farcs1)$ \citep{Plait:1995p25,Sugerman:2005p11362} and forms the waist of the hourglass. Exterior to the hourglass the density fades to the background density as ${s^{\prime}}^{-3}$ near the waist and as ${s^{\prime}}^{-4.5}$ at large $|z^{\prime}|$. 

\begin{figure}[h]
\includegraphics[width=9.5cm, angle=0]
{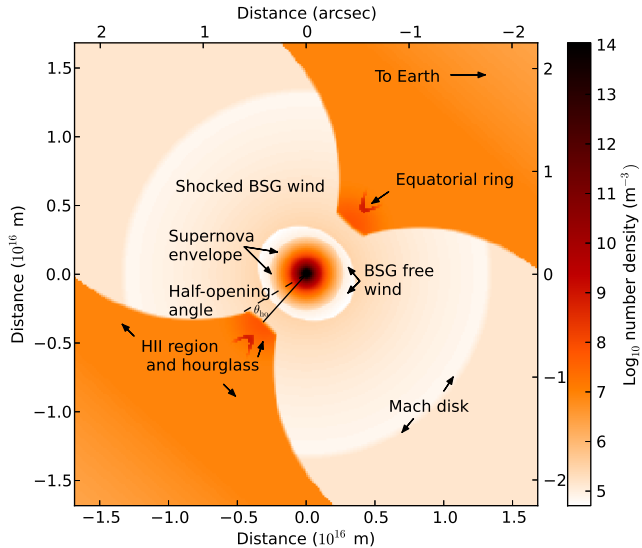}
\caption{Slice through the three dimensional volume taken halfway along the x axis. The variable shown is the base 10 log of the number density (m$^{-3}$). Earth is to the right and north is up. Features of the plot are the central supernova envelope and BSG free wind region, the H\textsc{ii} region, hourglass, and equatorial ring.}\label{log10_density_plot_0000}
\end{figure}

\begin{figure}[h]
\includegraphics[width=9.5cm, angle=0]
{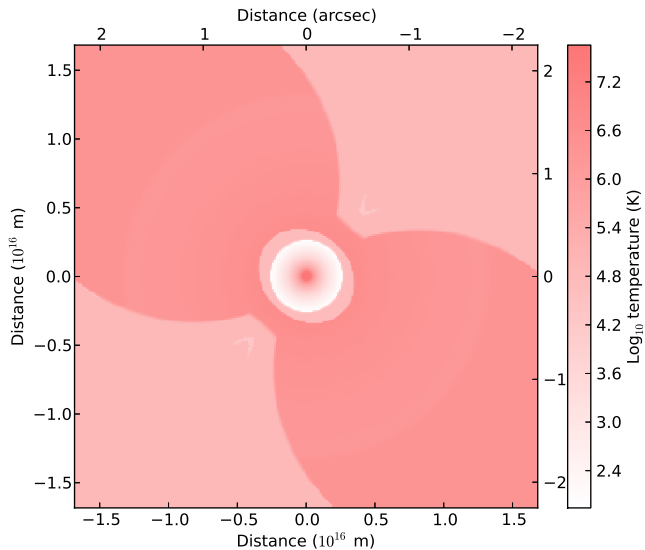}
\caption{Same slice as in Figure \ref{log10_density_plot_0000} but in the log of temperature. The highest temperature material is in the core of the progenitor and the shocked BSG wind.}\label{log10_temperature_plot_0000}
\end{figure}

\begin{figure}[h]
\includegraphics[width=9.5cm, angle=0]
{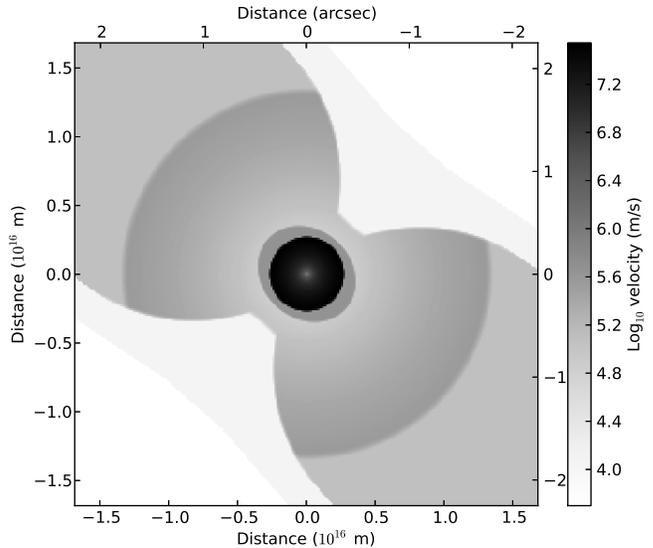}
\caption{Same slice as in Figure \ref{log10_density_plot_0000}, but in the log of velocity. The progenitor is the central core of highest velocity in the range $(10^{6}-10^{7})$ m/s.
}\label{log10_velocity_plot_0000}
\end{figure}

In Figures \ref{log10_density_plot_0000}, \ref{log10_temperature_plot_0000} and \ref{log10_velocity_plot_0000} are cross sections of the model environment, obtained by slicing halfway along the X axis in the log of particle number density, temperature, and velocity. In Figure \ref{log10_density_plot_0000} we have labelled the main features of the model, including the progenitor, free BSG wind, shocked BSG wind, mach disk, HII region, and equatorial ring. Details of how we arrived at the model will be discussed in forthcoming sections.  

\subsubsection{Properties of the plasma}


For all simulations we assume an ideal monatomic gas with $\gamma$, the ratio of specific heats, set to $\frac{5}{3}$. An average atomic mass per particle was derived using the MAPPINGS shock and photo-ionization code \citep{Sutherland:1993p17000} and the abundances of the inner ring from Table 7 of  \citet{Mattila:2010p18200}. The derived average atomic mass per particle, $\mu$, is 0.678 amu, and the average number of particles per electron, Hydrogen atom, ion, and nucleon is 2.62, 3.51, 3.00, and 1.62 respectively.

\subsubsection{Progenitor}\label{progenitor}

The progenitor star Sanduleak -69$^{\circ}$ 202 was observed to be a B3 I blue supergiant \citep{Rousseau:1978p18031,Walborn:1989p19593}, with an estimated surface temperature of 16,000 K \citep{Arnett:1989p9242}, a mass of $19 \pm 3$ $M_{\sun}$ an envelope mass of $5-10$ $M_{\sun}$ \citep{Woosley:1988p18039,Nomoto:1988p16} and an estimated wind velocity and mass loss rate of $450 \ \mathrm{km} \ \mathrm{s}^{-1}$ and $7.5 \times 10^{-8}$ $M_{\sun}$  $\mathrm{yr}^{-1}$ \citep{Chevalier:1995p4450}. 
As the initial stages of the explosion are too small to represent at our chosen given grid resolution, we used an analytic solution to evolve the supernova to a size large enough to represent within a sphere 20.25 voxels or $r_{sn}=2.7 \times 10^{15}$ m ($0\farcs36$) in radius. The self-similar analytical solutions in \citet{Chevalier:1976p645} and \citet{Truelove:1999p17810} describe the propagation of a supernova into a circumstellar environment with density profile of $\rho(r)\propto r^{-s}$. The density of the expanding supernova envelope varies with velocity as $\rho(v)\propto v^{-n}$. We use $n=9$, $s=2$ from \citet{Chevalier:1995p4450} and modify the analytic solution of \citet{Truelove:1999p17810} to include internal energy and introduce an asymmetric explosion in the east-west direction. Details of these modifications are described at length in Appendix \ref{progenitor_appendix}. The mechanics of the model supernova envelope are completely described by: $M_{\mathrm{env}}$ the mass of the supernova envelope; $E_{\mathrm{tot}}$, the total mechanical energy of the explosion (kinetic plus internal),  $\chi_{\mathrm{ke}}$, the ratio of kinetic energy to total; and $\chi_\mathrm{asym}$, the ratio of kinetic energy in the eastern hemisphere of the supernova envelope with respect to the western hemisphere. We adopt $E_{\mathrm{tot}}=1.5 \times 10^{44} $J, and $M_{\mathrm{env}}=10 \ M_{\sun}$, consistent with the results of \citet{Woosley:1988p18039, Arnett:1989p18352, Bethe:1990p5917, Shigeyama:1990p15496}. In order to keep internal energy low we set the initial ratio of kinetic energy to total mechanical energy to an arbitrary value of $\chi=0.9$ for our standard SN 1987A progenitor. The radio morphology was particularly sensitive to the asymmetry parameter $\chi_{\mathrm{asym}}$. It was fitted as $1.55 \pm 0.05$ using the radio morphology from the high-resolution October 2008 observations at 36 GHz \citep{Potter:2009p16343}, however it is likely that our fitted explosion asymmetry is an upper bound. In future work the asymmetry fit needs to be refined with more high resolution radio images. Using this solution a supernova explosion radius of $20.25$ voxels corresponds to a simulated time of around $850$ days after explosion. A cross section of the initial supernova envelope can be seen in Figures \ref{log10_density_plot_0000},  \ref{log10_temperature_plot_0000}, and \ref{log10_velocity_plot_0000}. 

\subsubsection{BSG wind region}


A star emitting a spherically symmetric and steady wind with a mass loss rate $\dot{M}$ and a wind velocity $v_w$ produces a wind whose  density profile scales with radius (r) as 

\beq
\rho(r)=\frac{\dot{M}}{4 \pi v_w r^2}.
\eeq

For our model BSG wind exterior to the progenitor we use $\dot{M}=7.5 \times 10^{-8}$ $M_{\sun}$  $\mathrm{yr}^{-1}$, $v_w=450 \ \mathrm{km} \ \mathrm{s}^{-1}$ as in \citet{Chevalier:1995p4450}. This is consistent with the density profile for the BSG wind derived from our environment formation simulation. From \citet{Lundqvist:1991p15005} there is evidence that the BSG wind was ionized by the shock breakout and attained a temperature in the range $(3.5-7.5) \times 10^4$ K. For the wind we assume an isothermal temperature of $5.5\times10^4$ K.

Previous theoretical work has shown that the BSG wind ends in a termination shock located around $(2.25-3.0) \times 10^{15}$ m $(0\farcs3-0\farcs5)$  from the central star \citep{Blondin:1993p14977,Zhekov:2010p17017, Berezhko:2011p19115}, (see also  Appendix \ref{remnant_formation}). The EFS shows that the termination shock is a prolate spheroid whose ratio of polar to equatorial minor axes is $1.18$ with an average radius of $4.19 \times 10^{15}$ m ($0\farcs56$) from the two semi-axes. We found that the turn on in radio emission around day 1200 is sensitive to the location of the termination shock. We used the same ratio of polar to equatorial axes as for the prolate sphere and found that an average radius of $R_{\mathrm{ts}}=(3.5\pm0.37) \times 10^{15}$ m $(0\farcs47 \pm 0\farcs07)$ provided the best fit to the return of radio emission around day 1200 as seen in Figure \ref{flux_vs_time}.

\subsubsection{Shocked BSG wind}

Exterior to the termination shock, but still within the hourglass and H\textsc{ii} region is a bubble of shocked BSG wind. Our EFS indicates that the bubble is hot, with a density of $1.5 \times 10^{-22} $ kg m$^{-3}$  a temperature of around $2.4 \times 10^6$ K and a velocity of $170$ km s$^{-1}$ on average. Outwards from the termination shock the outflowing material thins slightly and becomes supersonic again at a Mach disk, which we take to be at a radial distance of $1.3\times10^{16}$ m $(1\farcs8)$ from the progenitor. Outside the Mach disk the gas properties are approximately constant until the edge of the expanding bubble is reached.  We also use EFS results to fit polynomials to the pressure and density profiles of the shocked BSG wind. Details of the fitted polynomial and related constants are given in Table \ref{efstable}. The total mass for all BSG wind structures within the grid is $8.8\times10^{-4} M_{\odot}$.

\subsubsection{H\textsc{ii} region and hourglass}

Measurements of radial expansion show that around day $1800$ the shock slowed significantly from $30,000 \ \mathrm{km}~\mathrm{s}^{-1}$ to $3,000 \ \mathrm{km}~\mathrm{s}^{-1}$ at a radius of $4.82\times10^{15}$ m ($0\farcs{642}$) . This implies the shock encountered material of significantly lower sound speed than in either the free or shocked BSG wind \citep{StaveleySmith:1993p6793,Gaensler:1997p7998,Ng:2008}. \citet{Chevalier:1995p4450} predicted the existence of ionised H\textsc{ii} gas in the vicinity of the equatorial ring, swept there by either the expansion of the shocked BSG wind or evaporated from the ring. As neither the literature nor the EFS have any information on the morphology of the H\textsc{ii} region we model the innermost edge of the H\textsc{ii} gas by a convex circular profile in the toroidal $(z^{\prime},s^{\prime})$ plane and set it as the inner edge of the waist of the hourglass. The curve describing the inner edge of the H\textsc{ii} region is constructed using an arbitrary radius of $9.9 \times 10^{15}$ m ($1\farcs33$) with its origin placed in the equatorial plane and beyond the equatorial ring. The height of the waist (in $|z^{\prime}|$) is $2.53 \times 10^{15}$ m ($0\farcs338$). 

The simulation also shows that the shock slowed significantly after encountering the H\textsc{ii} region. The optimal fit to the radius observations places the inner edge of our model H\textsc{ii} region at $R_{\mathrm{H\textsc{ii}}}=(4.71\pm0.07) \times 10^{15}$ m $(0\farcs63 \pm 0\farcs01)$, which is within errors of the location where the supernova shock was observed to have slowed. We found that a sharp transition to the HII region of width no larger than $3.0\times 10^{14}$ m ($0\farcs04$) provided the best fit to turnover in shock velocity.

From  \citep{Sugerman:2005p11362} the hourglass is defined between a cylindrical radius $(1.0-1.3) \times 10^{16}$ m ($1\farcs34-1\farcs73$) and a maximum height of $2.37\times 10^{16}$m ($3\farcs16$) above the equatorial plane. In order to form the inner edge of the hourglass above the waist we use an exponential profile in $|z^{\prime}|$ such that it asymptotes to the outer rim of the hourglass at large $|z^{\prime}|$. The parameters of the exponential were chosen such that it completes $3$ e-folding lengths between corners of the inner waist and the outer rim of the hourglass.

We anticipate that the H\textsc{ii} region gradually merges with the density of the hourglass at large $s^{\prime}$ and $|z^{\prime}|$, such that the boundary between the H\textsc{ii} region and hourglass is undefined. For the density and pressure profiles of the H\textsc{ii} region we use a truncated two-dimensional raised Gaussian in the $(s^{\prime},z^{\prime})$ plane. Fits to the evolving supernova shock radius place the peak density of the HII region at its innermost edge. The FWHM of best fit in the  $s^{\prime}$ direction was $s^{\prime}_{\mathrm{FWHM}}=(2.25\pm 0.37 ) \times10^{15}$ m ($0\farcs30 \pm 0\farcs05$). The simulated radio morphology of the remnant is sensitive to the half-opening angle of the HII region. In terms of the model, the half-opening angle $\theta_{\mathrm{ho}}$ is defined as $\theta_{\mathrm{ho}}=\tan^{-1} \left  (\frac{z^{\prime}_{\mathrm{FWHM}}}{2R_{\mathrm{H\textsc{ii}}}} \right )$. Fits to the observed radio morphology of the radio emission \citep{Potter:2009p16343} show that the best-fit half-opening angle is $15^{\circ}\pm5^{\circ}$, which yields $z^{\prime}_{\mathrm{FWHM}}=(2.52 \pm 0.88)\times10^{15}$ m ($0\farcs34 \pm 0\farcs12$). The best-fit peak particle number density of the Gaussian is  $(7.11 \pm 1.78) \times10^{7}$ m$^{-3}$ which is consistent with the results from \citet{Zhekov:2010p17017}.

For the hourglass \citet{Sugerman:2005p11362} measured a Hydrogen gas density of $(2-3) \times 10^6$ m$^{-3}$ out to a cylindrical radius of $s^{\prime}=1.51\times10^{16}$m ($2\farcs02$) . Given the abundances in use we set the gas number density of the hourglass to $8.77\times10^6$ m$^{-3}$ and have the H\textsc{ii} region gas properties asymptote to this value at large $s^{\prime}$ and $z^{\prime}$. Outside the hourglass in the radial direction we model the equatorial belt and outer walls described in \citet{Sugerman:2005p11362} using a density profile that scales as ${s^{\prime}}^{-3}$ for $|z^{\prime}|<9.46\times10^{15}$ m and as ${s^{\prime}}^{-4.5}$ elsewhere. The particle density of all the hourglass structures is limited to a floor value of $3.5\times 10^5$ m$^{-3}$, consistent with \citet{Sugerman:2005p11362}. The upper and lower boundaries of the hourglass at $|z^{\prime}|=2.37\times 10^{16}$ m $(3\farcs16)$ are smoothed using a logarithmic ramp functions in the pressure and density profiles, and a transition width of $2.43\times10^{15}$ m $(0\farcs32)$. The total mass for the hourglass structures within the grid is approximately $5.2\times10^{-2} M_{\odot}$.

The EFS shows that the temperature of the H\textsc{ii} region is about $10^4$ K, approximately a factor of 2 higher than the estimated $4,500$ K for the H\textsc{ii} region prior to the UV flash, and an order of magnitude less than the $10^5$ K for post supernova models \citep{Lundqvist:1999p20069}. For the pressure profile of the H\textsc{ii} region we adopt an isothermal temperature of $8.0\times 10^4$ K \citep{Lundqvist:1999p20069}
for the hourglass, H\textsc{ii} region, equatorial belt and outer walls.

\subsubsection{Equatorial ring}

Ionisation of the equatorial ring by the supernova UV flash has enabled accurate distance measurements to the supernova \citep{panagia:2005}. Assuming a circular ring, the radius $r$ and width $w$ of the ionized ring has also been determined as $r_{\mathrm{er}}=(6.4 \pm 0.8) \times 10^{15} $ m $(0\farcs86 \pm 0\farcs01)$ and $w_{\mathrm{er}}=(9.0 \pm 1.6) \times 10^{14}$ m ($0\farcs12\pm0\farcs02$) \citep{Plait:1995p25,Sugerman:2005p11362}. While the geometry of the non-ionized portion is unknown, model fits to HST radial profiles of the ring \citep{Plait:1995p25} suggest a crescent torus geometry for the ionized region. There is also convincing evidence that over-dense clumps of material reside within the ring and form hot spots of optical and radio emission when the shocks encounter it \citep{Pun:2002p24, Sugerman:2002p40, Ng:2011p18225}.

Spectroscopic optical and $u-v$ line emission measurements of the equatorial ring between days 1400 and 5000 show that the characteristic atomic number density for the ionized gas varies in the range $(1\times10^8-3\times10^{10})$ atoms m$^{-3}$ ($1.8\times10^{-19}-5\times10^{-17}$ kg m$^{-3}$ ), giving a total ionized mass of around $5.8 \times 10^{-2} M_{\odot}$ and a ring temperature of around $(3-8) \times 10^4$ K \citep{Lundqvist:1991p15005, Mattila:2010p18200}. 

In similar fashion to \citet{Dewey:2012p25187} we use a two-component ring model consisting of a smooth equatorial ring interspersed with dense clumps. The smooth ring begins at $s^{\prime}_{\mathrm{er,inner}}=5.95\times10^{15}$ m ($0\farcs8$), and is centred on $(s^{\prime}_{\mathrm{er}}=6.4 \times 10^{15}, z^{\prime}=0)$ m ($0\farcs86, 0\farcs0 $). In order to approximate a crescent torus, as suggested in \citet{Plait:1995p25}, we adopt a raised Gaussian profile for the inner edge, where the cylindrical radius delineating the inner edge, $s^{\prime}_{\mathrm{inner}}$, is a function of the height $z^{\prime}$ from the equatorial (ring) plane.
The width (in the $z^{\prime}$ direction) of the Gaussian is $w_{\mathrm{er}}=9.4\times10^{14}$ m ($0\farcs12$) at a height of $z^{\prime}=w_{\mathrm{er}}/2$, and the Gaussian asymptotes to the inner edge of the hourglass $s^{\prime}=1.0\times10^{16}$ m $(1\farcs3)$ for large $z^{\prime}$.  As the filling factor of the ring is uncertain, we delineate the outer edge of the smooth ring using the same Gaussian profile as for the inner edge but translated outwards by a width $w_{\mathrm{er,eff}}=4.6\times10^{14}$ m ($0\farcs06$) in  $s^{\prime}$.  

Within the boundaries of the inner and outer edges of the smooth ring we specified the pressure and density profiles using another raised Gaussian function as a function of the minimum distance from the ring locus at $s^{\prime}_{\mathrm{er}},z^{\prime}=0$. The FWHM was set to $w_{\mathrm{er}}$ and the floor of the Gaussian was set to the density and temperature of the hourglass. The density and pressure were truncated at the values of the surrounding H\textsc{ii} material to ensure a smooth transition. The peak number density and temperature of the smooth ring was set to 
$8.0\times10^8$ m$^{-3}$ and $2.0\times10^4$ K. At the innermost edge of the ring the number density and temperature are 
$4.0\times10^8$ m$^{-3}$ and $2.1\times10^4$ K. At the outermost edge of the ring in the equatorial plane, the density and temperature are at peak values. The total mass of the smooth ring is $6\times10^{-3} M_{\odot}$.
Within the smooth ring we place 20 dense clumps of material, centred on the ring at a radius of $6.4\times10^{15}$ m and evenly distributed in azimuth. Each clump has a diameter of $4.5\times10^{14}$ m, a peak density of 
$3.1\times10^{10}$ m $^{-3}$ and a peak temperature of $2.0\times10^{4}$ K. For the density and pressure profile we choose the FWHM of the Gaussian such that at the periphery, the density of each clump is 
$3.4\times10^8$ m$^{-3}$ and has a temperature of $2\times10^{4}$ K. The total mass of the dense clumps is $3.5\times10^{-2} M_{\odot}$. Along with the mass of the smooth ring, this is consistent with the $~5.8 \times 10^{-2} M_{\sun}$ currently estimated for the ionized material in the ring \citep{Mattila:2010p18200}.

\subsection{Summary of parameters}\label{parameters}

In Tables \ref{fixed_parameters} and \ref{fitted_parameters} and \ref{efstable} is a summary of fixed and fitted parameters describing the environment of the final model. Error estimates on the fitted parameters are based on the discretisation of parameter space used in the model search.

\begin{table}
\caption{Key fixed parameters} 
\centering
\begin{tabular}{ll}
\hline \hline
Description & Parameter \\
\hline
Length of the grid (m) &  $3.36\times10^{16}$  \\
Inclination of the environment &  $i_x=41^{\circ}$  \\
& $i_y=-5^{\circ}$ \\
& $i_z=-9^{\circ}$ \\
Ratio of specific heats & $\gamma=5/3$ \\
Plasma particle mass (amu) & $\mu=0.678$ \\
\hline
Initial supernova radius (m) & $r_{sn}= 2.7\times10^{15}$ \\
Index on BSG wind density profile & $s=2$ \\
Index on supernova envelope & \\
density profile & $n=9$ \\
Supernova energy (J) & $E_{\mathrm{tot}}=1.5\times10^{44}$ \\
Supernova envelope mass (kg) & $M_{\mathrm{env}}=1.99\times10^{31} $  \\
Ratio of kinetic to total energy & $\chi=0.9$ \\
\hline
BSG mass loss rate (kg s$^{-1}$) & $\dot{M}=4.7\times10^{15}$  \\
BSG wind velocity (m/s) & $v_w=4.5\times10^5$ \\
Ratio of polar to equatorial distances & \\
for BSG wind termination shock & 1.18  \\
Distance to Mach disk (m) & $1.3\times10^{16}$ \\ 
\hline
Radius describing inner profile \\
 of HII region (m) & $9.9\times10^{15}$ \\ 
Height (above equatorial plane) &  \\
of inner profile of HII region (m) & $2.53 \times10^{15}$ \\
Temperature of the HII region  & \\
and hourglass (K) & $8.0\times10^4$ \\
\hline
Hourglass number density m$^{-3}$ & $8.77\times10^{6}$  \\
Minimum background & \\
number density (m$^{-3}$) & $3.5\times10^5$  \\
\hline
Equatorial ring radius (m) & $r_{er}=(6.4\pm0.8) \times 10^{14}$  \\
Equatorial ring width (m) & $w_{er}=(9.0\pm1.6) \times 10^{14}$  \\
Equatorial ring number density, & \\
(smooth component) (m$^{-3}$) & $8.0\times10^8$  \\
Equatorial ring temperature (K) & $2.0\times10^4$  \\
Equatorial ring clump \\
peak number density m$^{-3}$ & $3.1\times10^{10}$ \\
Equatorial ring clump \\
peak peak temperature (K) & $2\times10^4$ \\
Total mass of 
ring clumps (kg) & $7.0\times10^{28}$ \\
\hline
\end{tabular} \label{fixed_parameters}
\end{table}

\begin{table}
\caption{Fitted parameters} 
\centering
\begin{tabular}{ll}
\hline \hline
Description & parameter \\
\hline
Supernova envelope asymmetry & $\chi_{\mathrm{asym}}=1.55 \pm 0.05$ \\
BSG wind & \\
termination shock (m) & $R_{\mathrm{ts}}=(3.5 \pm 0.37) \times10^{15}$ \\
Inner boundary & \\
of HII region (m) & $R_{\mathrm{HII}}=(4.71 \pm 0.07) \times 10^{15}$  \\
Peak number density & \\
 of HII region (m$^{-3}$) & $(7.11\pm1.78) \times 10^7$  \\
$z^{\prime}$ FWHM of HII region (m) & $z^{\prime}_{\mathrm{FWHM}}=(2.52\pm0.88) \times 10^{15}$ \\
$s^{\prime}$ FWHM of HII region (m) & $s^{\prime}_{\mathrm{FWHM}}=(2.25\pm0.37) \times 10^{15}$ \\
HII region half opening angle & $15\pm5^{\circ}$ \\
\hline
\end{tabular} \label{fitted_parameters}
\end{table} 

\subsection{Modelling radio and thermal emission processes} \label{model_radio_thermal}

We assume a population of ultra-relativistic particles is produced at both forward and reverse shocks via diffusive shock acceleration, where particles gain energy by repeatedly sampling the converging flows at a strong shock front. Frequent scattering on magnetic fluctuations maintains a near isotropic distribution, ensuring that, on average, a particle will cross the shock many times before escaping downstream. In the absence of non-linear effects, this results in a uniform power-law spectrum in momentum space $f(p) = \kappa p^{-b}$, where $\kappa$ is a normalisation term. The distribution extends over several decades in energy. These ultra-relativistic electrons cool via synchrotron radiation, and the emission can typically be observed in the radio band. In our simulations, we determine the synchrotron radio emission, by calculating the volume emissivity $J(\nu)$ and absorption coefficient $\chi(\nu)$ directly from the particle momentum distribution $f(p)$ at shocked voxels. We assume a randomly-oriented magnetic field, and inject it at the shock using the analytic estimates for cosmic-ray driven magnetic field amplification. Then we follow its adiabatic evolution downstream. For the emissivity and absorption we use the expressions for $J(\nu)$ and $\chi(\nu)$ given in \citet{longair_vol2:1994}. The synchrotron emissivity, in units of $\mathrm{Watts} \ \mathrm{m}^{-3} \ \mathrm{Hz}^{-1} \ \mathrm{sr}^{-1}$ is
 
 \beq \label{syn_emiss}
 J_{\mathrm{syn}}(\nu)=\frac{\sqrt{3}e^3 B \kappa c^{(b-4)}}{4\pi \epsilon_0 m_e } \left ( \frac{3 e B}{2 \pi \nu m_e^3 c^4} \right )^{\frac{b-3}{2}} a_{1}, 
 \eeq
 
 where $a_{1}$ is given in terms of the Gamma function $\Gamma$ as
 
 \beq
 a_{1}(b)=\frac{\sqrt{\pi}}{2} \frac{\Gamma \left( \frac{b-2}{4} + \frac{19}{12} \right )\Gamma \left( \frac{b-2}{4} - \frac{1}{12} \right )\Gamma \left( \frac{b+3}{4} \right )}{(b-1)\Gamma \left( \frac{b+5}{4} \right )}.
 \eeq
 
 The synchrotron absorption coefficient, in units of $m^{-1}$ is
 
 \beq
 \chi_{\mathrm{syn}}(\nu)=\frac{\sqrt{3 \pi} e^3 \kappa c^{(b-2)} B^{\frac{b}{2}} }{16 \pi \epsilon_0 m_e} \left ( \frac{3 e}{2 \pi m_e^3 c^4} \right )^{\frac{b-2}{2}} a_{2} \nu^{-\frac{b+2}{2}}, 
 \eeq
 
 where $a_{2}$ is
 
 \beq
 a_{2}(b)=\frac{\Gamma \left( \frac{b-2}{4} + \frac{11}{6} \right )\Gamma \left( \frac{b-2}{4} + \frac{1}{6} \right )\Gamma \left( \frac{b+4}{4} \right )}{\Gamma \left( \frac{b+6}{4} \right )}.
 \eeq

The parameters $\kappa$ and $b$ are calculated at the shock location at each time-step, using the dynamically determined shock-jump conditions. The magnetic field intensity $B$ is estimated from the local shock parameters, using the saturated magnetic field amplification estimates and evolved downstream, together with the distribution of shocked particles. Details of the model are discussed in the following subsections. In order to produce an effective comparison with monitoring observations we calculate synchrotron emission and absorption at frequencies $843$ MHz and $1.38$ GHz. Synchrotron cooling can be safely neglected, as the loss timescale for microwave emitting electrons is on the order of $10^4$ years, much longer than the dynamical timescale being studied here.

\subsubsection{Shock localisation}

Within the diffusion approximation, the shape of the power-law spectrum produced from shock acceleration depends solely on the compression ratio of the shock, which can be determined from the shock velocity and the upstream plasma conditions. The ability to accurately locate shock positions in the hydrodynamical simulation is clearly a necessity. In the shock rest-frame, fluid of density $\rho_1$, pressure $P_1$ enters from upstream with velocity $v_1$  and exits down-stream with $\rho_2$,$P_2$ and velocity $v_2$. The compression ratio of a shock, $\zeta=\rho_2/\rho_1$, can be related to the pressure ratio ${P_2}/{P_1}$ using the Rankine-Hugoniot shock relations \citep{Landau2007Fluid}. 
\beq \label{comprat}
\zeta=\frac{(\gamma-1)+(\gamma+1)\frac{P_2}{P_1}}{(\gamma+1)+(\gamma-1)\frac{P_2}{P_1}}
\eeq
For $\gamma={5}/{3}$ and a strong shock ${P_2}/{P_1}>>1$, and the compression ratio asymptotes to 4.

 In hydrodynamical simulations the shock is not a thin discontinuity, but is spread over a region several cells wide. In order to locate shocks within the simulation we have adapted the shock locator that FLASH 3.2 uses to switch on a hybrid Riemann scheme in the presence of a shock \citep{Fryxell:2000p299}. It works by finding (via the velocity divergence) voxels where fluid is being compressed . If the local pressure gradient is greater than a threshold value then a voxel is deemed to be in a shock. This is not sufficient to find points outside a shock, so the pressure gradient is followed upstream and downstream until the gradient relaxes at points $\rho_1, \rho_2, P_1, \mathrm{and} P_2$. Details of this technique are in Appendix \ref{advection_shock_localise}. Once the upstream and down-stream variables have been determined, the shock compression ratio associated with a shocked voxel is derived from Equation \ref{comprat}. As $\zeta=v_1/v_2$ from the shock relations, the inbound fluid velocity in the shock frame $v_1$, is obtained in terms of the lab frame velocities $v_{1L}$ and $v_{2L}$

\beq
v_1=\left | (v_{1L}-v_{2L})\frac{\zeta}{\zeta-1} \right | \label{v1_eq}.
\eeq
 
\subsubsection{Magnetic field amplification}


We assume the magnetic field energy density upstream of the shock is amplified via the Bell instability \citep{Bell:2004p12496}, which has been shown in numerical simulations to amplify fields by more than an order of magnitude. This is achieved through the stretching of magnetic field lines, driven by the cosmic-ray current $\textbf{j}_{\rm cr}$, which accelerates the background plasma via the $\textbf{j}_{\rm cr}\times\textbf{B}$ force. Hence, the free energy available to amplify magnetic fields, is some fraction of the cosmic-ray energy density $U_{\rm cr}$. 

If $\mu_0$ is the permeability of free space (in SI units) then \citet{Bell:2004p12496} relates the magnetic field energy density to the cosmic ray energy density as
\beq
\frac{B_{sat}^2}{2 \mu_0} \approx \frac{1}{2}\frac{v_1}{c}U_{cr} \label{bmag_energy_dens},
\eeq
We define an efficiency factor $\eta_{\mathrm{cr}}$ 
\beq
U_{\mathrm{cr}} v_2 =\frac{\eta_{\mathrm{cr}}}{2} \rho_1 v_1^3\label{ucr_eq}.
\eeq
such that
\beq
B_{sat} \approx \sqrt{\frac{1}{2} \mu_0 \eta_{\mathrm{cr}} \rho_1 \frac{v_1}{v_2} \frac{v_1^3}{c}}=\sqrt{\frac{1}{2} \mu_0 \eta_{\mathrm{cr}} \rho_2 \frac{v_1^3}{c}} \label{B_vs}.
\eeq
In practice we found that $v_1$, as calculated from Equation \ref{v1_eq}, is not very stable due to the finite width of the numerical shock. This consequently dampens the response to changes in shock speed from abrupt changes in the upstream environment. We therefore adopt a more conservative approach where $v_1$ is approximated from the lab-frame shock velocity  $\textbf{v}_{2,L}$ and the shock normal \textbf{n} by $v1\approx \textbf{v}_{2,L} \cdot \hat{\textbf{n}} \frac{\zeta}{\zeta-1} $. The shock normal is derived from the pressure gradient, and the saturated magnetic field is approximated by
\beq
B_{sat} \approx \sqrt{\frac{1}{2} \mu_0 \eta_{\mathrm{cr}} \rho_2 \left [ (\textbf{v}_{2,L} \cdot \hat{\textbf{n}}) \frac{\zeta}{\zeta-1} \right ]^2 \frac{v_1}{c}} \label{b_sat_v1}.
\eeq




Supernova remnants are generally thought to be the primary source of Galactic cosmic rays, which requires an acceleration efficiency for protons and other heavy nuclei of $\eta_{\mathrm{cr}}\approx0.1$  \citep{Bell:2004p12496,Volk:2005p19073}, in order to satisfy current measurements. We adopt this value for all our calculations in the paper. 
%

\subsubsection{Acceleration of electrons at the shock}

A precise treatment of diffusive shock acceleration over the entire remnant is not possible, and we are forced to use a reduced model for the acceleration of electrons at the shock front. 
The standard theory of shock acceleration predicts an acceleration time \citep{Drury83}
\beq 
t_{\rm acc} = \frac{3}{v_1-v_2}\left(\frac{D_1}{v_1}+\frac{D_2}{v_2}\right)
\eeq
where $D_{\rm 1,2}$ are the shock-normal spatial diffusion coefficients in the upstream and downstream regions. These coefficients are
typically taken to be Bohm-like, i.e. on the order $c^2/\Omega_{\rm g}$, where $\Omega_{\rm g}=eB/\gamma m$ is the electron relativistic gyro frequency.
Given that the peak in the synchrotron spectrum emitted from particles at a given Lorentz factor $\gamma$ is
\beq 
\nu_{\rm syn} \approx \frac{1}{4\pi} \gamma^3 \Omega_{\rm g}\enspace,
\eeq
it follows that the characteristic acceleration time for radio emitting electrons in the GHz range is shorter than our numerical time steps. This allows us to 
update the electron spectrum at every timestep in our simulations, such that a new spectrum is deposited at the shock location at each update.
In the simplest theory of Diffusive Shock Acceleration (DSA), the power-law index of the distribution, $b$ is related to the compression ratio by
\beq
b=\frac{3\zeta}{\zeta-1}.
\eeq

The index $b$ is related to the spectral index of radio emission $\alpha$ $(F(\nu) \propto \nu^{-\alpha})$ by $\alpha=\frac{b-3}{2}$. For a strong shock in our ideal monatomic gas the compression ratio is 4 and the spectral index from shock acceleration is $0.5$. Interestingly, this is not the case with the observed radio spectral index from SNR 1987A. Following the return of radio emission the spectral index was approximately $0.9$ around day $1500$ as the shock encountered the H\textsc{ii} region. It reached a peak of $1.0$ around day $2300$ and has since been hardening linearly, attaining $0.7$ at day $8000$ \citep{Zanardo:2010p17425}. A possible explanation is the that the compression ratio has been lowered due to the influence of cosmic rays on the upstream material. This hypothesis was investigated in \citet{Duffy:1995p18382, Berezhko:2000p56}, however there are problems such as arbitrary injection, stability of modified solutions and the effect of self consistent field amplification on cosmic rays. Alternatively, if cosmic ray pressure is not important, the electrons may be sub-diffusing. In a tangled magnetic field the mean square distance a particle sub-diffuses is instead proportional to time $t$ as $t^{1/2}$ \citep{Kirk:1996p18122}. The resulting index on the momentum is modified to
 
\beq
b=\frac{3\zeta}{\zeta-1} \left (1+\frac{1}{2\zeta} \right ).
\eeq
  
A tangled magnetic field is consistent with observations as significant polarisation is yet to be observed in SNR 1987A \citep{Potter:2009p16343}. 

The fraction of available electrons that were injected into the shock, $\chi_{el}$, while distinct from the acceleration efficiency, can be estimated with radio observations given assumptions about the injection momentum. 
Assuming the electrons are injected into the shock at a single momentum $\delta(p-p_0)$, it can be shown (e.g \citep{Melrose:2009p11879} ) that the isotropic downstream power law distribution of electrons, $f(p)=\kappa p^{-b}$, (where $\kappa$ is a constant) is defined between $p_0$ and the maximum momentum, which is taken to be indefinite. If the downstream density of energised electrons is a fraction $\chi_{el}$ of the electron number density $n_2$, then conservation of mass requires that $\chi_{el}n_2$=$\int_{p_0}^{\infty} 4 \pi p^2 f(p) dp$. Solving for $f(p)$ shows that
 
\beq
 f(p)=  \frac{\chi_{el} n_2 (b-3)}{4\pi} {p_0}^{b-3}p^{-b} .
\eeq

and therefore

\beq
\kappa = \frac{\chi_{el} n_2 (b-3)}{4\pi} {p_0}^{b-3}.
\eeq

We assume electrons are injected into the shock from downstream. The injection momentum is derived by assuming the electrons are in thermal equilibrium with the downstream plasma and are injected into the shock at the electron thermal velocity. By equating thermal energy to relativistic kinetic energy, then the injection momentum is given in terms of the  temperature at the downstream point $T_2$
\beq 
p_0(T_2)= m_{e} c \sqrt{\left (\frac{ \frac{3}{2}k_b T_2}{m_e c^2} + 1 \right )^2-1}\label{p_inj_temp_eq} \label{injection_momentum}.
\eeq

\subsubsection{Advection of the magnetic field}
\label{sec:advb}



Once the magnetic field has been amplified by the shock, we assume that it is frozen into the background flow, satisfying
\beq
\frac{d}{dt}\left(\frac{\textbf{B}}{\rho}\right) = \left(\frac{\textbf{B}}{\rho} \cdot \nabla\right)\textbf{u}\enspace.
\eeq 
Following \citet{Kirk:1994p18186}, we assume an homologous expansion inside the remnant, 
i.e. $\textbf{u} \propto r \hat{r}$, for which the ratio $\psi=B/ \rho^{2/3}$ is constant for a given fluid element. 

The method implemented to track $\psi$ is given in Appendix \ref{advection_appendix}. As the density is calculated in the main part of the hydro-code, the magnetic field can be reconstructed at a later time $t$ simply by multiplying $\psi$ by $\rho(t)^{2/3}$. 

\subsubsection{Advection of the particle distribution}
\label{sec:advf}

In order to track the evolution of the electron distribution in the downstream we follow the method of \citet{Duffy:1995p18382},
where the electrons are assumed frozen to the flow (i.e. diffusion is neglected). This allows us to simplify the transport 
equation
\beq
\label{eq:trans_nodiff}
\frac{\partial f}{\partial t}+\textbf{u} \cdot \nabla f - \frac{1}{3} (\nabla \cdot \textbf{u})p \frac{\partial f}{\partial p}=0.
\eeq

While this equation can in principle be solved using the method of characteristics, with
\beq \label{momentum_characteristic}
\frac{dp}{dt}=-\frac{1}{3} (\nabla \cdot \textbf{u})p.
\eeq
to reduce the numerical effort, we choose instead to replicate the approach used for the magnetic field advection. 
 
Equation (\ref{eq:trans_nodiff}) can be re-written in the form
\beq \label{full_fp_eq}
\frac{\partial f}{\partial t}+\nabla \cdot (\textbf{u} f) = (\nabla \cdot \textbf{u})f \left (1+ \frac{1}{3}  \frac{ \partial \ln f}{\partial \ln p} \right ).
\eeq
where it is immediately noticed that $\frac{ \partial \ln f}{\partial \ln p}$ is just the index of our power law $-b$. We update each component, $i$, of the two-point power law $f(p_{i})$ using the advection scheme in Appendix \ref{advection_appendix}. To preserve conservation of particle number we update the injection momentum $p_0$ by evolving it along the characteristic implied by Equation \ref{momentum_characteristic}.

 \subsubsection{Radiative cooling and thermal emission}\label{radiative_cooling}
 
\begin{figure}[htbp]
\epsfig{file=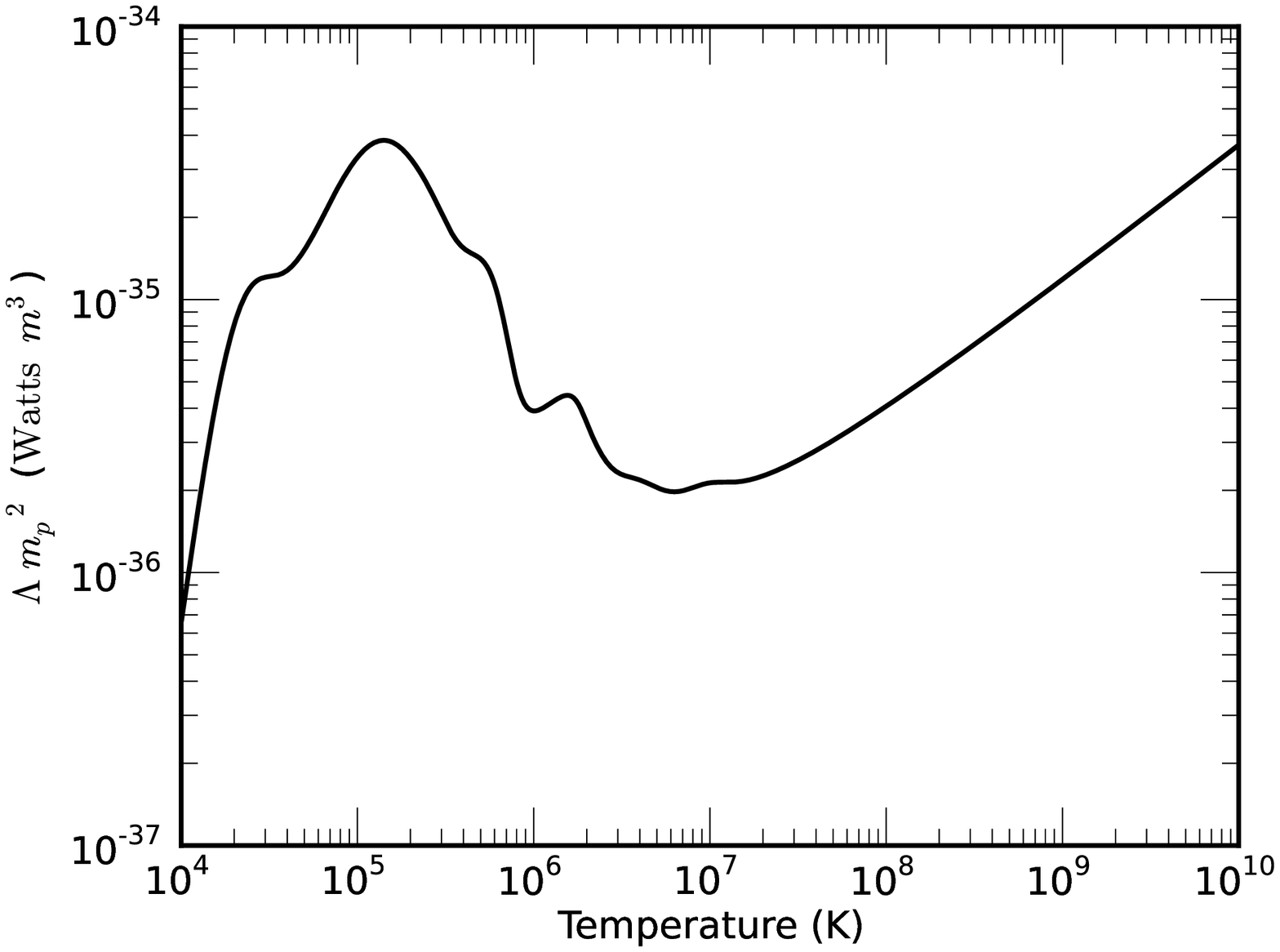,width=9.0cm}
\caption{The cooling function $\Lambda(T)$ multiplied by the square of the particle mass $\mu$. The function was derived from the abundances in \citet{Mattila:2010p18200} using the MAPPINGS shock and photoionization code \citep{Sutherland:2007p15906,Sutherland:2003p19568, Sutherland:1993p17000}}\label{cooling_curve}
\end{figure}

Unlike synchrotron emission, radiative cooling is implemented by converting a small fraction of the  available internal energy to thermal energy  as the simulation evolves. We constructed a temperature-dependent cooling function using  MAPPINGS \citep{Sutherland:1993p17000,Sutherland:2003p19568,Sutherland:2007p15906}. Figure \ref{cooling_curve} shows a plot of the cooling function multiplied by the particle mass squared ($\mu^2$).
 
If $\epsilon$ is the specific internal energy such that $P=(\gamma-1)\rho \epsilon$,  and $\Lambda(T)$ is the cooling function, then the evolution equation for the internal energy is given by 
\beq
\frac{d \epsilon}{dt}=\epsilon(\gamma-1)\frac{d \rho}{dt}-\rho \Lambda(T).
\eeq
As the $d\rho/dt$ term is already handled within FLASH through operator splitting, we complete the update to the internal energy through the first order ODE
\beq
\frac{d \epsilon}{dt } =- \rho \Lambda(T),
\eeq

which is solved using a fourth order Runge Kutta scheme. The thermal X-ray emissivity $J_{\mathrm{therm}}(T)$ from material is then 

\beq
J_{\mathrm{therm}}(T)= \frac{1}{4\pi}\rho^2 \Lambda(T).
\eeq
  
 \subsubsection{Radiative transfer}
 
Radiative transfer is implemented using the analytic solution of the radiative transfer equation to propagate the brightness across the grid in the direction of the observer. If $\Delta x$ is the width of each voxel then the analytic solution gives the brightness at the edge of each voxel in terms of the volume emissivity $J_{\rm syn}(\nu)$ and absorption coefficient $\chi_{\rm syn}(\nu)$ discussed in Section \ref{model_radio_thermal}
   
 \beq
 I_{v}=\frac{J_{\rm syn}(\nu)}{\chi_{\rm syn}(\nu)}[1-\exp{(-\chi_{\rm syn}(\nu)\Delta x)}].
 \eeq
 
After propagation across the grid, the flux density is obtained by multiplying by the apparent angular size of the voxel face as seen from Earth.

\section{Results and discussion}
 
\begin{figure*}[htbp]
\begin{center}
\begin{math}
\begin{array}{cc}
\includegraphics[width=8cm]{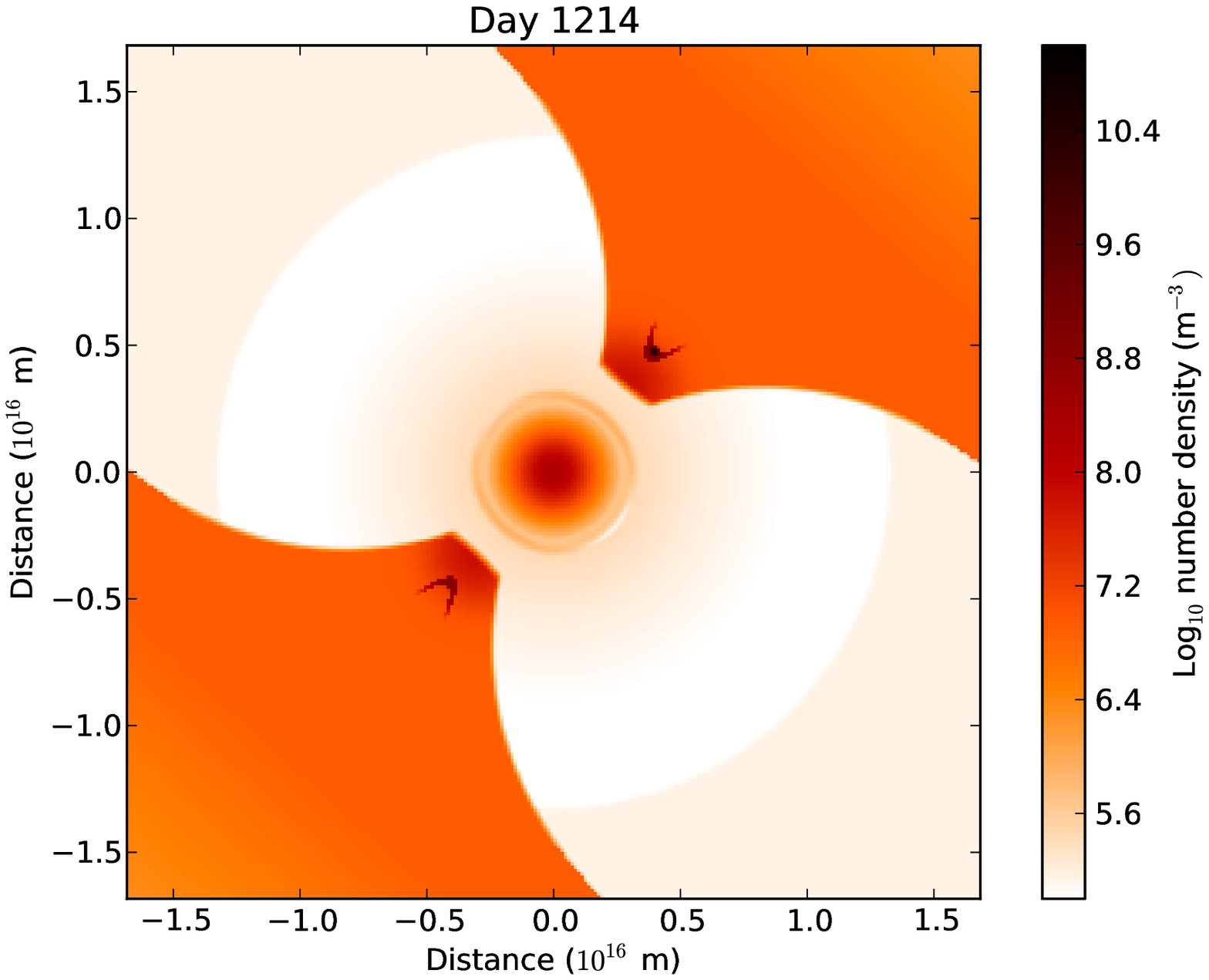} &
\includegraphics[width=8cm]{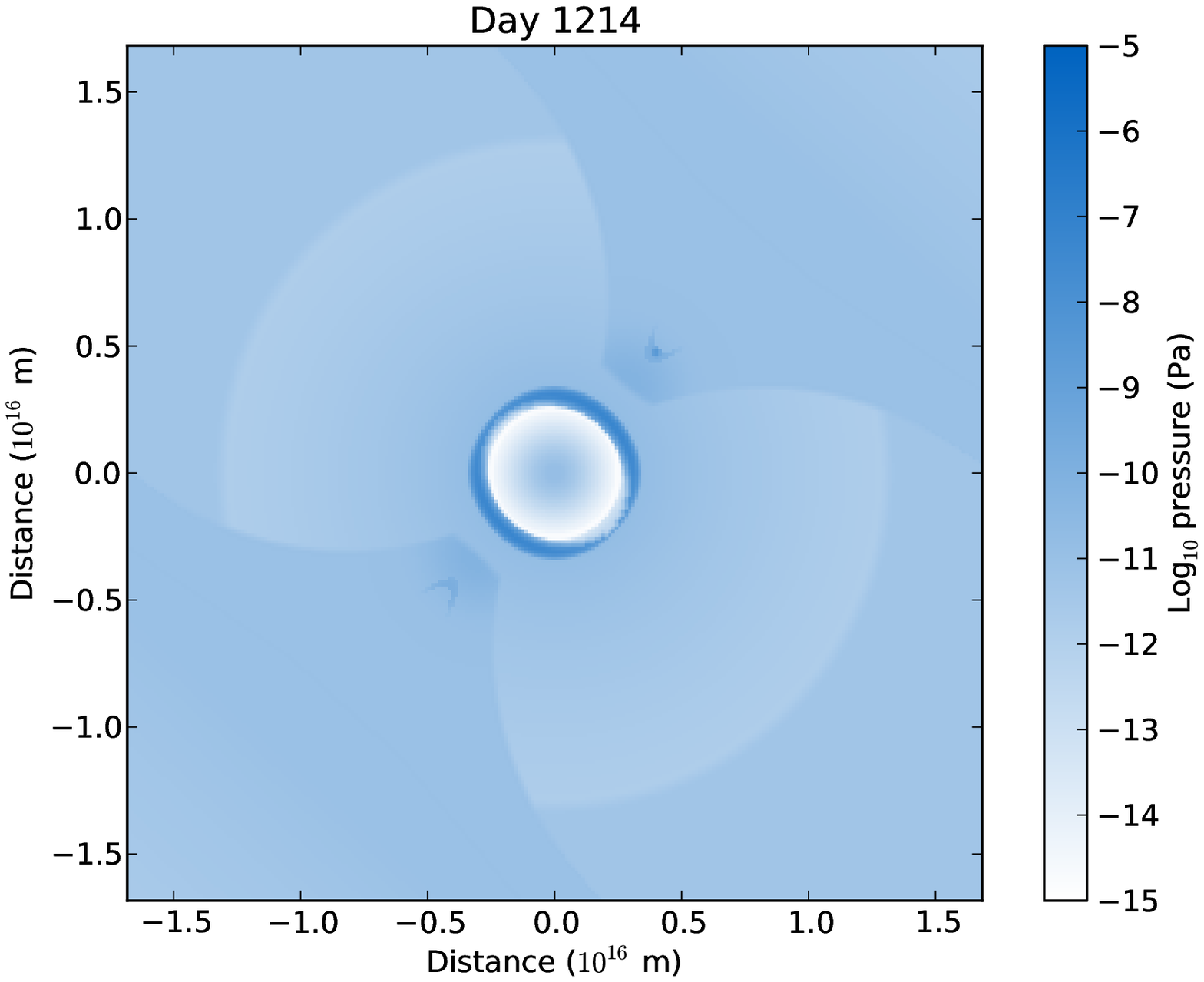} \\
\includegraphics[width=8cm]{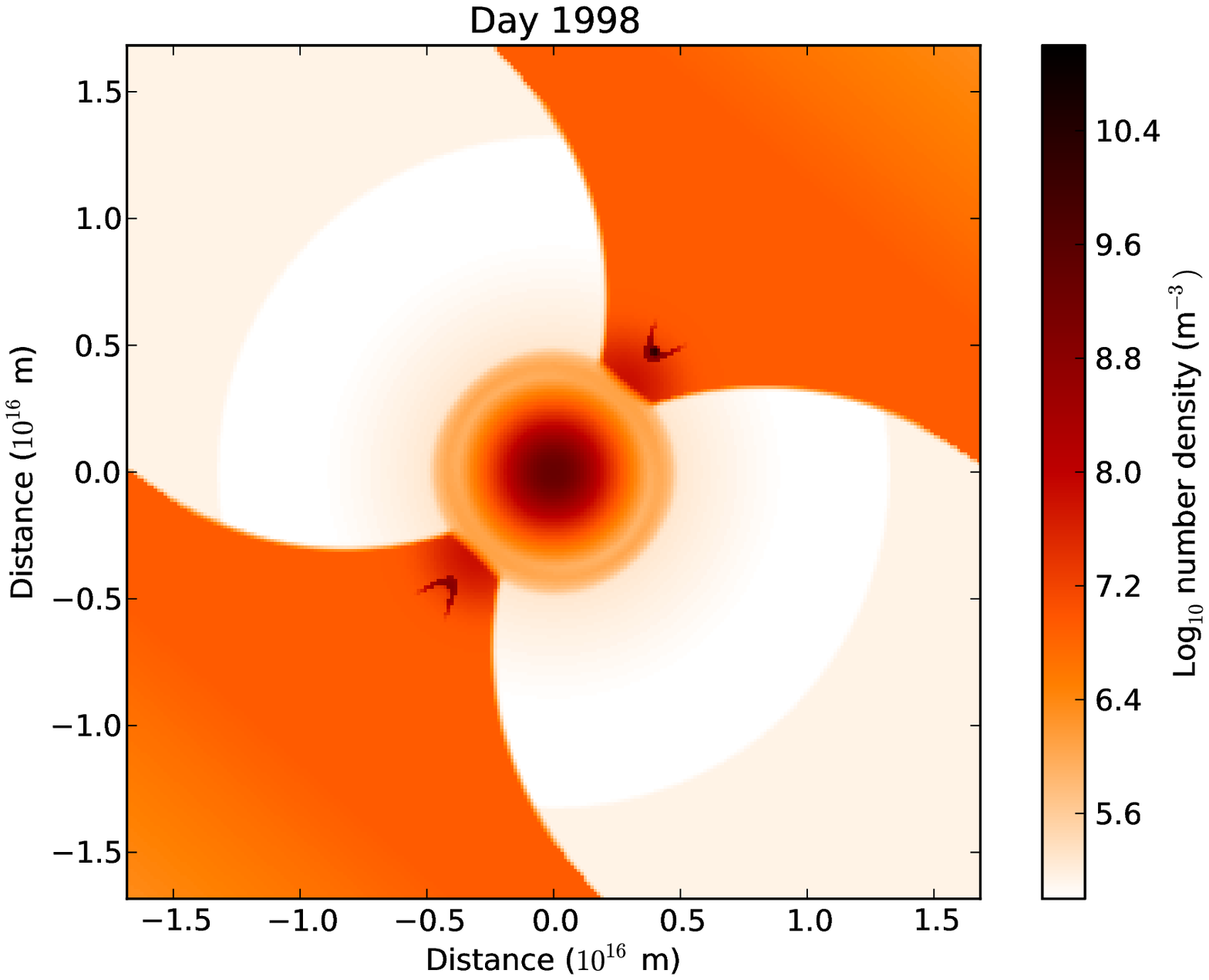} &
\includegraphics[width=8cm]{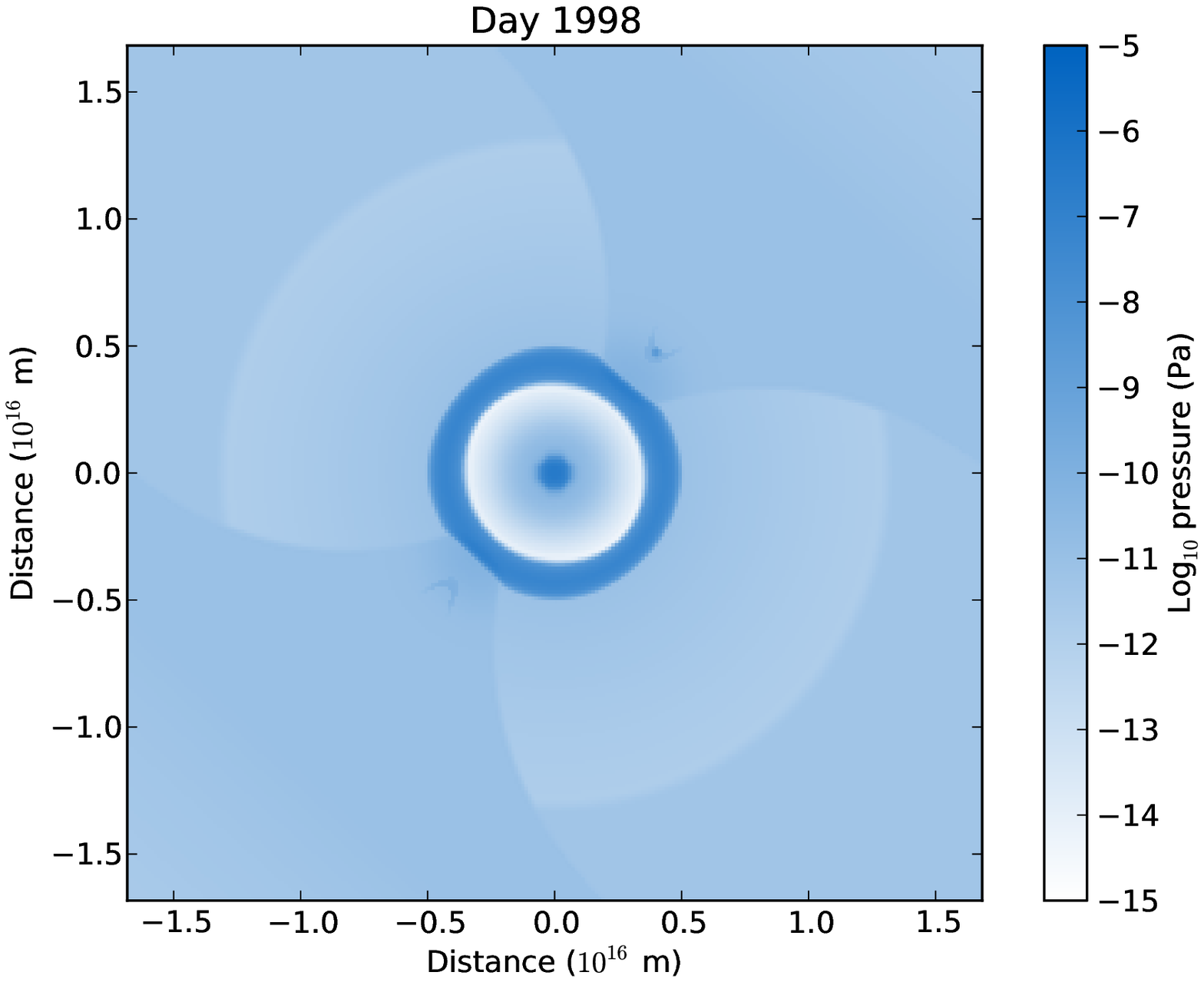} \\
\includegraphics[width=8cm]{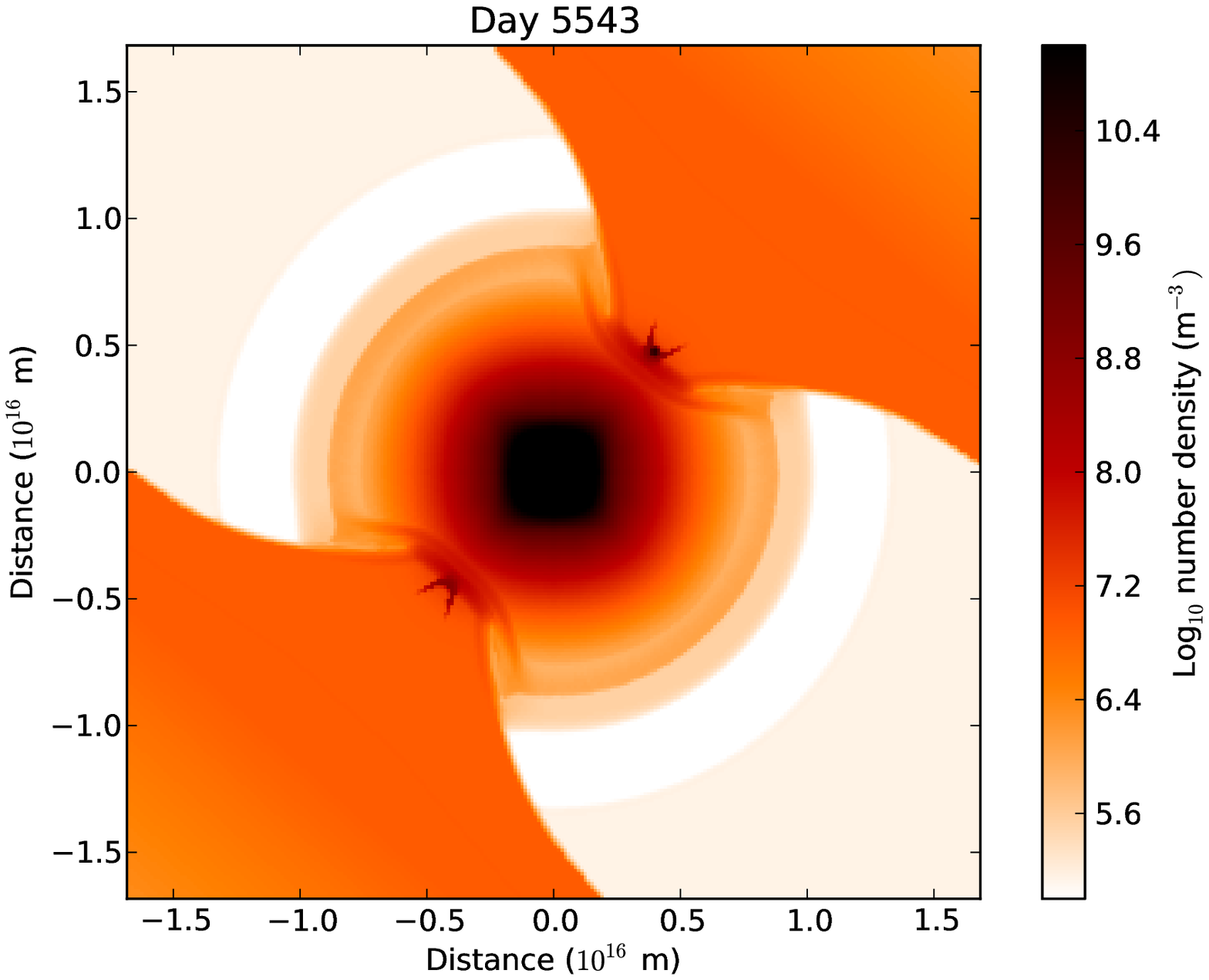} &
\includegraphics[width=8cm]{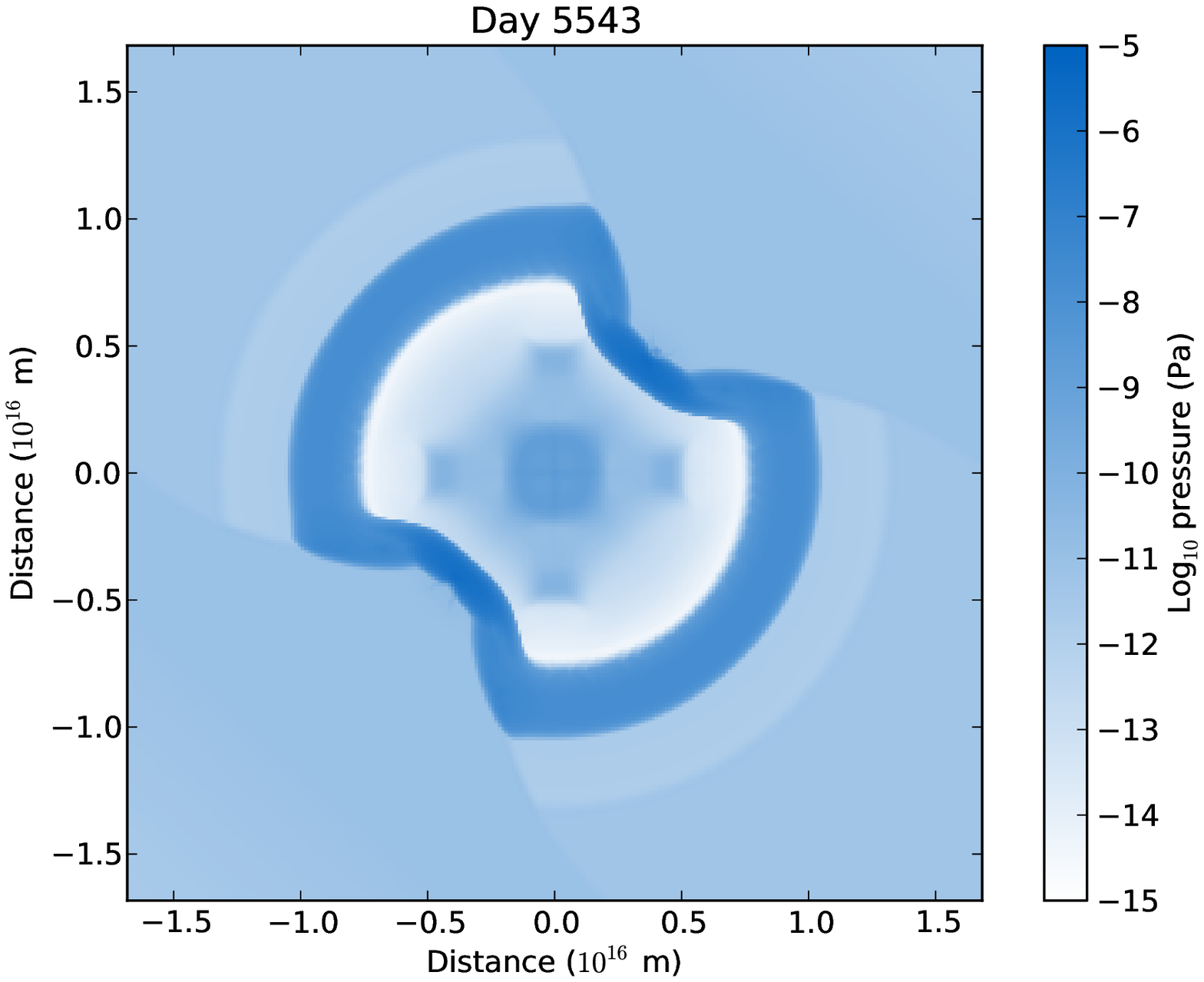} \\
\end{array}
\end{math}
\end{center}
\caption{Early epochs of the evolving shockwave from SN 1987A. Shown is the log of number density and pressure for a slice of the computational domain. Earth is to the right and the slice has been taken at 45\% along the $X$ axis in order to intersect one of the dense blobs in the ring.  In the top row is the interaction of the supernova shock with the inner edge of the hot BSG wind around day 1200. Around day $2000$ (middle row) the shocks begin to interact with the inner edge of the H\textsc{ii} region. By day 5500 (bottom row) the supernova shocks have begun interacting with the dense blobs within the ring.}\label{evolution_sequence1}
\end{figure*}

\begin{figure*}[htbp]
\begin{center}
\begin{math}
\begin{array}{cc}
\includegraphics[width=8cm]{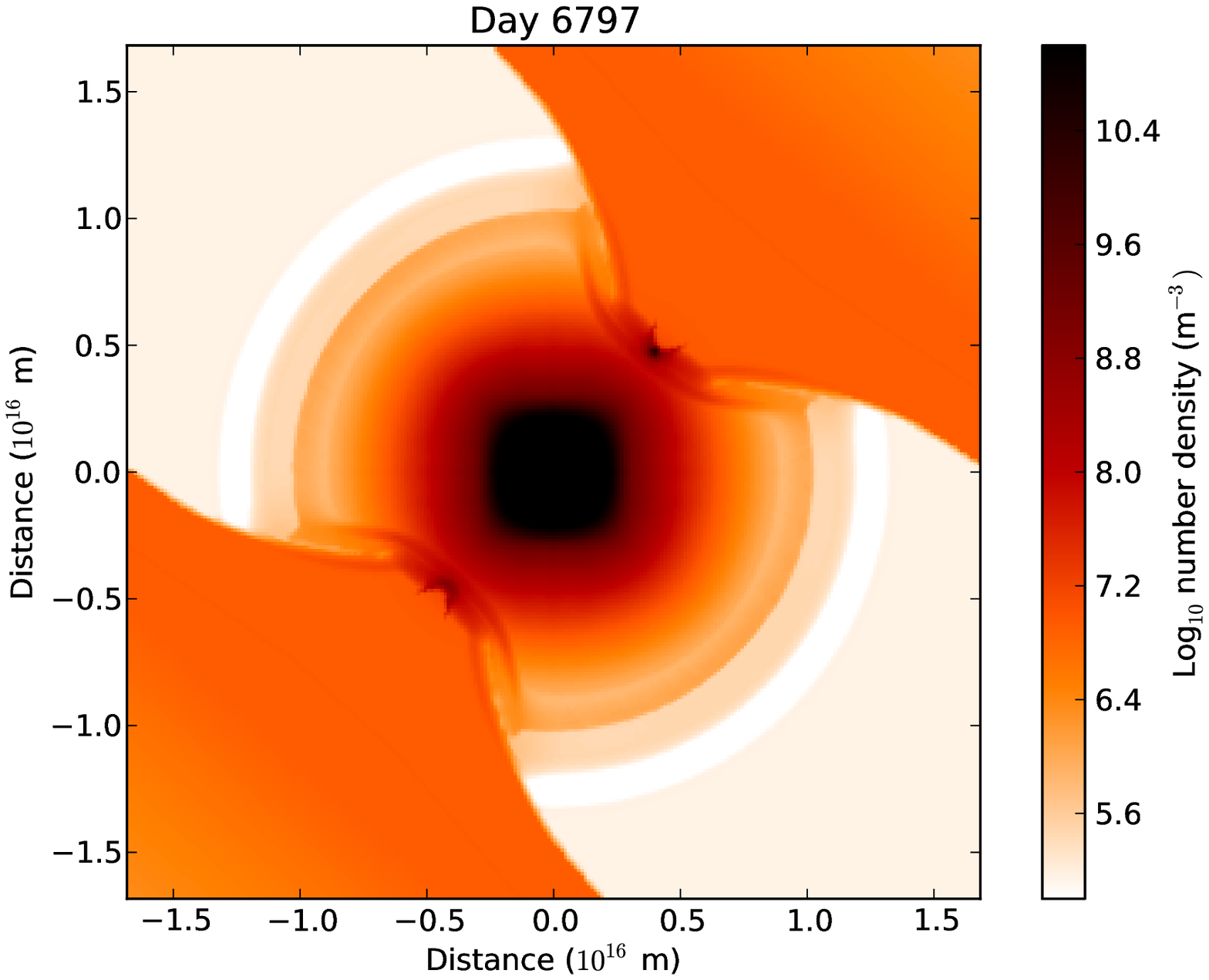} &
\includegraphics[width=8cm]{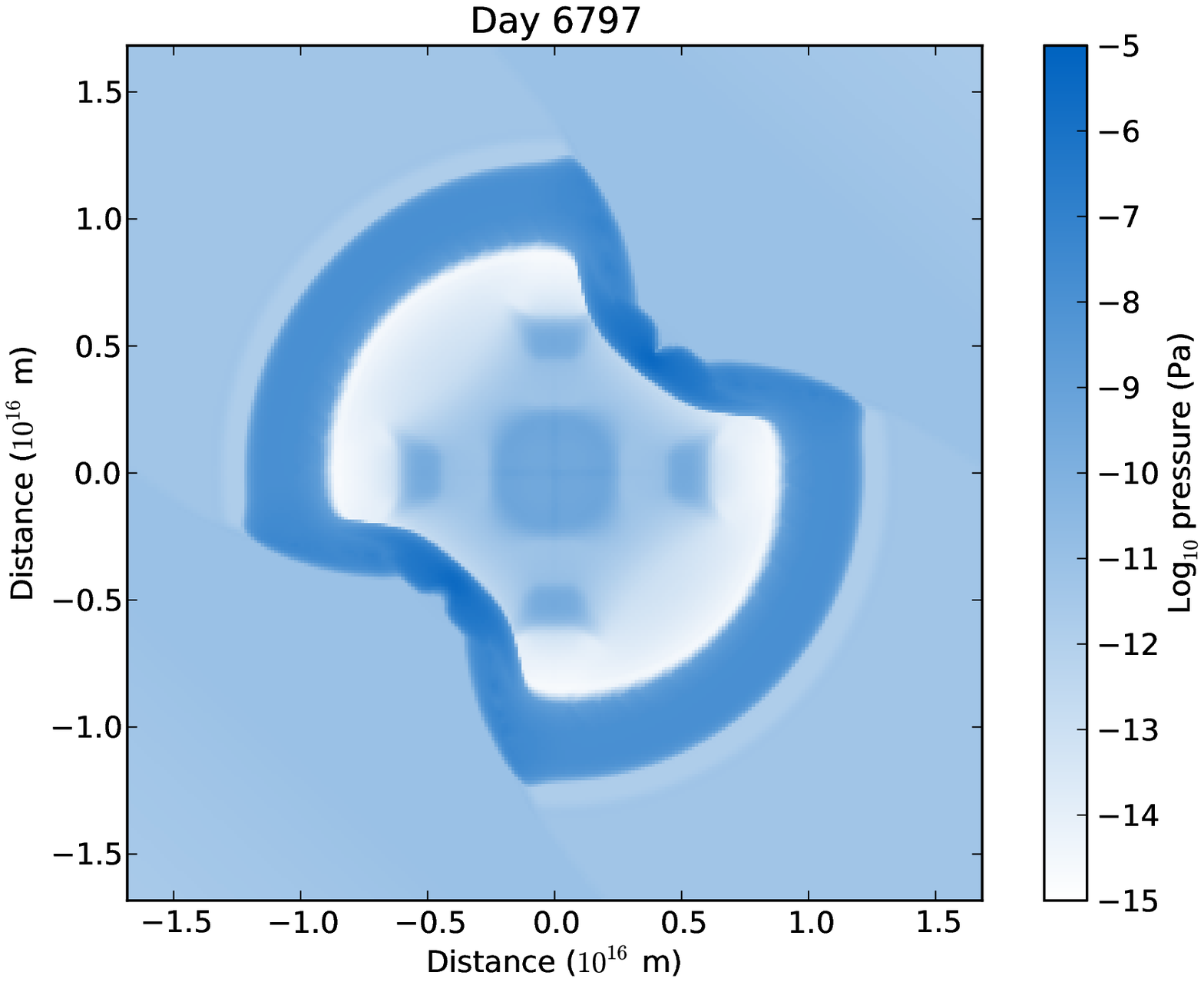} \\
\includegraphics[width=8cm]{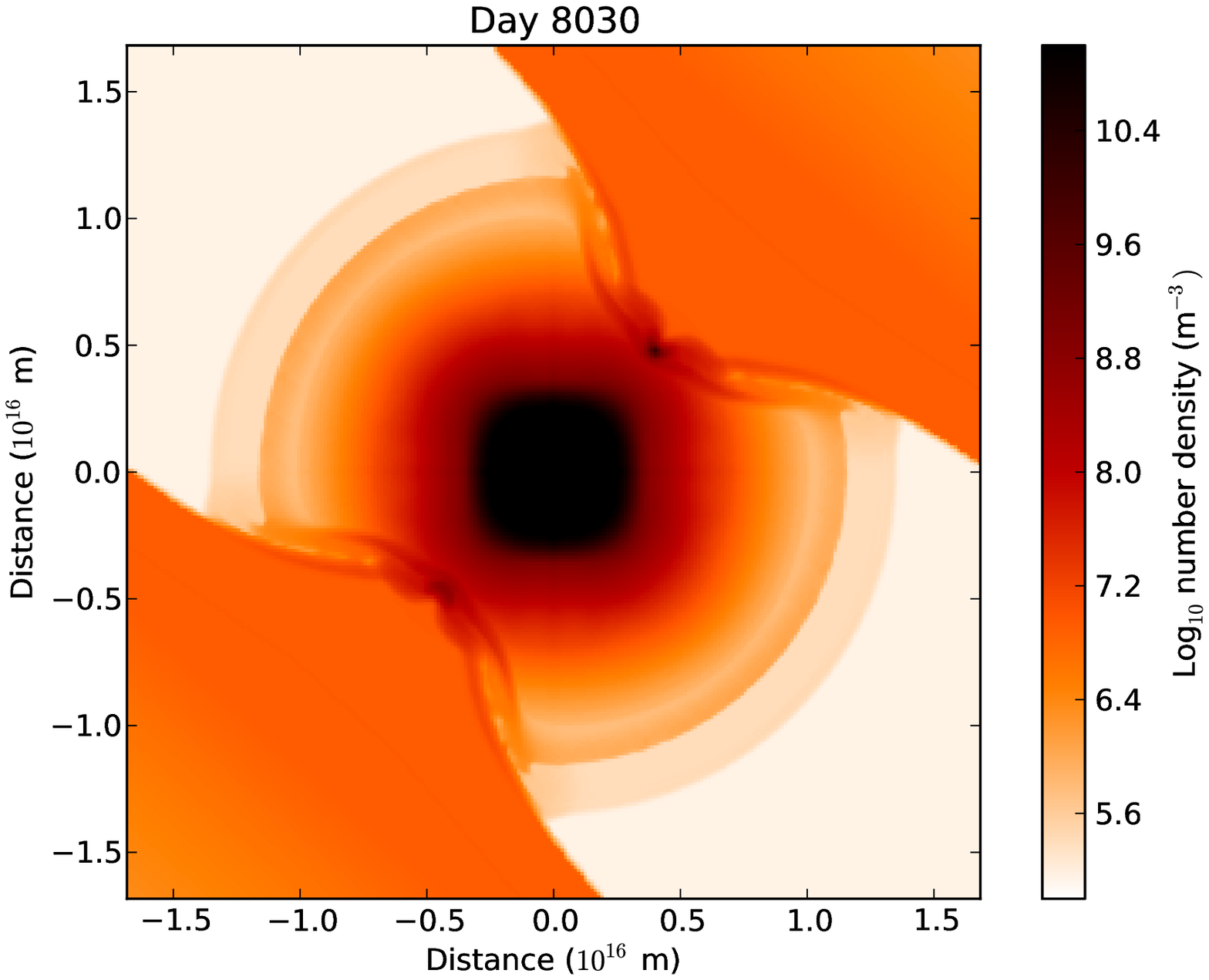} &
\includegraphics[width=8cm]{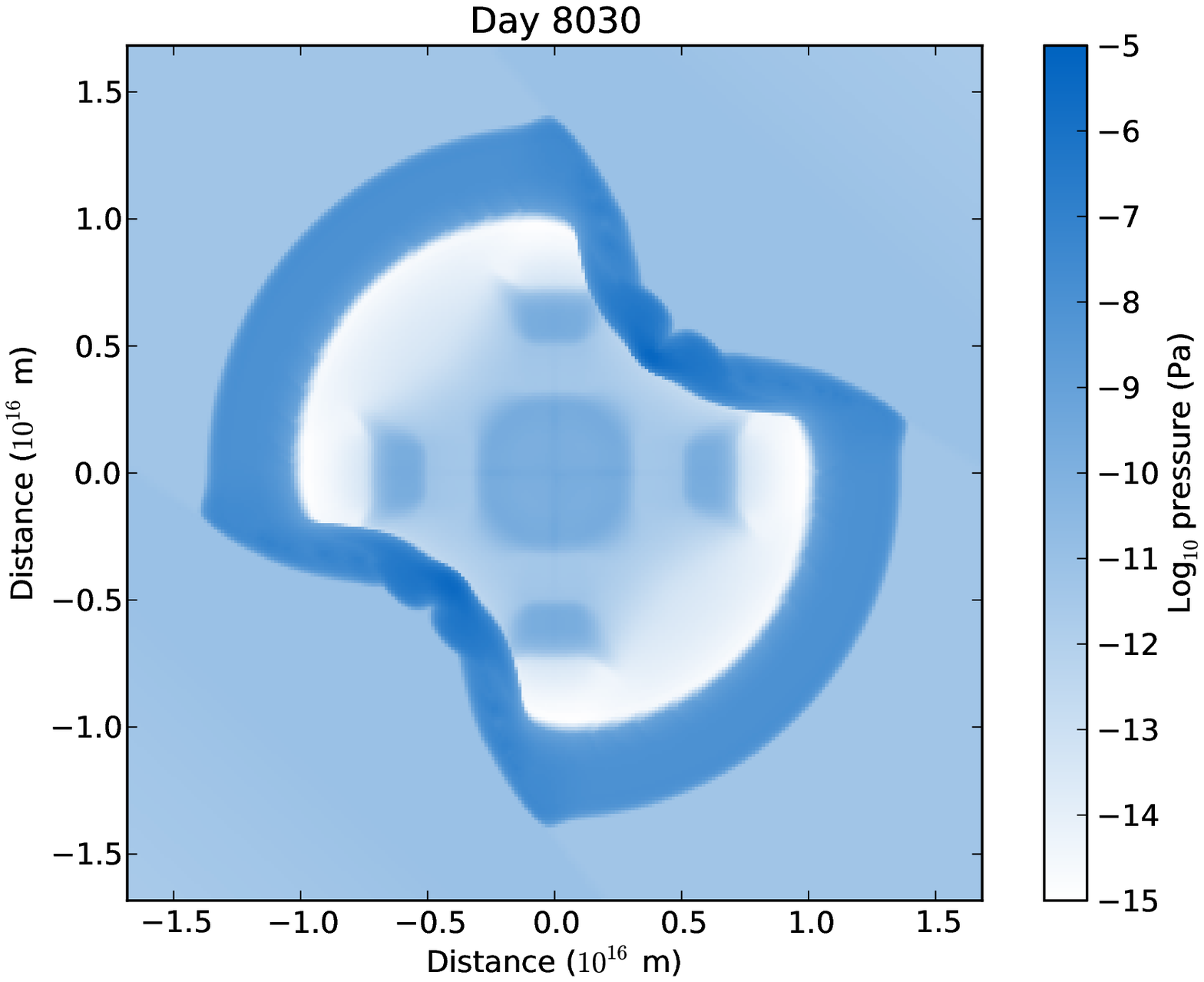} \\
\includegraphics[width=8cm]{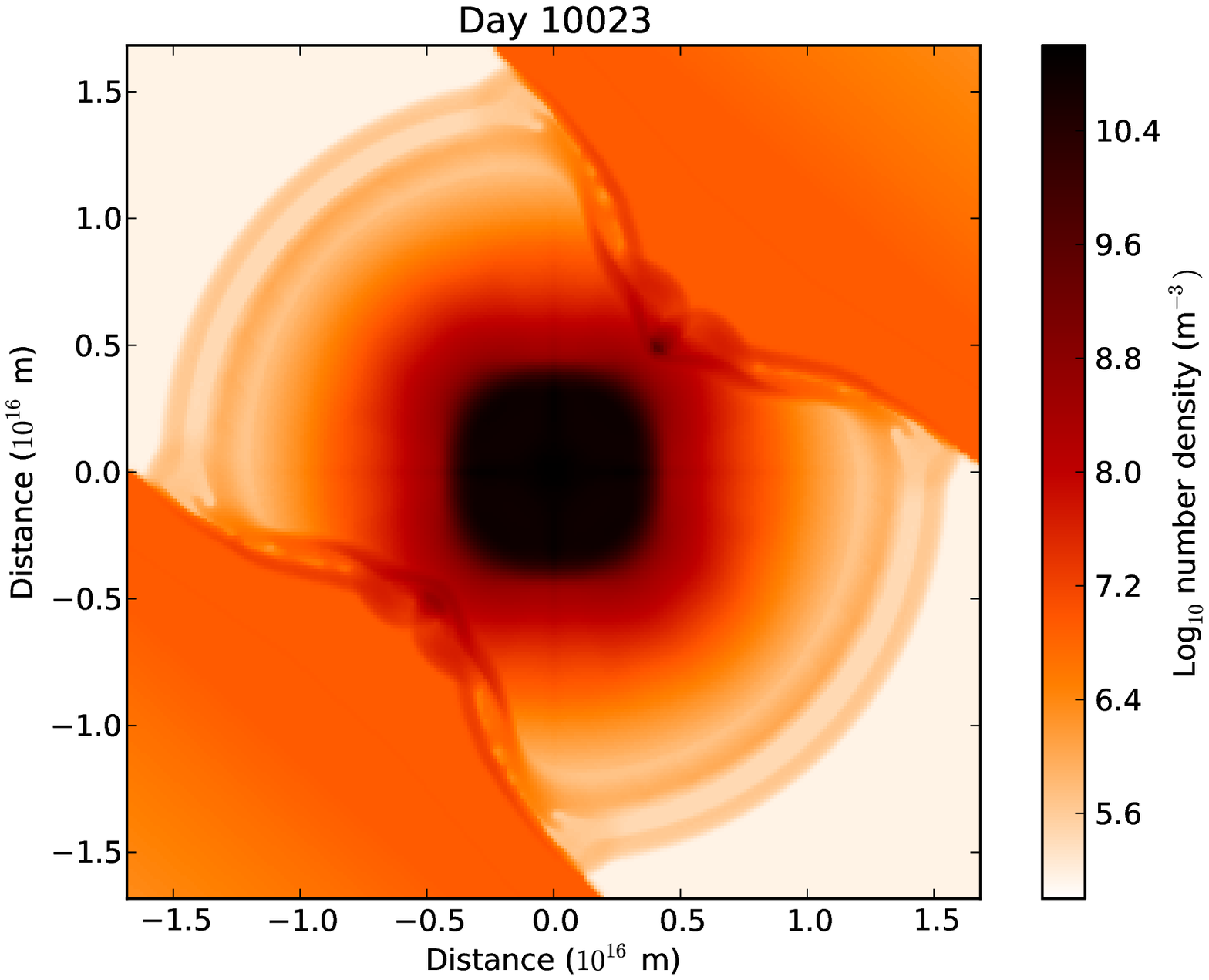} &
\includegraphics[width=8cm]{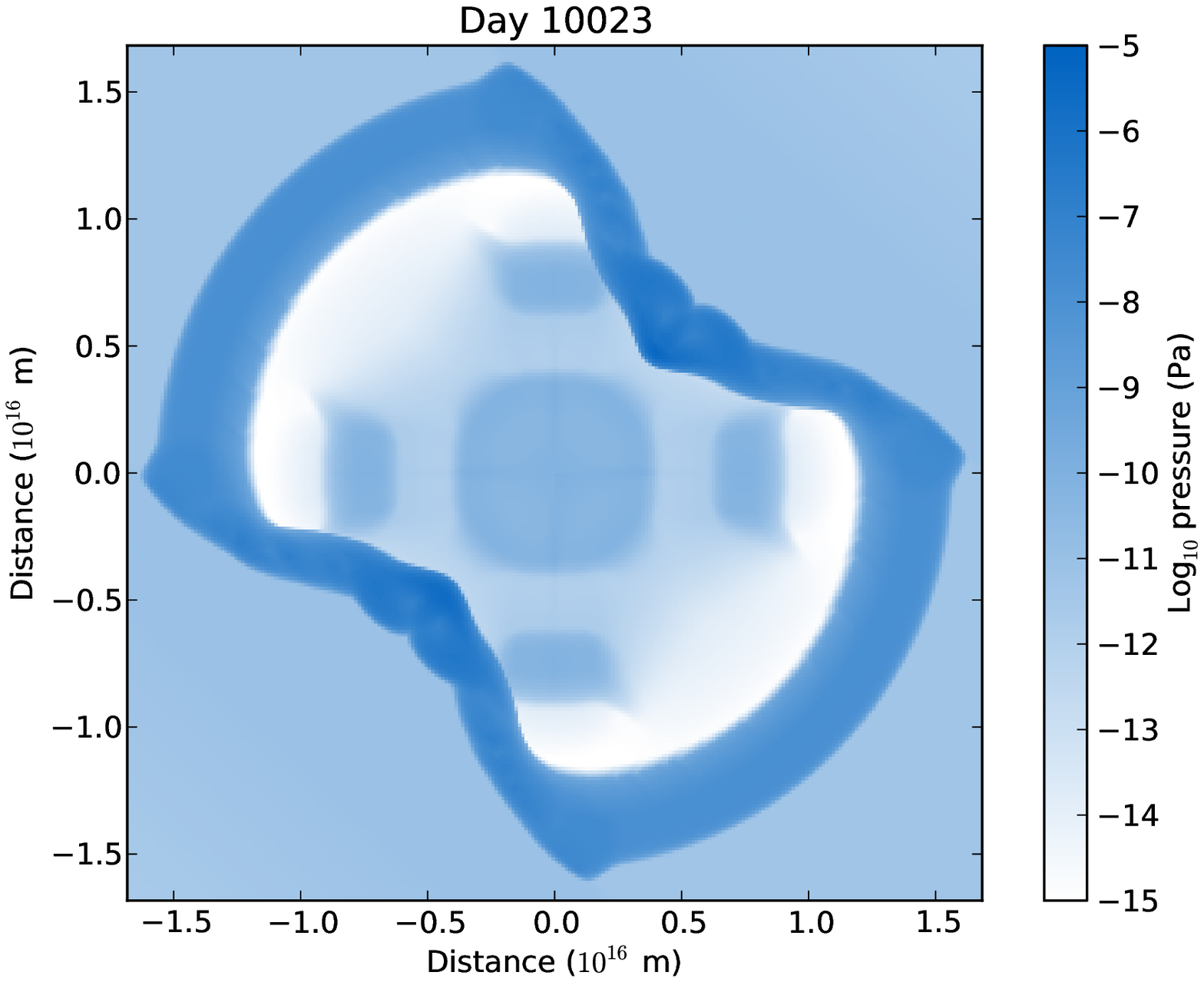} \\
\end{array}
\end{math}
\end{center}
\caption{Later epochs of the evolving shockwave from SN 1987A. Around day 6800 (top row), the supernova forward shock has almost completed its crossing of the ring. The reverse shock continues to interact with the highest density blobs within the ring. By day $8000$ (middle row), the forward shock has completely left the equatorial ring. The reverse shock continues to interact with the equatorial ring until after the end of the simulation at day $10,023$ (bottom row).}\label{evolution_sequence2}
\end{figure*}

The hydrodynamical simulations were evolved to day $10,023$ after the explosion, with an average timestep of 33 simulated days. In post-processing, distributions of accelerated electrons were placed in the downstream flow of the forward and reverse shocks and were advected with the flow. Radio emissivity at each timestep was generated from the electron distributions. At a resolution of $256^3$ simulations took approximately 4 hours to complete with 8 cores. In Figures \ref{evolution_sequence1} and \ref{evolution_sequence2} are slices of the log of density and pressure at a number of different simulated epochs. 

The plots were formed by taking a cut plane at around 45\% of the X axis total length. In the top row of Figure \ref{evolution_sequence1} the supernova shock reaches the hot BSG wind around day $1200$. After this encounter the shock splits into a forward shock, contact discontinuity, and a reverse shock. Around day 2000 (middle row) the forward shock encounters the H\textsc{ii} region and around day $5500$ (bottom row) the supernova forward shock begins encountering the equatorial ring. In the top row of Figure \ref{evolution_sequence2} the forward shock the forward shock has almost completed its crossing of the ring around day 6800. The reverse shock is beginning to encounter the highest density blobs within the ring. By day $8000$ (middle row), the forward shock has completely left the dense ring. The reverse shock continues to interact with the densest part of the ring until after the end of the simulation at day $10,023$ (bottom row).

\subsection{Shock radius}

\begin{figure*}[h]
\begin{center}
\epsfig{file=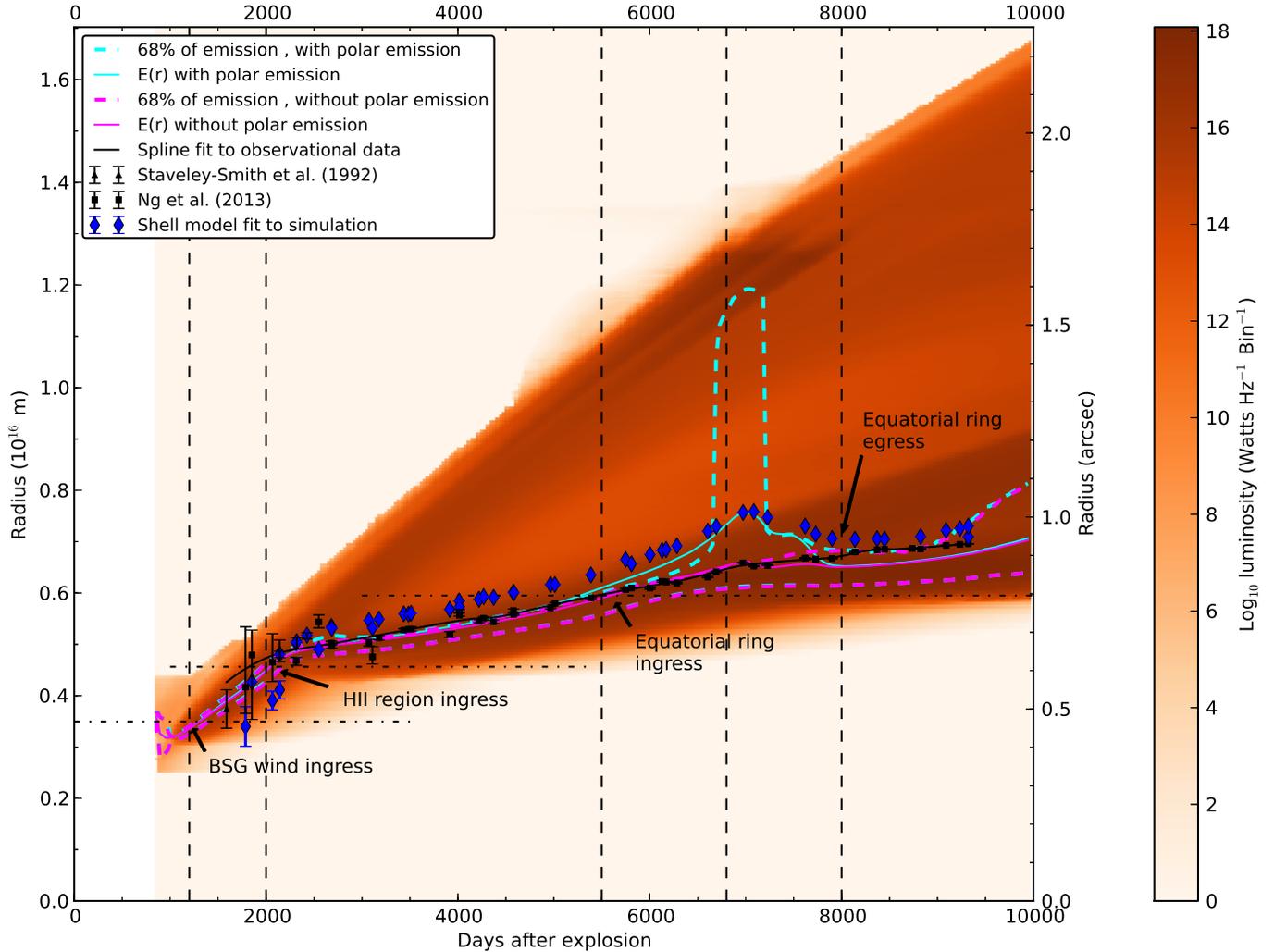,width=18.0cm} 
\caption{The evolving radial distribution of radio luminosity from the final simulation. In orange is the radial distribution at each timestep. Overlaid is the expectation of radius $E(r)$, and a boundary containing $68\%$ of the luminosity. This is plotted for distributions both with and without high latitude luminosity above a half-opening angle of $45^{\circ}$. The observed radius from \citet{StaveleySmith:1992p680,Ng:2008,Ng:2013p25184} and a spline fit to the observed data is also shown for comparison. Shown in blue diamonds is the shock radius formed by fitting the truncated shell model of \citet{Ng:2013p25184} to the simulated radio emission. The vertical lines correspond approximately to epochs where the forward shock encountered various hydrodynamic structures; the BSG wind at day $1200$, the HII region at day $2000$, the equatorial ring ingress at day $5500$, egress from the eastern and western lobes of the ring around days $6800$ and $8000$.} \label{shock_radius_vs_time} 
\end{center}
\end{figure*}

The expanding radius from the simulation was calculated from the 3D expanding morphology by deriving the radial distribution of radio luminosity at each timestep. The expectation value of the distribution forms an estimate of the radius. In this way we hope to determine the measured shock radius in the most general way possible that approximates model fits to the observational data from \citet{StaveleySmith:1992p680,Ng:2008,Ng:2013p25184}. The resulting radius curves are shown in Figure \ref{shock_radius_vs_time}. The orange background in the plot is the time-varying radial distribution of radio luminosity. The bin width for the distribution has been normalised by its representative width of $5.69 \times 10^{13}$ m. A sum over all dimensionless bins in the distribution produces the total luminosity of the remnant at each timestep. We used radio emission at a simulated frequency of $1.4$ GHz for the distribution of radio emission and calculations of the radius. 

Overlaid on the radio luminosity distribution in Figure \ref{shock_radius_vs_time} is the expectation of radius $E(r)$ for the distribution, along with  upper and lower bounds containing $68\%$ of the radio luminosity. The radius and bounds are plotted both with and without polar emission above a half-opening angle of $45^{\circ}$ from the equatorial plane. Atop this is plotted the radius from $u-v$ domain models fitted to the observations \citep{StaveleySmith:1992p680,Ng:2008,Ng:2013p25184}. We have also plotted a second-degree smoothing spline fitted to the radii from the $u-v$ domain models.
The spline fit at each point was weighted by the inverse of the error of the observed radius and the sensitivity of the spline was adjusted to produce a reasonably smooth function for the noisy data around day 2000. The radial position and width of the H\textsc{ii} region and equatorial ring are also delineated by horizontal lines within the plot. In order to compare the simulated radius with the truncated shell model of \citet{Ng:2013p25184}, we also fit a truncated shell to the simulation data at the same epochs as the observations. The midpoint radius and accompanying errors of the shell model are overlaid as blue diamonds. From the plot it is clear that the radius from the truncated shell is systematically larger than the expectation of radius. It appears that the truncated shell model fit to the simulation more closely follows the forward shock of the simulated data, however caution is advised in applying the same interpretation for the truncated shell fit to the observations as it is still largely unknown how the radio emission is distributed between the real forward and reverse shocks. For the simulation we have assumed that radio emission is generated at both forward and reverse shocks. This assumption may not be accurate.

Overall, the fitted radius from observations is well approximated by the expectation of radius from the simulation. Prior to the collision with the H\textsc{ii} region around day 2000, the shape of the luminosity distribution is a broad and steep line, a clear signature of spherical expansion. Beyond the H\textsc{ii} region the time varying distribution of radio luminosity is clearly aspherical, as indicated by the bi-modality in the distribution after day 2250. The encounter of the forward and reverse supernova shocks with high latitude material above the plane of the ring is responsible for the concentration of radio luminosity at radii greater than $8.5\times10^{15}$ m  ($1\farcs14$). When  high latitude emission is included in the computation of radius, it introduces a large upward bias toward large radii around day 7000. As the model fits to the observational data are not sensitive to high latitude emission, we do not expect a similar effect to be observed in the observational results. When high latitude emission is not included in the calculation, then the expectation of radius fits the observational data to within the region formed by $68\%$ of the simulated luminosity. Interestingly, the shock encounter with the Mach disk at a radius of $1.32 \times 10^{16}$ m ($1\farcs77$) produces a dramatic reduction in the production of radio luminosity due to a lowering of shock velocity as the shocks restart at that interface. 

At lower latitudes, radio luminosity is dominated by the interaction of the supernova shocks with the equatorial ring and H\textsc{ii} region. From the plot it appears that the forward shock began to encounter the ring around day 5400 and the reverse shock began to encounter the ring around day 6200. Around day 7000 there is a distinct turnover in radius for all estimates. This is more likely to be the result of a change in the distribution of radio emitting material than a real deceleration. The apparent deceleration might be due to the forward shock leaving the equatorial ring. This seems likely to be true, as the distribution of radio emission, and therefore the expectation of radius, is biased toward the reverse shock after the forward shock leaves the ring. 

\begin{figure*}[htbp]
\begin{center}
\begin{math}
\begin{array}{cc}
\includegraphics[width=8cm]{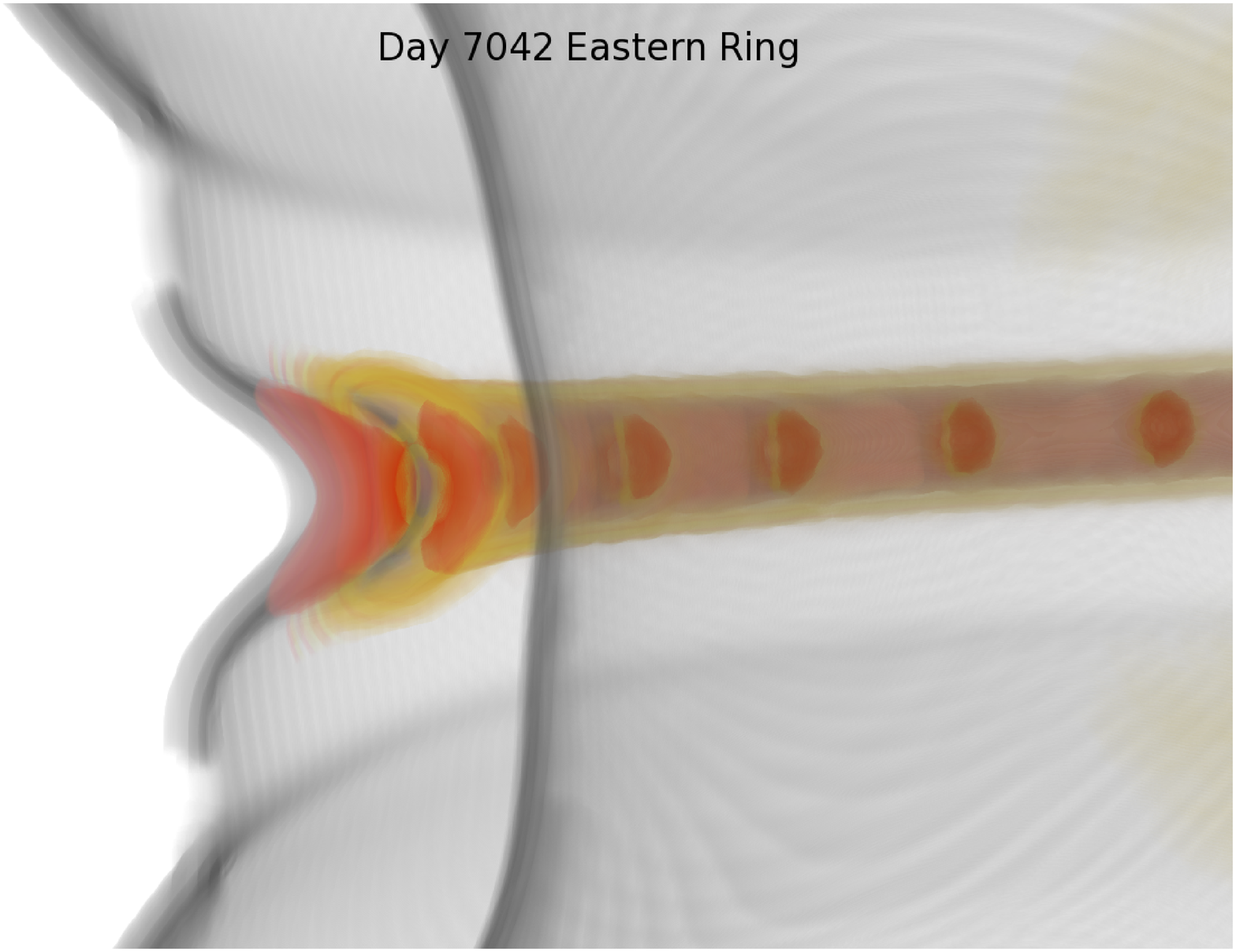} &
\includegraphics[width=8cm]{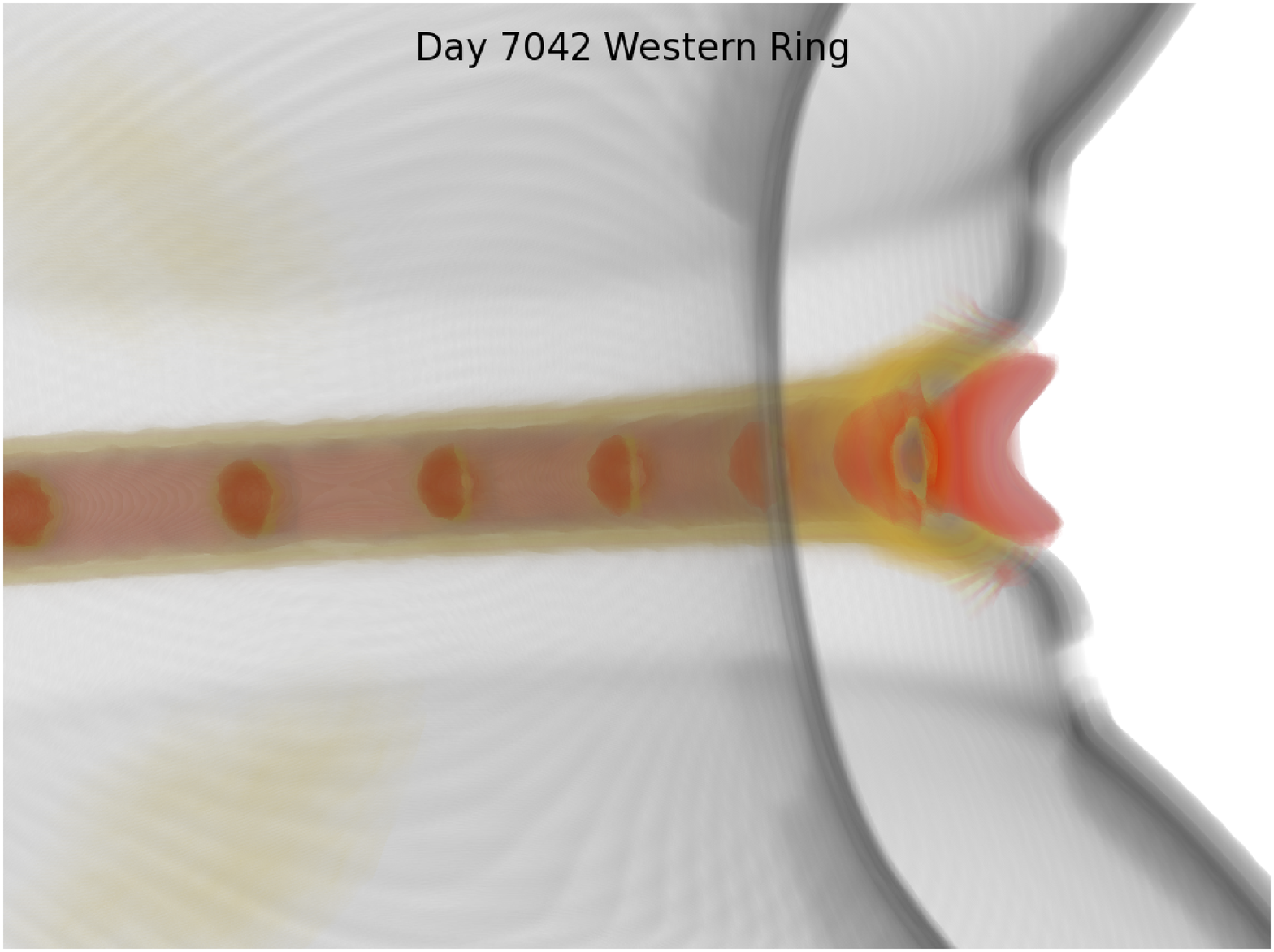} \\
\includegraphics[width=8cm]{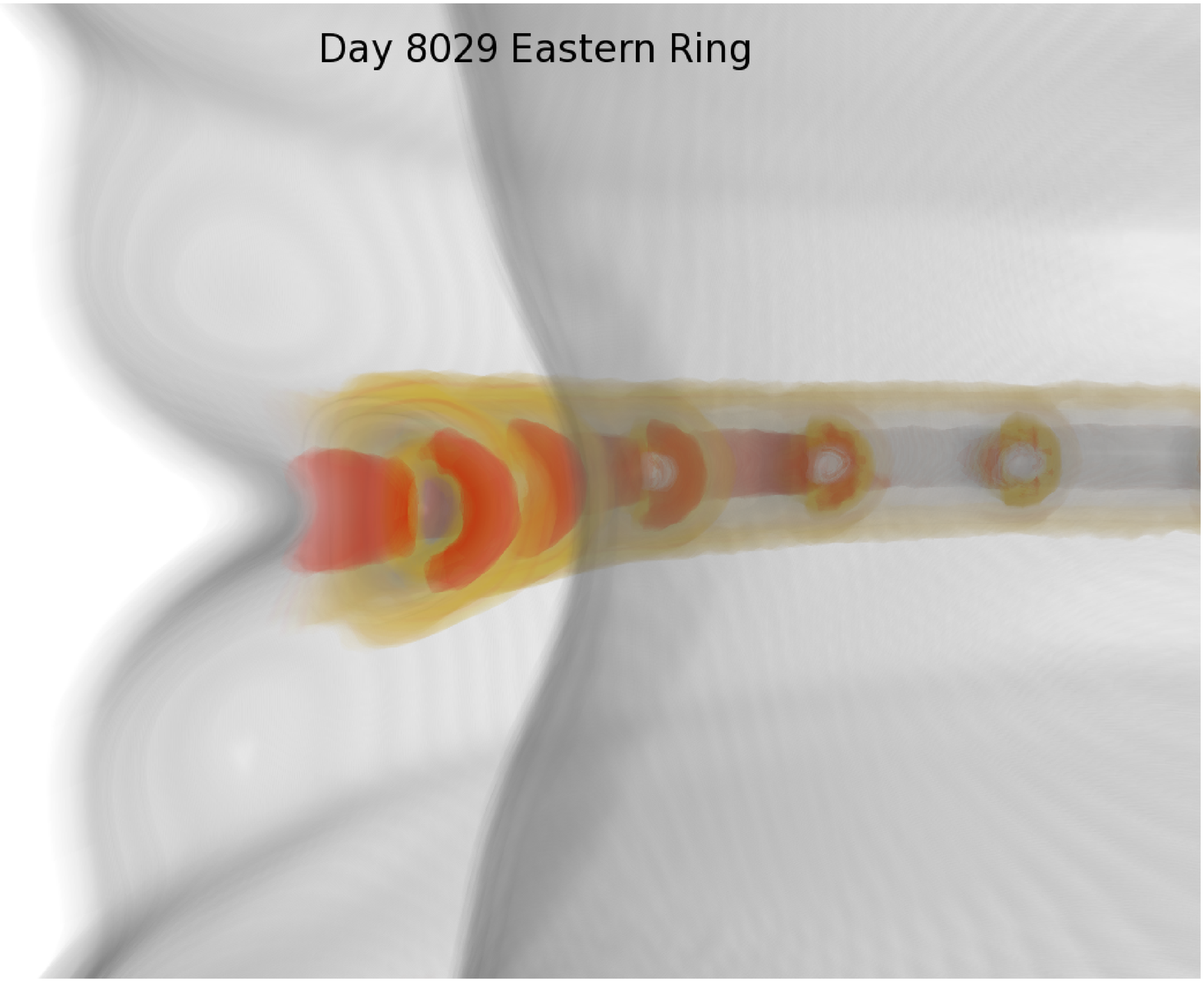} &
\includegraphics[width=8cm]{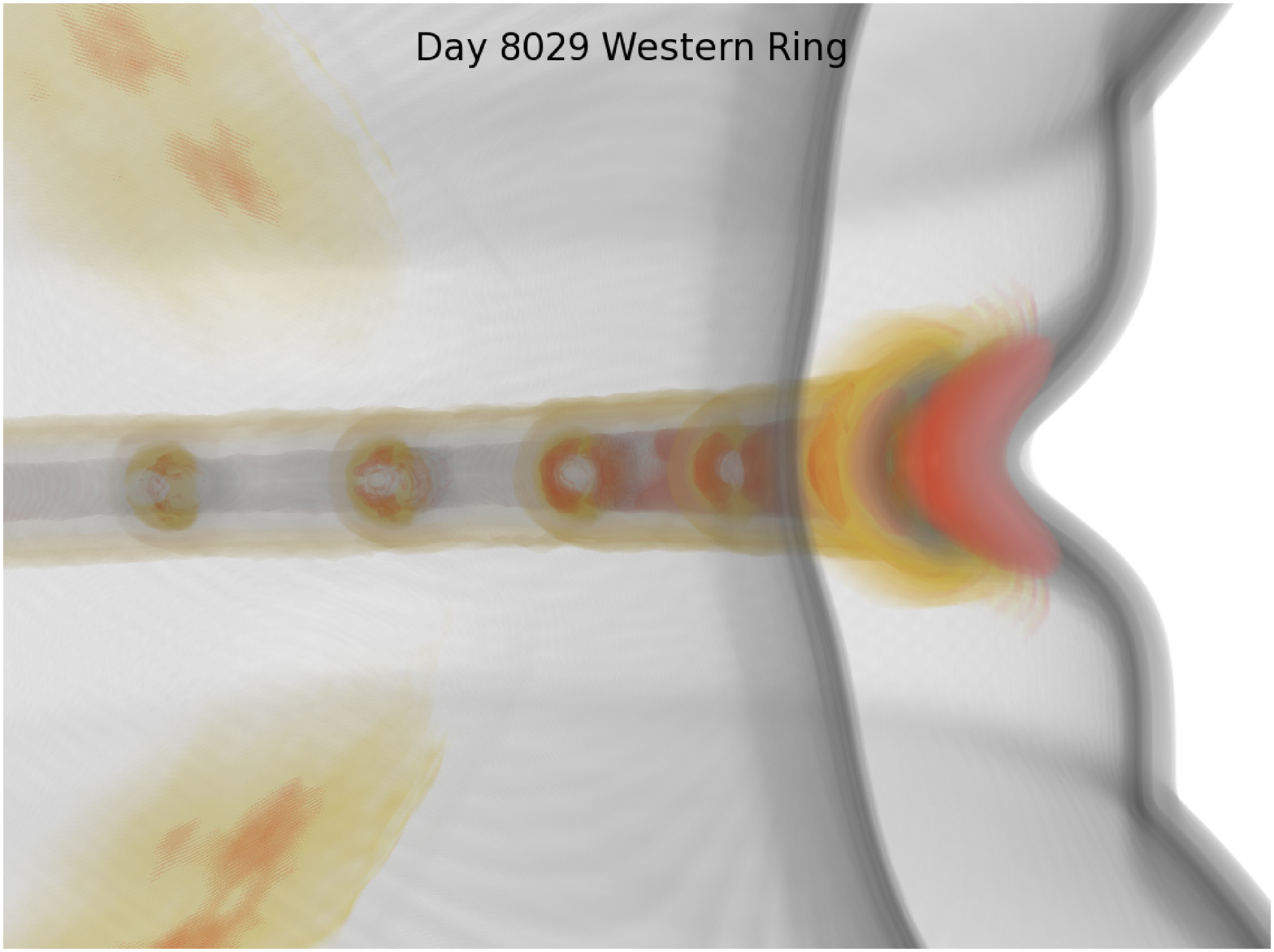} \\
\end{array}
\end{math}
\end{center}
\caption{Volumetric renderings of the shock interaction with the equatorial ring at days 7042 (top row) and 8029 (bottom row). In the left column is a cross section of the easternmost part of the equatorial ring, on the right is a similar cross section of the western ring. Rendered in greyscale is the log of entropy to reveal the reverse and forward shocks (in that order) from the centre of the figure. In red and tan is the equatorial ring in the log of particle density. From the Figure it is clear that the forward shock has almost left the eastern equatorial ring by day 7000. By day 8000 the forward shocks on both sides of the ring have almost completed traversing the ring, as indicated in Figure \ref{evolution_sequence2}. }\label{volume_rendering_ring}
\end{figure*}

The Drishti \citep{limaye:2006} volume renderings in Figure \ref{volume_rendering_ring} confirm that the forward shock does indeed leave the equatorial ring between days 7000-8000. Shown in the figure is a volume rendering of the shock interaction with a cross-section of two sides of the equatorial ring at day 7000 and 8000. Plotted in greyscale is the log of entropy, which is particularly sensitive to the reverse and forward shocks. Contrasted with this is the equatorial ring and ring blobs, rendered as red and tan features. The figure shows that the forward shock has almost left the eastern equatorial ring by day 7000. By day 8000, the forward shock has completely left the eastern ring, and has almost completed its crossing of the western ring. 

After day 8000 the radius appears to again accelerate as the relative amount of radio luminosity in the forward shock begins to increase relative to the luminosity in the ring. 
 
\subsection{Shock velocity }
 
\begin{figure}[h]
\centering
\epsfig{file=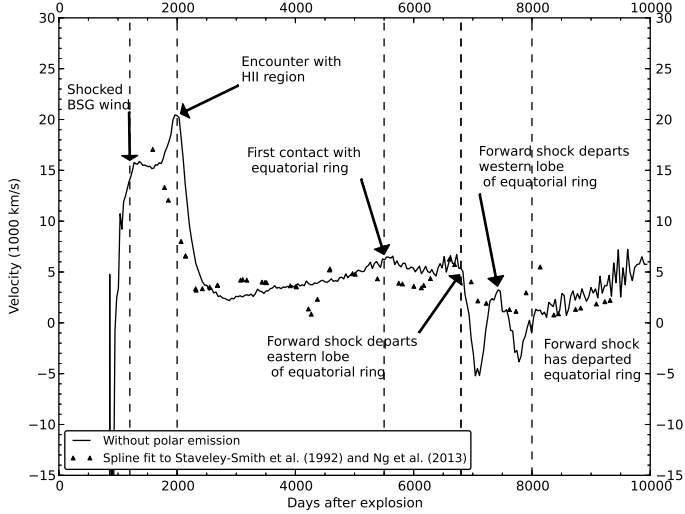,width=9.0cm} 
\caption{Time-varying shock velocity computed from the smooth radius curves in Figure \ref{shock_radius_vs_time}. Also included is the velocity derived from a spline fit to the to the observations from \citet{StaveleySmith:1992p680} and \citet{Ng:2013p25184}. Overlaid are markers showing the assorted interaction events from the forward shock and its effect on the distribution of radio emission and hence the derived shock velocity. The vertical lines are at the same epochs for the hydrodynamical events as discussed in Figure \ref{shock_radius_vs_time}.} \label{shock_velocity_vs_time} 
\end{figure}

In Figure \ref{shock_velocity_vs_time} is the average shock velocity computed as the time derivative of the smoothed radius curves in Figure \ref{shock_radius_vs_time}. Also shown is the velocity derived from the spline fit to the observational data, obtained by obtaining the slope of the fitted spline at the observed epochs.
From the plot we see that prior to day $2000$ after explosion, the average supernova shock expansion velocity of around $(1.7-1.9) \times 10^4$ km $s^{-1}$ is due to the forward shock propagating through the BSG wind. After the shock encounters the H\textsc{ii} region, the average velocity is dramatically slowed to around $2300$ km s$^{-1}$. Following this, the apparent shock velocity climbs steadily. For radio luminosity within $45^{\circ}$ of the equatorial plane, the velocity reaches a peak around $6500$ km s$^{-1}$ around day $5600$. 
As the forward shock leaves portions of the the ring, both estimates of radio emission experience a sequence of rapid drops in average shock velocity between days $6700$ and $8000$. The double dip structure in the shock velocity may arise due to the forward shock leaving the eastern lobe first, around day $7000$, then the western lobe at day $7800$. These events cause a large change in the average position of radio emitting material, and result in a perceived reversal in the shock velocity. After day $8000$, the average shock velocity appears to gradually accelerate as the forward shock  encounters material with a greater sound speed. At day $8000$ the average shock velocity is around $1000$ km s$^{-1}$. By day $10,000$ the average shock speed has accelerated to around $6,000$ km s$^{-1}$. The velocity derived from a spline fit to the observations appears to follow the general trend from the simulations, however caution is advised in interpreting high frequency oscillations from the general trend, as the fit to the expanding shock radius is determined by the sensitivity of the least squares spline fit. The rate of increase in the average shock speed for the observations appears slower after day $8000$, this may mean that the real sound speed in the ring and/or beyond the ring may be lower than we expected, indicating it may have a lower temperature than the temperature of $8\times10^4$ K we had set for the ring and H\textsc{ii} region. 

 
\subsection{Flux density and spectral index}

\begin{figure}[h]
\centering
\epsfig{file=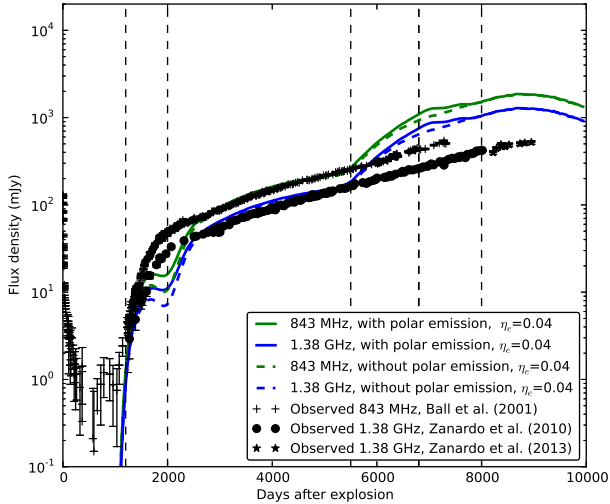,width=8.0cm} 
\caption{Simulated flux density plotted against observations at 843 MHz (green) and 1.38 GHz (blue). The flux density both with and without polar emission is plotted for comparison. In black is the observed fluxes at 843 MHz from \citet{Ball:2001p3450} and at 1.4 GHz from \citet{Zanardo:2010p17425} and Zanardo et al. (2013). The vertical lines correspond to the same hydrodynamical epochs as in Figure \ref{shock_velocity_vs_time}.}  \label{flux_vs_time}
\end{figure}

 We calculate flux density by summing over radial bins in the time-varying radio luminosity distribution in Figure \ref{shock_radius_vs_time}. The free parameter of the model, the fraction of available electrons swept up by the shock $\chi_{el}$, was fitted to the observations by chi-square minimisation of the simulated flux density scaled by $\chi_el$. We performed this fit for 843 MHz and 1.38 GHz. In Figure \ref{flux_vs_time} is the fitted flux densities plotted against the observed flux densities at the two frequencies.
 From the plot the shape of the simulated flux density provides a good fit to the observational data at both frequencies between days 1200 and 5400. The scaling factor, $\chi_{el}=4\%$, provides an optimal fit to the flux density, assuming the electrons are in thermal equilibrium with the ions and are injected into the shock from the downstream region at a momentum consistent with their thermal velocity. This fraction is consistent with the range of ($1-4\%$) obtained by \citet{Berezhko:2000p56}, but higher than $\chi_{el}=6 \times 10^{-6}$ obtained in \citet{Berezhko:2011p19115}. The scaling factor is derived using the additional assumption that a constant fraction of the electrons are injected into the shock at all times. Since magnetic field amplification is highly non-linear and is still an area of active research, these assumptions may not be correct and we consider our derived value of $\chi_{el}$ a preliminary result. The sudden increase in flux density around day 1200 is particularly sensitive to the location of the termination shock in the relic BSG wind. In our simulation the termination shock was placed at distances in the range $(3.2-3.8) \times 10^{15}$ m ($0\farcs43-0\farcs51$) from the progenitor, with corresponding gas densities in the range $(7.6-5.4) \times 10^{-23}$ kg m$^{-3}$. The flux densities around day 1900 are discrepant with the observational data because the shock velocity (and hence the magnetic field) slows considerably at the H\textsc{ii} region prior to restarting. This is probably an artefact of a comparatively large numerical shock width, and might be resolved with an increase in grid resolution in future studies. Another possibility is that radius as reported from the truncated shell model fits may be overestimated, as suggested by the truncated shell model fits to the simulated data in Figure \ref{shock_radius_vs_time}. We note that the spurious dip in flux density disappears in some of our models if we move the HII region and BSG termination shock closer to the progenitor. Around day $5500$ there is an even greater discrepancy between the observed and simulated fluxes. It is interesting that the observed flux density does not also display a similar marked jump as the shock encounters the ring. There are many possible reasons for this. The mass or filling factor of the simulated ring may be overestimated. Alternatively, the velocity of the simulated shock or the injection efficiency may be overestimated during the crossing of the ring, thus more radio emission is produced than is observed. It may also be that the ambient magnetic field within the equatorial ring is lower than expected, hence the shock encounter with the ring is not producing as much synchrotron emission as the simulations predict. A spectral index was calculated from the two frequencies. However we see little deviation from $\alpha=0.75$, which is expected for a strong shock and sub-diffusive shock acceleration without cosmic ray feedback. 

\subsection{Morphology and opening angle}

The exact reason for the persistent asymmetry in the radio morphology of the remnant has been a longstanding mystery. Magnetic field amplification may provide a solution to the problem by explaining the asymmetric radio morphology as a consequence of an asymmetric explosion. From Equation \ref{syn_emiss}, we see that synchrontron emissivity is a nonlinear function of $b$, $B$ and $\kappa$. If we employ shock acceleration to generate the particle distribution $f(p)$ and magnetic field amplification to obtain $B$, then radio emissivity should scale with shock velocity $v_s$ as
\beq
J_{\nu} \propto \rho_{1}^{(b+3)/4}  v_s^{3(b-1)/4} \nu^{-(b-3)/2}.
\eeq
From \citet{Landau2007Fluid} the shock velocity of a strong forward shock propagating into a stationary medium is proportional to downstream pressure $P_2$ and upstream density $\rho_1$ as $(P_2/\rho_1)^{1/2}$. Radio emissivity then scales as
\beq
J_{\nu} \propto P_2^{3(b-1)/8} \rho_{1}^{(9-b)/8}  \nu^{-(b-3)/2}.
\eeq
For a strong shock, $b$ is in the range $4-4.5$, and radio emission is more sensitive to shock strength than density. Conversely, thermal X-ray emission is more sensitive to density. Observations of thermal X-rays from SN 1987A on day 7736 \citep{Ng:2009p18119} show that the east-west asymmetry is around $3-5\%$, which is an order of magnitude less than the observed radio asymmetry. Thus an asymmetric circumstellar environment appears to be an unlikely cause for the radio emission. Under the assumption of magnetic field amplification, radio emissivity is highly responsive to the downstream pressure. If the eastern shock is stronger than the western shock, such as from an asymmetric explosion, then magnetic field amplification provides a plausible mechanism for a corresponding asymmetry in the radio remnant. 

\subsubsection{Morphological comparisons with the observations}

\begin{figure*}[htbp]
\begin{center}
\begin{math}
\begin{array}{cc}
\epsfig{file=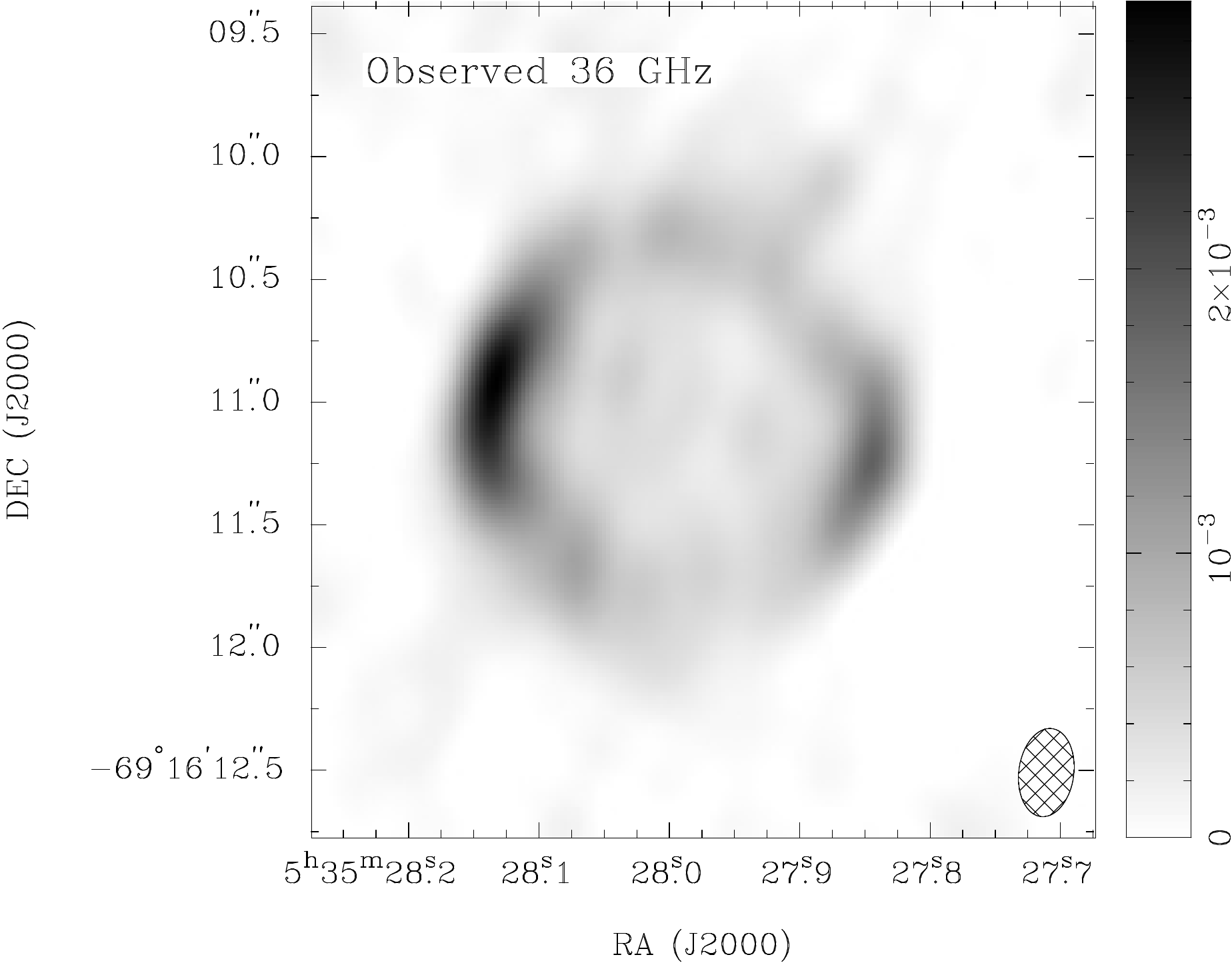,width=8.0cm} &
\epsfig{file=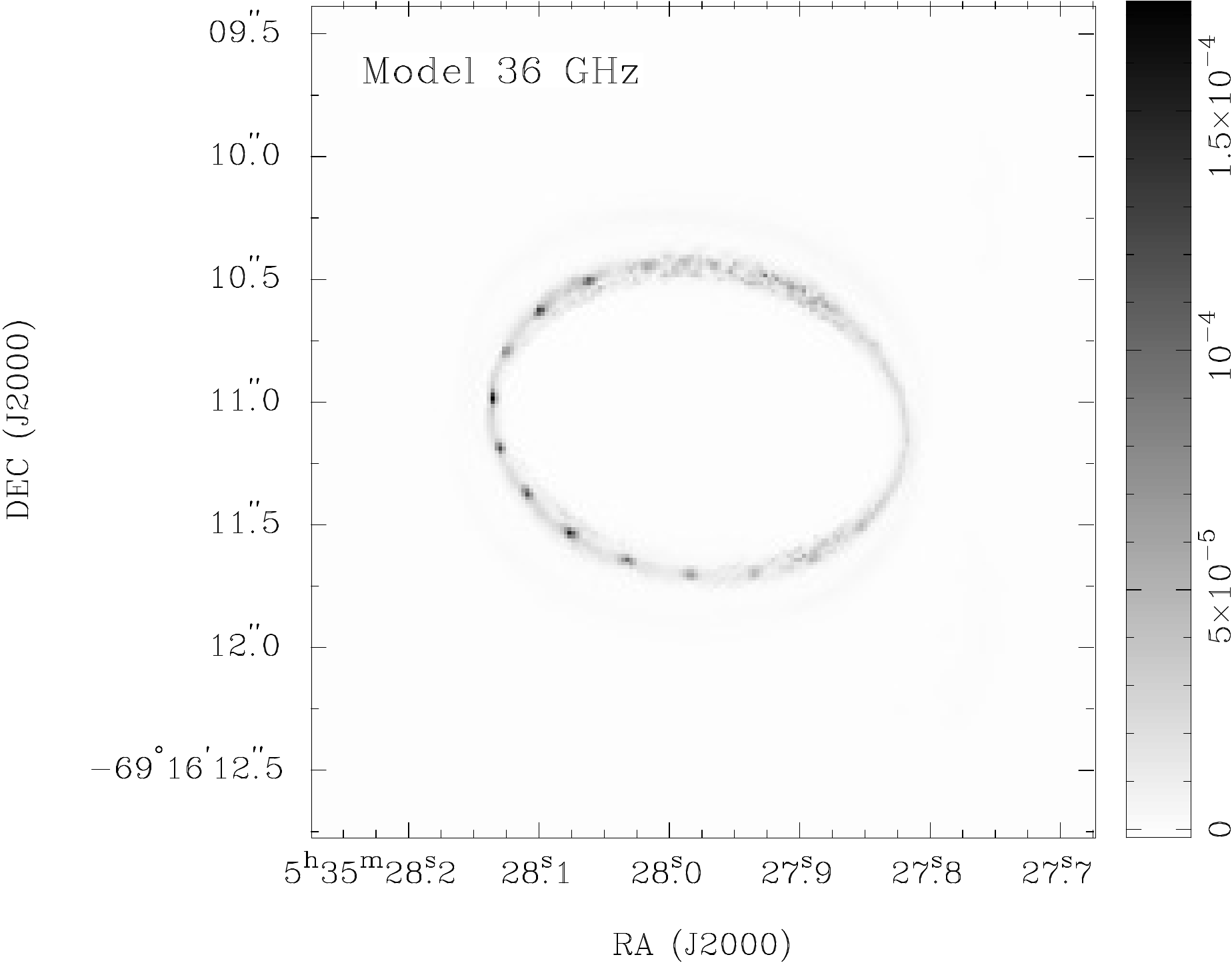,width=8.0cm} \\
\epsfig{file=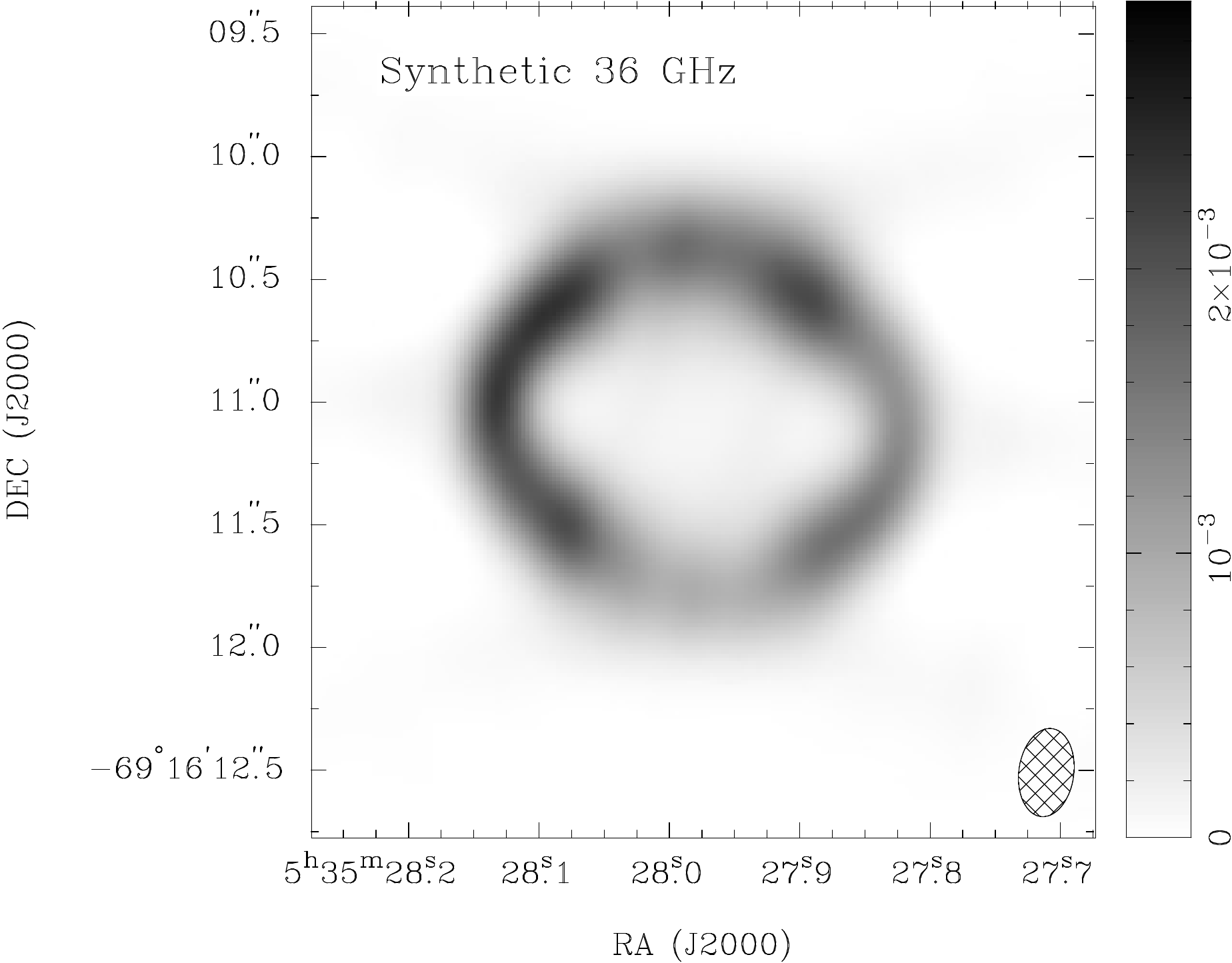,width=8.0cm} &
\epsfig{file=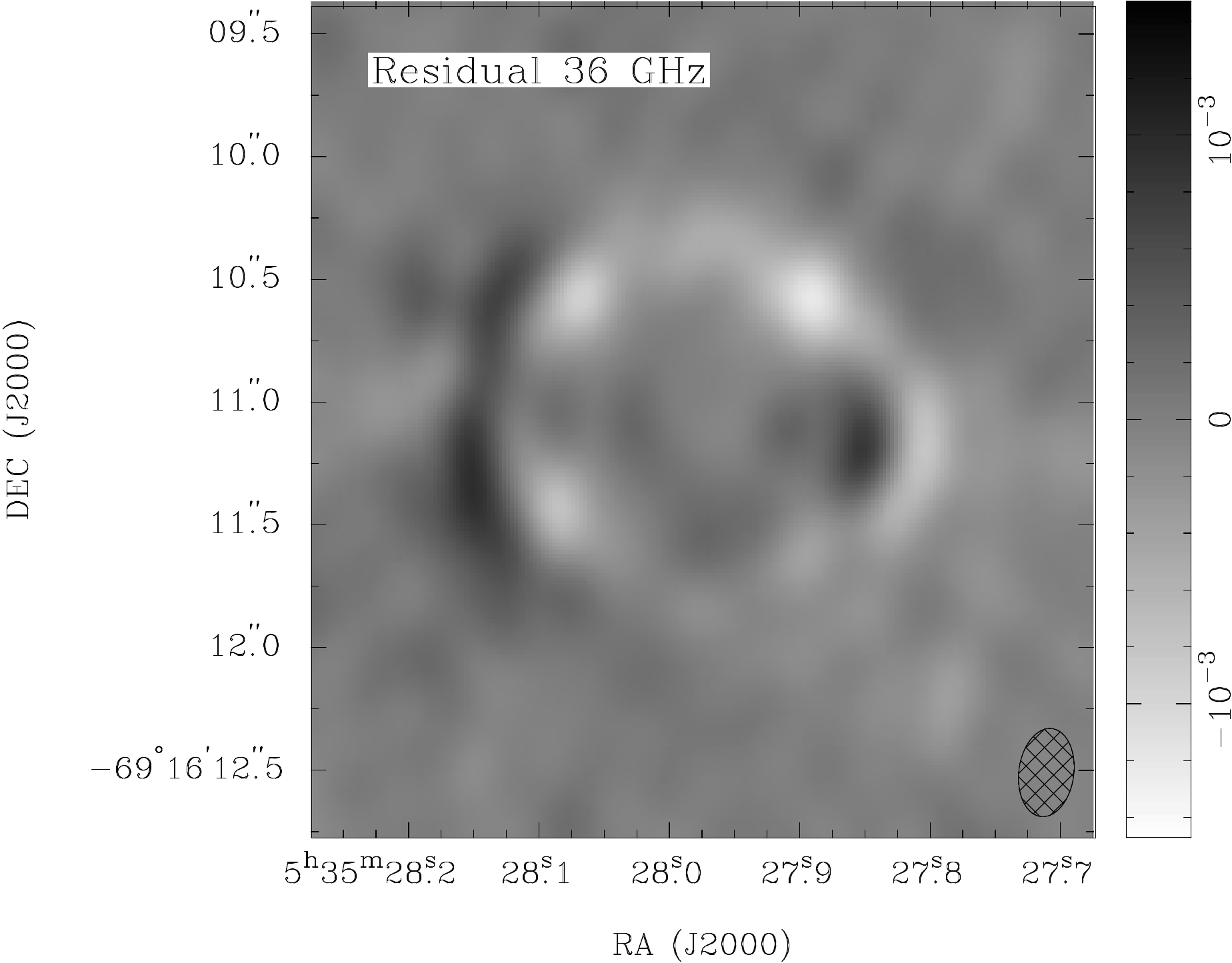,width=8.0cm} \\
\end{array}
\end{math}
\end{center}
\caption{Comparison of the real 36 GHz image with a corresponding synthetic model at day 7900.  At top left is the 36 GHz image at day 7900 from \citet{Potter:2009p16343}. At top right is the model image formed from radio emission in the simulation. At lower left is the model imaged with the same beam as the observations. At lower right is the residual image formed by subtracting synthetic $u-v$ data of the model from the $u-v$ data of the observation and imaging the result. Units are in Jy/Beam for the images and Jy/pixel for the model. }
\label{compare_model_obs}
\end{figure*}

In order to test the magnetic field amplification hypothesis we derived $36$ GHz synthetic images of the radio morphology at day $7900$ and compared them with observations at the same epoch \citep{Potter:2009p16343}. Figure \ref{compare_model_obs} contains the result. At the top left is the observed image. At top right is the imaged model at day 7900. The flux density of the model has been scaled to match that of the observations. At bottom left is the imaged model where the model has been transformed to the $u-v$ domain and the $u-v$ data of the transformed model is used to replace corresponding $u-v$ data of the observations. The result has been convolved with the $(0\farcs4 \times 0\farcs2)$ beam from the October 2008 (day 7900) observation in \citet{Potter:2009p16343}. At the lower right is the imaged residual, where the $u-v$ data of the model has been subtracted from the observation prior to imaging.

From the plot it is clear that the simulated image and model show similar morphologies. Due to the faster eastern shock,  radio emission in the eastern lobe of the model has more radio emission than the western lobe. 
The residual images shows that extra radio emission from the model occurs at the eastern lobe outside the position of the ring. This is unlikely to be due to image registration as the difference in position between the dark patches on the eastern lobe is greater than the registration error of $0\farcs03$. It may be that the speed of the eastern shock is faster than expected, or that the real forward shock is interacting with more high-latitude material than the simulation indicates. 

\subsubsection{The evolving asymmetry}

\begin{figure}[htbp]
\begin{center}
\begin{math}
\begin{array}{c}
\epsfig{file=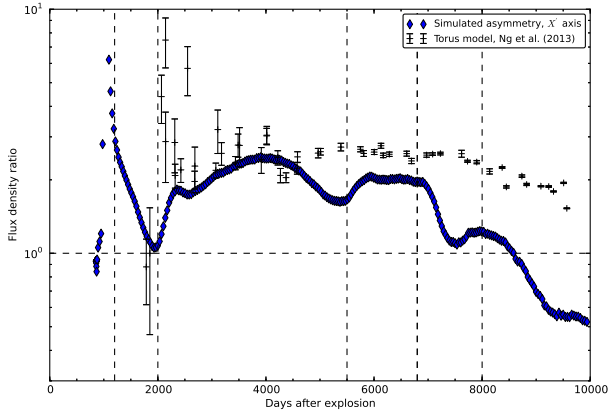,width=8.0cm} \\
\end{array}
\end{math}
\end{center}
\caption{The evolving asymmetry of the supernova model. In black is the asymmetry from the model fit in \citet{Ng:2013p25184}. In blue is the the asymmetry obtained by summing flux densities either side of the origin in the rotated $X^{\prime}$ coordinate. }
\label{asymmetry}
\end{figure}

In order to track the evolution of asymmetry in the simulation we obtained the ratio of the total integrated flux density either side of the origin in the rotated $X^{\prime}$ coordinate. The resulting evolution in asymmetry for the simulation is shown as blue diamonds in Figure \ref{asymmetry}. In similar fashion we integrated the flux density over truncated shell models that were fitted to both the observed and simulated data. The results are shown as black and blue points.

From the figure we see that the evolution in asymmetry provides a reasonably good fit between the asymmetry measured from the simulation and the asymmetry obtained via a truncated shell model fit to the observations in \citet{Ng:2013p25184}. The truncated shell model fit to the simulation appears to have a very high level of asymmetry. We suspect the truncated shell model is biased by high-latitude components of radio emission. Overall, the eastern lobe in both simulated and observed remnants has consistently more flux than the western lobe from day 2000 to day 7000. This shows that an asymmetric explosion combined with magnetic field amplification at the shock is a viable physical model for reproducing the asymmetry in the remnant. The sudden positive jumps in the simulated asymmetry in Figure \ref{asymmetry} appear to be correlated with hydrodynamical events such as the interaction with the H\textsc{ii} region around day 2000 and the encounter with the equatorial ring around days (5000-6000). Such behaviour indicates that the 3D model may not be smooth enough. 

Both observed and simulated asymmetries experience a decline around day 7000. This suggests that either eastern lobe of the remnant loses a large portion of its flux density relative to the western lobe at that time. Such a decline may be due to the forward shock exiting the equatorial ring first, as is expected for an asymmetric explosion. It may also indicate that the real shock has encountered a significant overdensity in the western lobe of the ring. However the consequent X-ray emission from a shock encounter with such an overdensity has not been observed in X-ray images taken around the same time \citet{Ng:2009p18119}. 

The rapid decline in the simulated asymmetry around day 7000, in contrast to that derived from observation, is likely to be the result of placing the ring at points equidistant from the progenitor. The decline in asymmetry from fits to the observations is more gradual. This might indicate the ring is more broadly distributed in radius than we have simulated. The plot shows that the timing for the simulated events is sooner than the observed events and that we may have overestimated the asymmetry in the simulated explosion.

\subsubsection{Morphological predictions}

\begin{figure*}[htbp]
\begin{center}
\begin{math}
\begin{array}{cc}
\epsfig{file=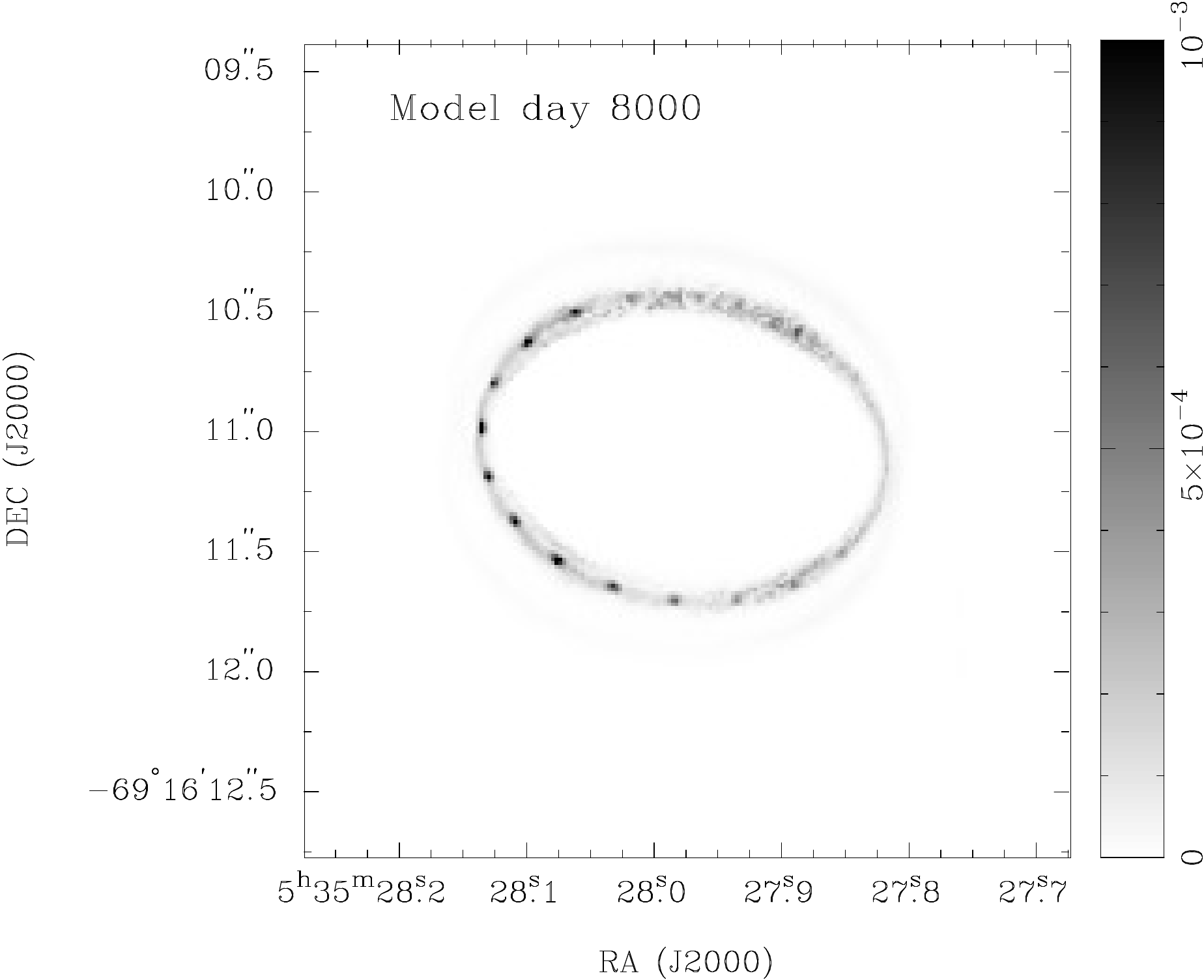,width=8.0cm,angle=0} &
\epsfig{file=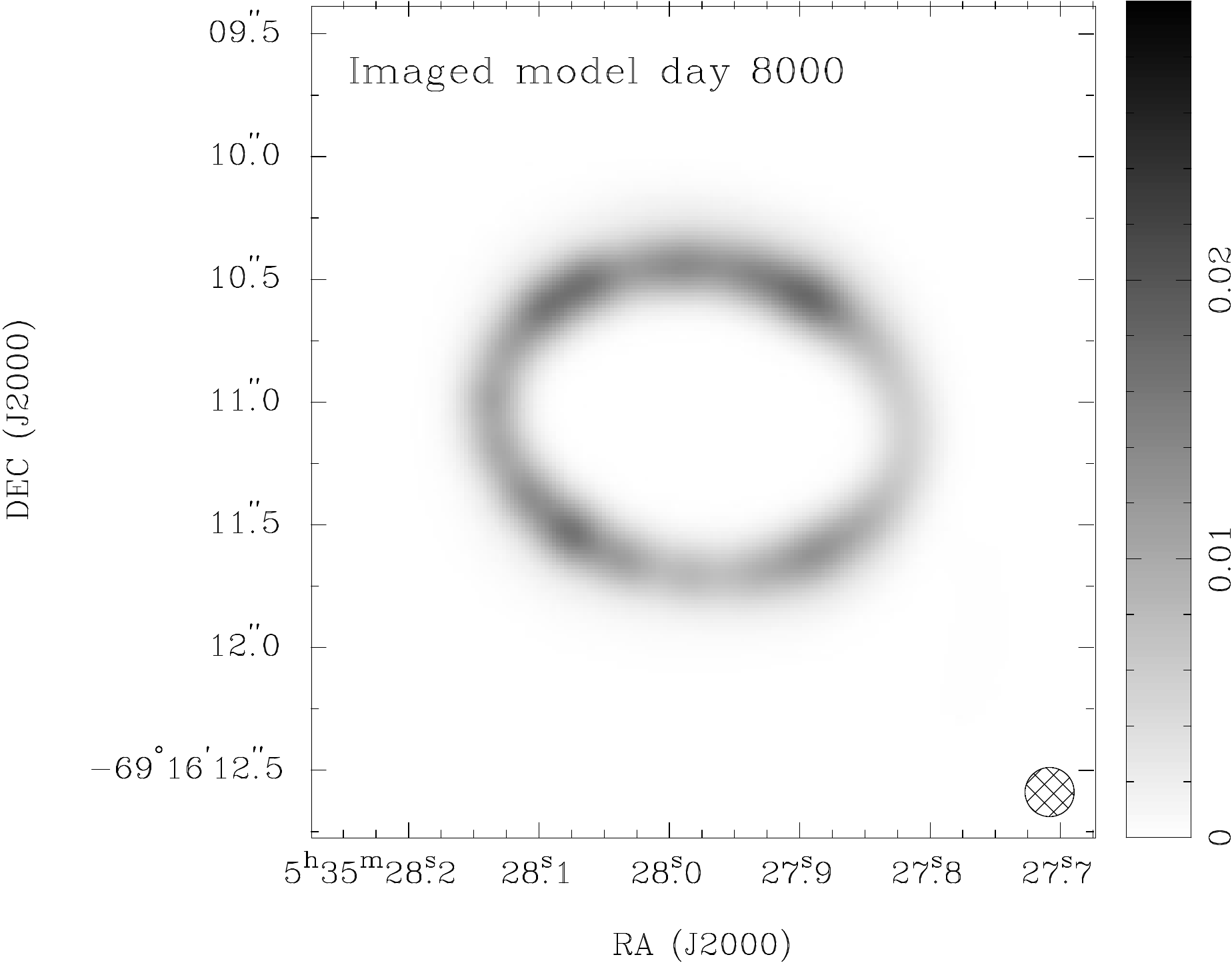,width=8.0cm,angle=0} \\
\epsfig{file=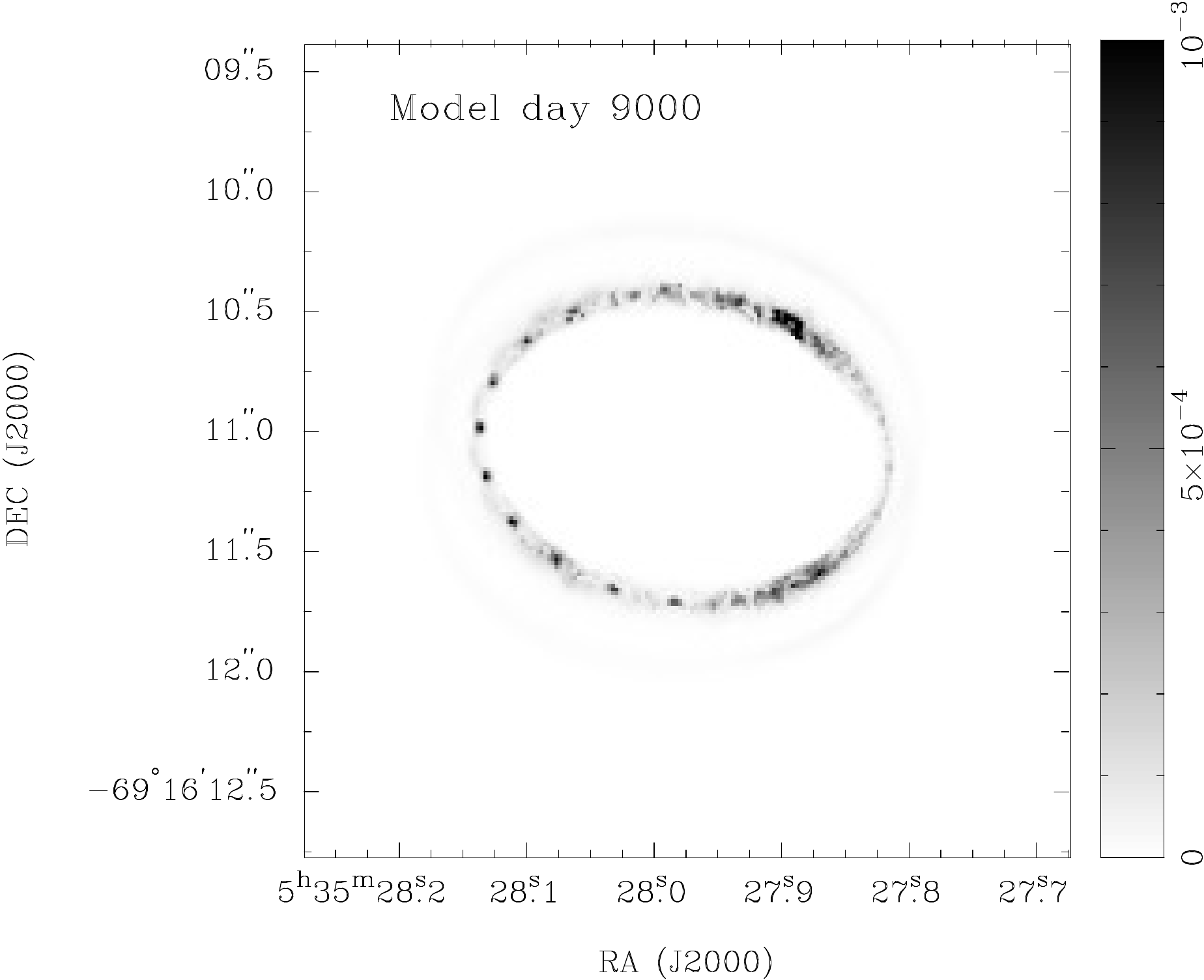,width=8.0cm,angle=0} &
\epsfig{file=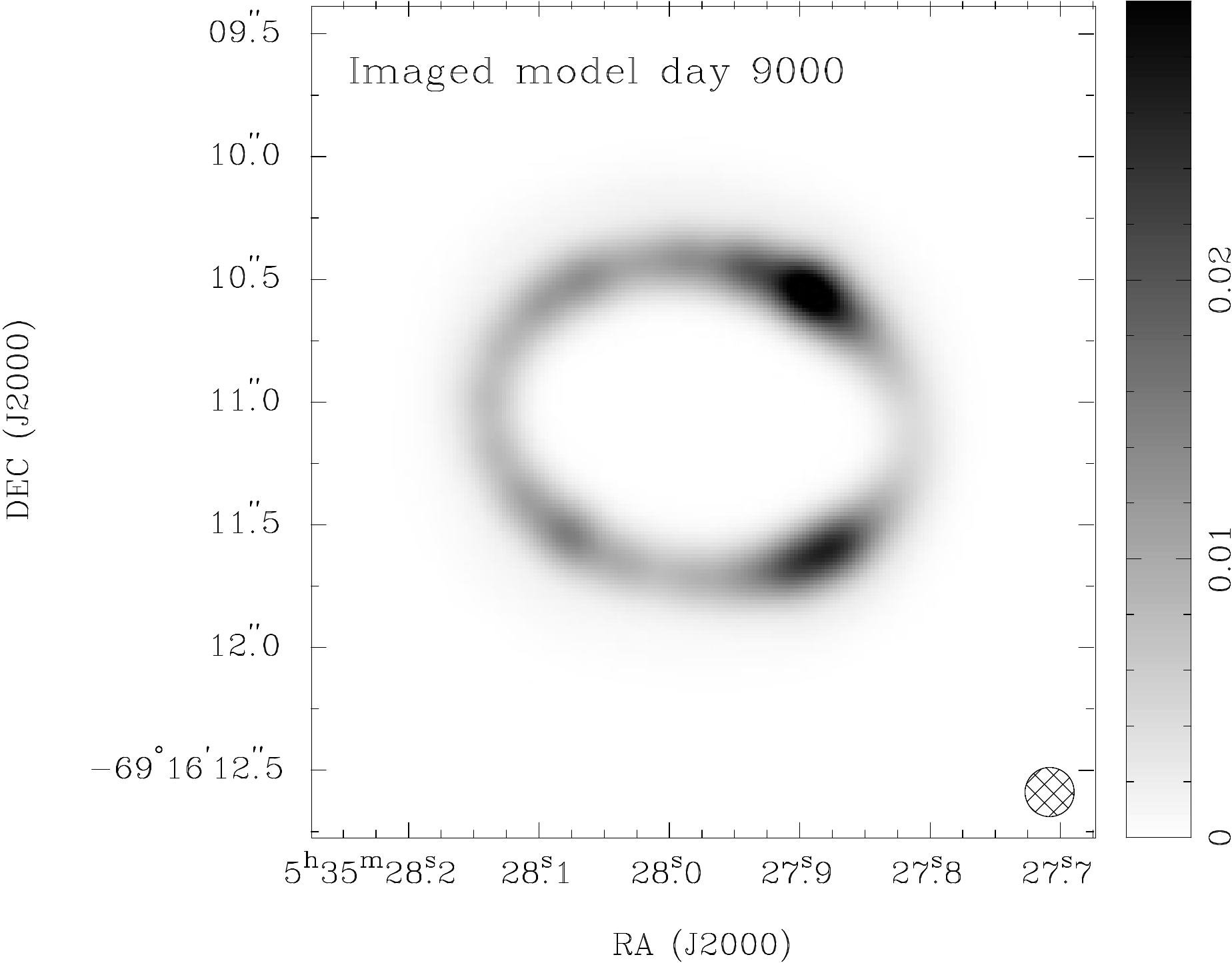,width=8.0cm,angle=0} \\
\epsfig{file=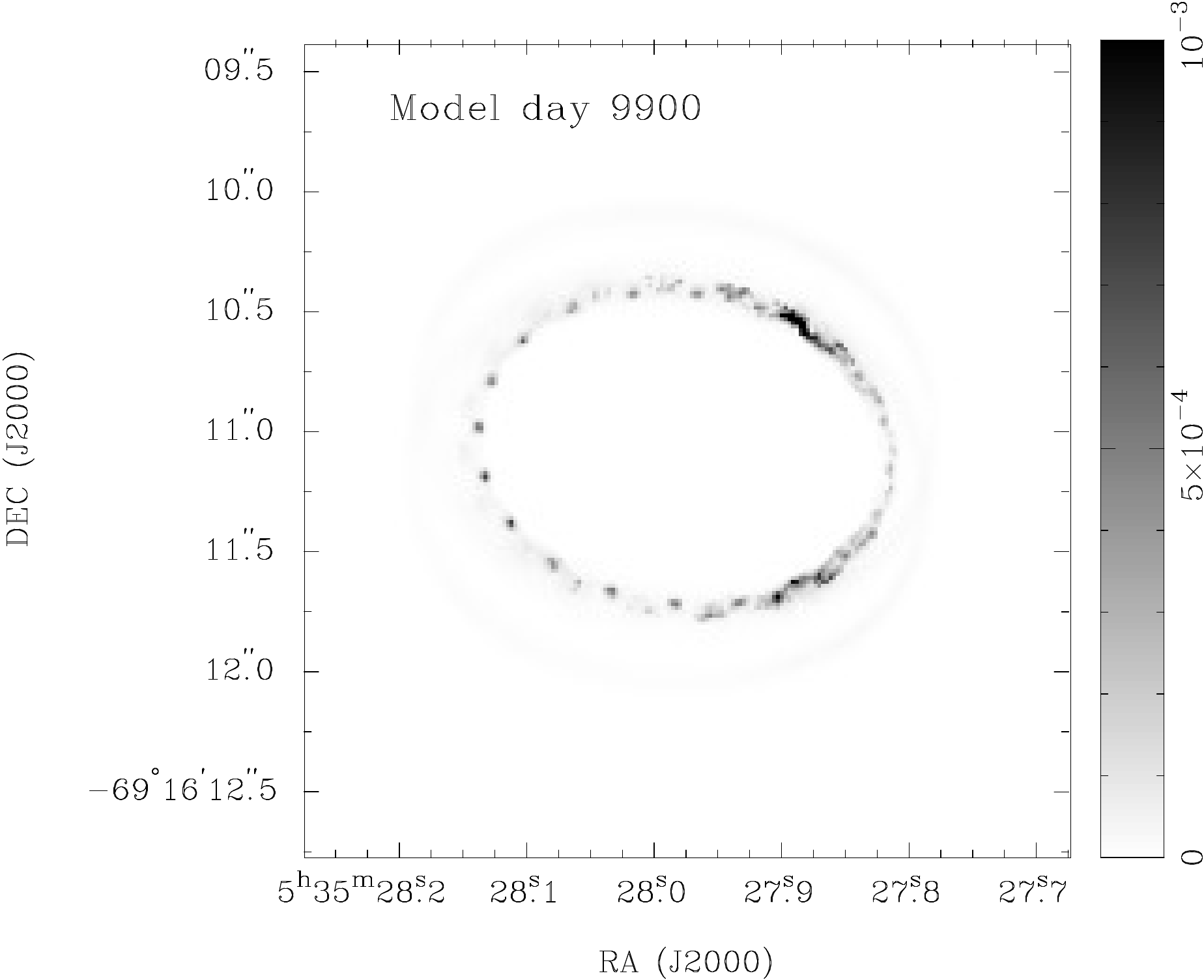,width=8.0cm,angle=0} &
\epsfig{file=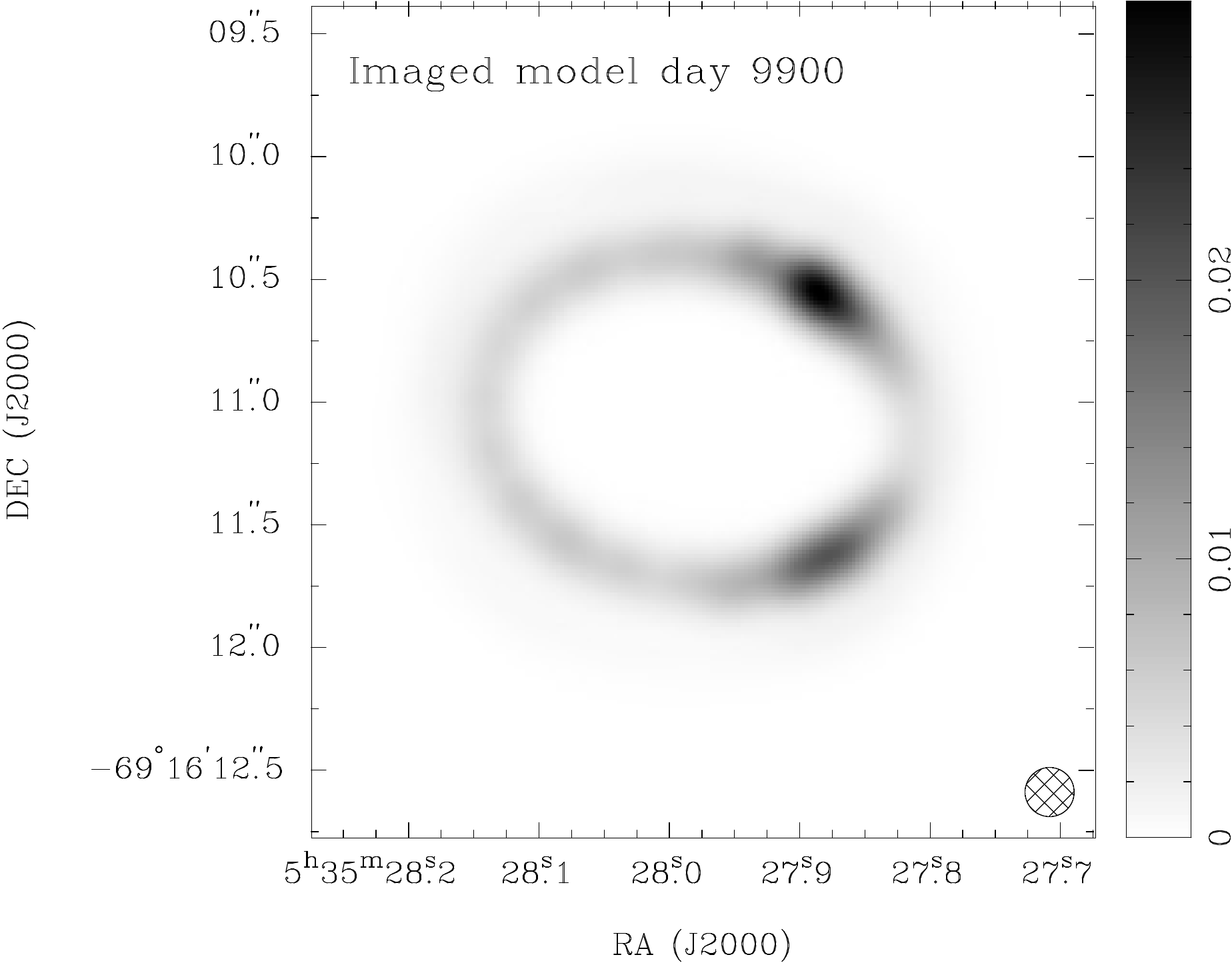,width=8.0cm,angle=0} \\
\end{array}
\end{math}
\end{center}
\caption{Simulated 8.7 GHz radio images of SNR 1987A, between days 8000-9900. The model images (left column) were convolved with $0\farcs1$ restoring beam to form the images in the right column. It is anticipated that the asymmetry in the radio morphology is will swap sides as the faster eastern shock leaves the ring first.}\label{morph_images_future}
\end{figure*}

An interesting prediction from the simulation is that the asymmetry will at least temporarily reverse direction in coming years, as evidenced by the asymmetry of the simulation dipping below parity after day 8000. In Figure \ref{morph_images_future} is synthetic images of the radio morphology between days 8700 and 9900. The images show that the western lobe of the ring will dim more slowly than its eastern counterpart due to the lower shock speed. 


Measurements of the radius obtained through truncated shell modelling have shown that the radius curve shows an apparent deceleration at that time, \citep{Ng:2013p25184}, suggesting that the shock has slowed down or the relative contribution of radio emission from the forward shock has decreased. Fits to the radius obtained with ring and torus models suggest that the radio emission is becoming more ringlike with age. These observations are consistent with the hypothesis put forward in \citet{Ng:2013p25184} - that day 7000 corresponds to the time the forward shock left the ring, leaving the reverse shock buried within the ring. 

\subsubsection{Radio luminosity at different half-opening angles}

\begin{figure*}[htbp]
\begin{center}
\begin{math}
\begin{array}{c}
\epsfig{file=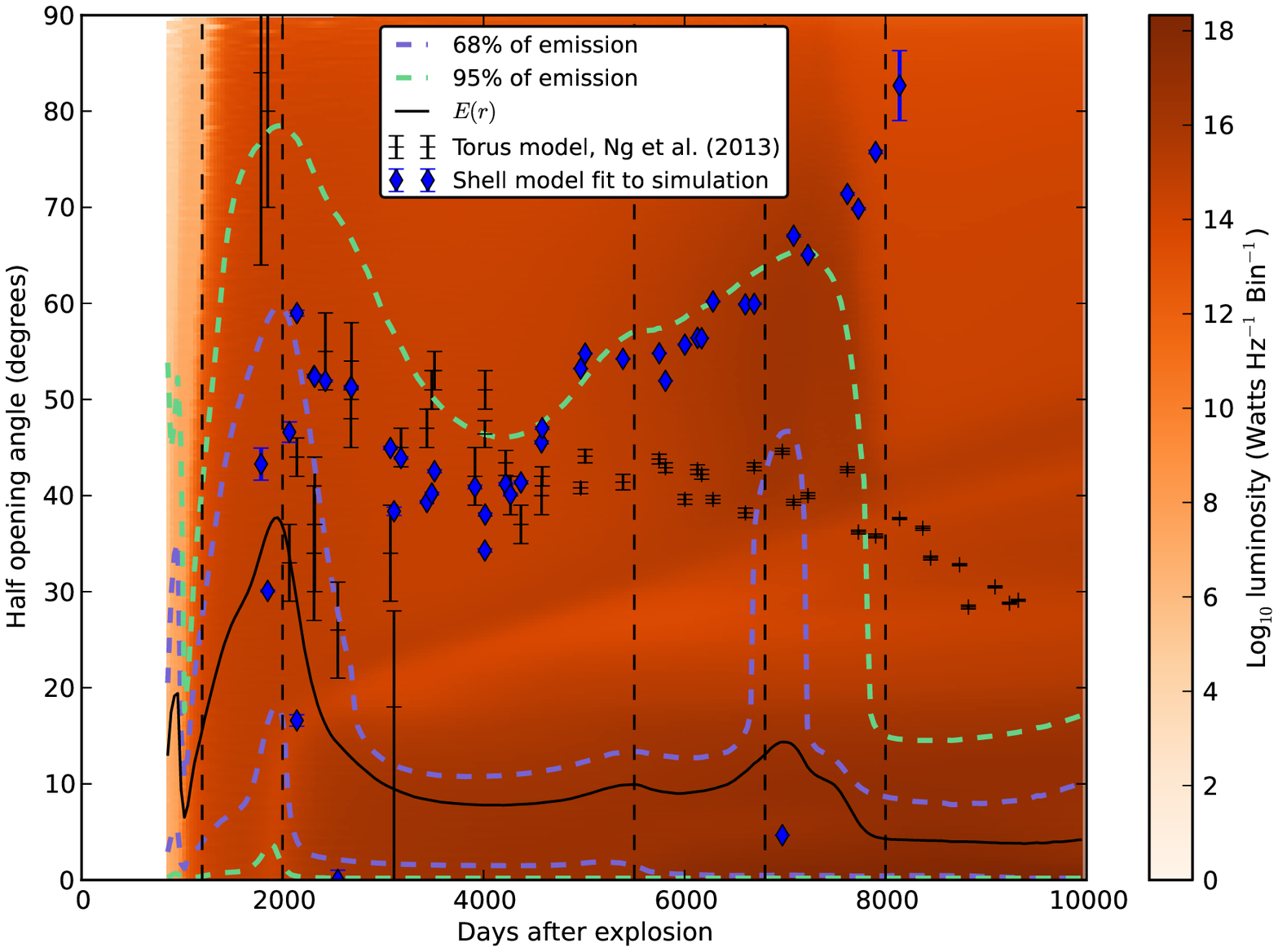,width=13.0cm} \\
\end{array}
\end{math}
\end{center}
\caption{The evolving distribution of radio emission as a function of half opening angle. Overlaid is the expectation of half opening angle and the contours containing $68$ and $95 \%$ of the radio luminosity. Also plotted is the half-opening angle from the truncated shell model fit to synthetic images made from the simulation (in blue), and truncated shell model fits to observations (in black) from \citet{Ng:2013p25184}. The vertical lines at days 1200, 2000, 5500, 6800 and 8000 delineate the shock interaction events discussed in Figure \ref{shock_velocity_vs_time}. }
\label{half_opening_angle}
\end{figure*}

In Figure \ref{half_opening_angle} is the evolving distribution of radio emission as a function of half-opening angle from the equatorial plane. Overlaid on the distribution is the expectation of half-opening angle from the distribution, along with contours containing $68$ and $95$\% of the radio emission. For comparison, we have overlaid the half-opening angle derived from truncated shell model fits to both the simulation (in blue) and the observations (in black) from \citet{Ng:2013p25184}.
At early times the supernova shock is spherical, as evidenced by a half-opening angle around $90^{\circ}$ seen before day 2000.  After day 2000, the half-opening angle from the observations appears to separate into two distributions representing components from high latitude material and the shock interaction with the H\textsc{ii} region. The expectation value of the simulated half-opening angle follows the radio emission near the H\textsc{ii} region and drops sharply, reaching a half-opening angle of $8^{\circ}$ by day 4000.  Between days 4000 and 6000, both simulated and fitted models show a flat slope for the evolving half-opening angle. The $95\%$ boundary of the simulated distribution appears to diverge from the fitted model after day 4000. This is due to radio emission from high latitude material between days $4000-8000$. Around day 7000 both the simulated and fitted models show a turnover in half-opening angle. This indicates that the relative fraction of radio emission from the ring itself is increasing after day 7000. The simulated half-opening angle after this drops to its minimum value of around $3-4^{\circ}$ between days 8000 and 10000. Conversely, points from the truncated shell model fits to the simulation appear to be scattered around the expectation of half-opening angle at early times, however they soon diverge from the expected half-opening angle around day 2000 and appear to follow the 95\% confidence contour from the distribution, presumably as a result of high-latitude radio emission. The truncated shell failed to converge to a solution after day 8000. It is suspected this is caused by hotspots in the in the simulated western ring at late times. The truncated shell model fits to the observed data appear to track corresponding fits to the simulation until around day 4000. This may be because high-latitude emission may not be present in the observations or is lost in the noise. Both expectation of radius  from the simulation and truncated shell model fits to the observations suggest that a hydrodynamical event occurs after day 7000. We suggest it is most likely the exit of the forward shock from the eastern lobe of the equatorial ring. 

\subsection{Injection parameters}

\begin{figure*}[h]
\begin{center}
\begin{math}
\begin{array}{cc}
\epsfig{file=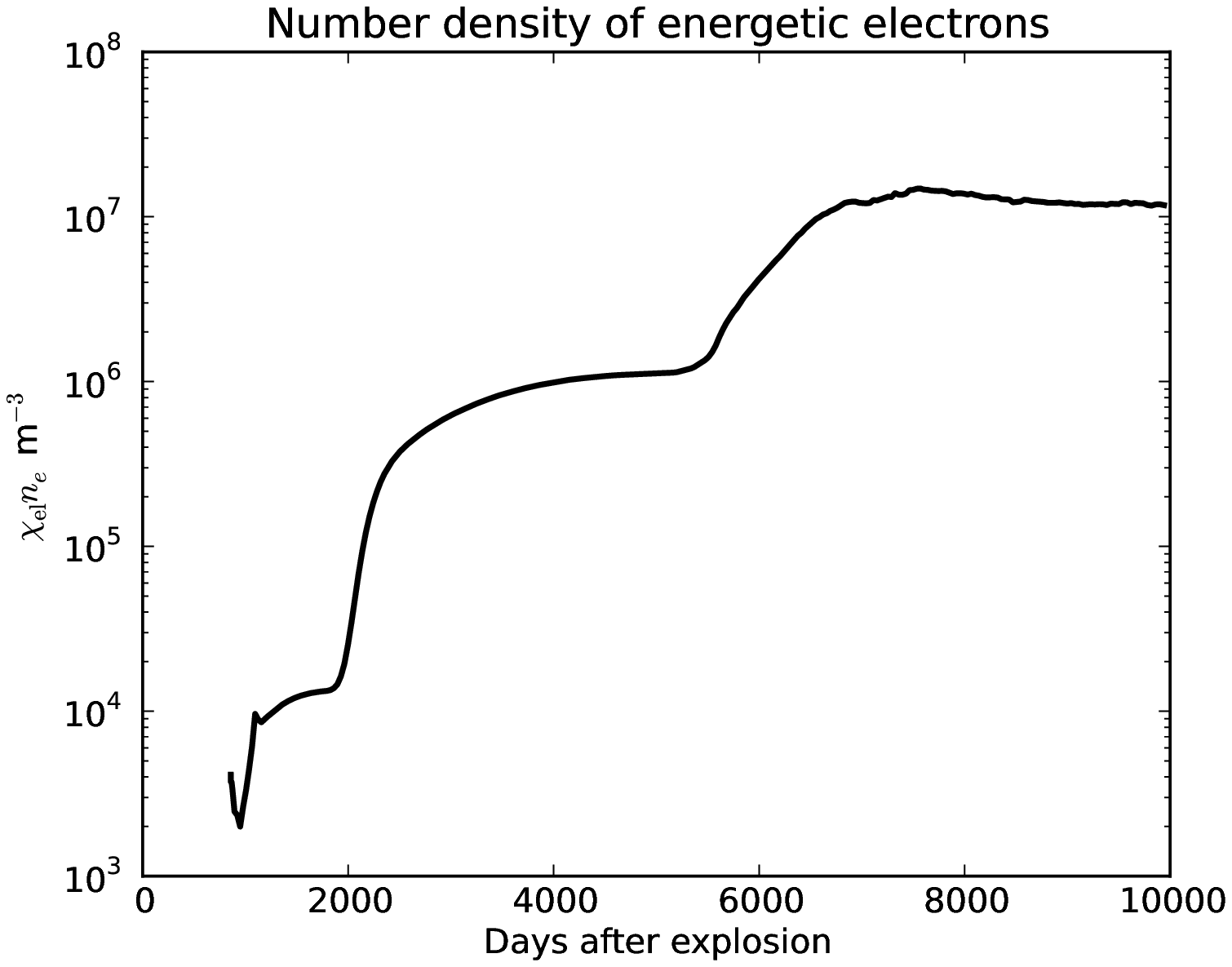,width=7.5cm,angle=0} &
\epsfig{file=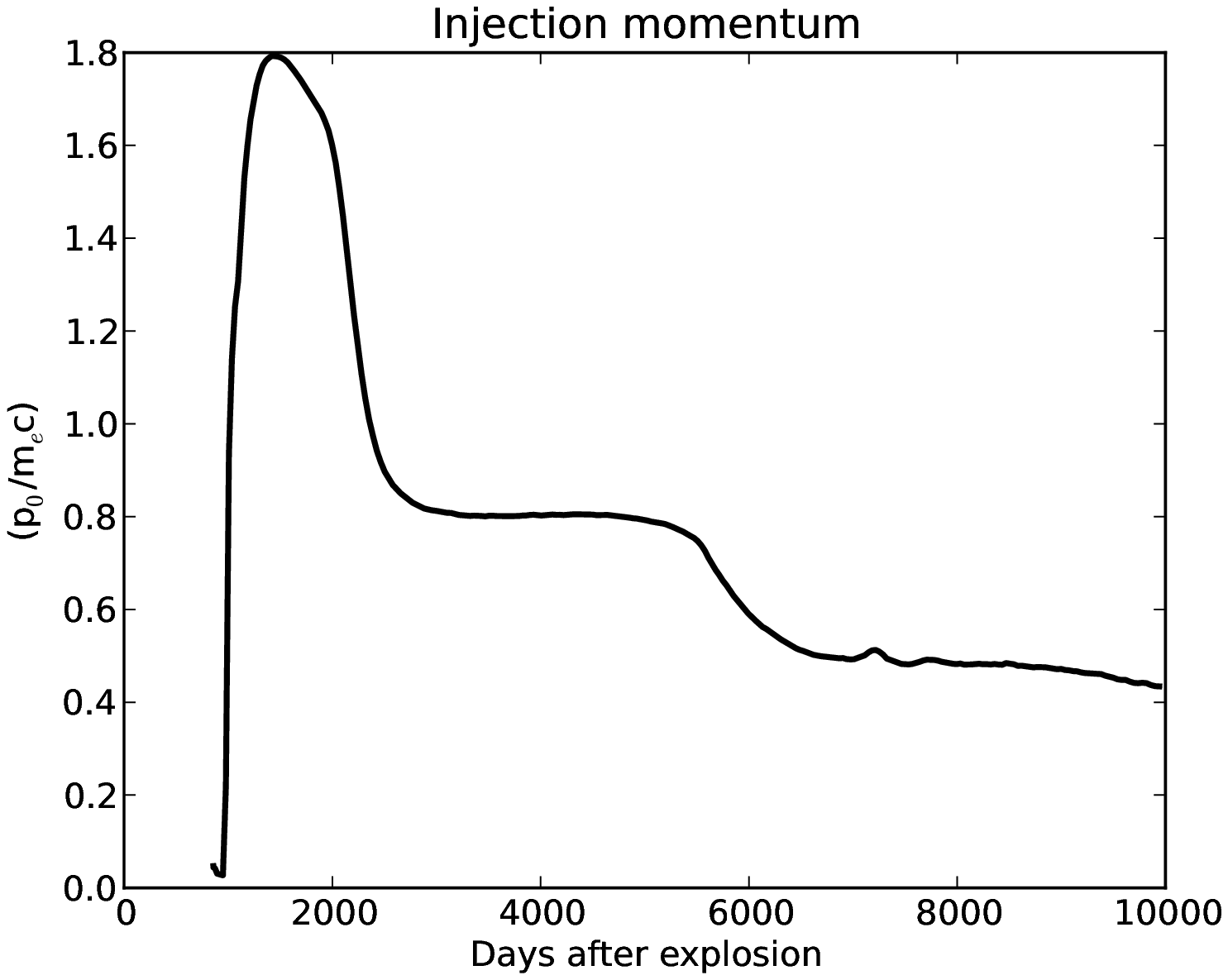,width=7.5cm,angle=0} \\
\epsfig{file=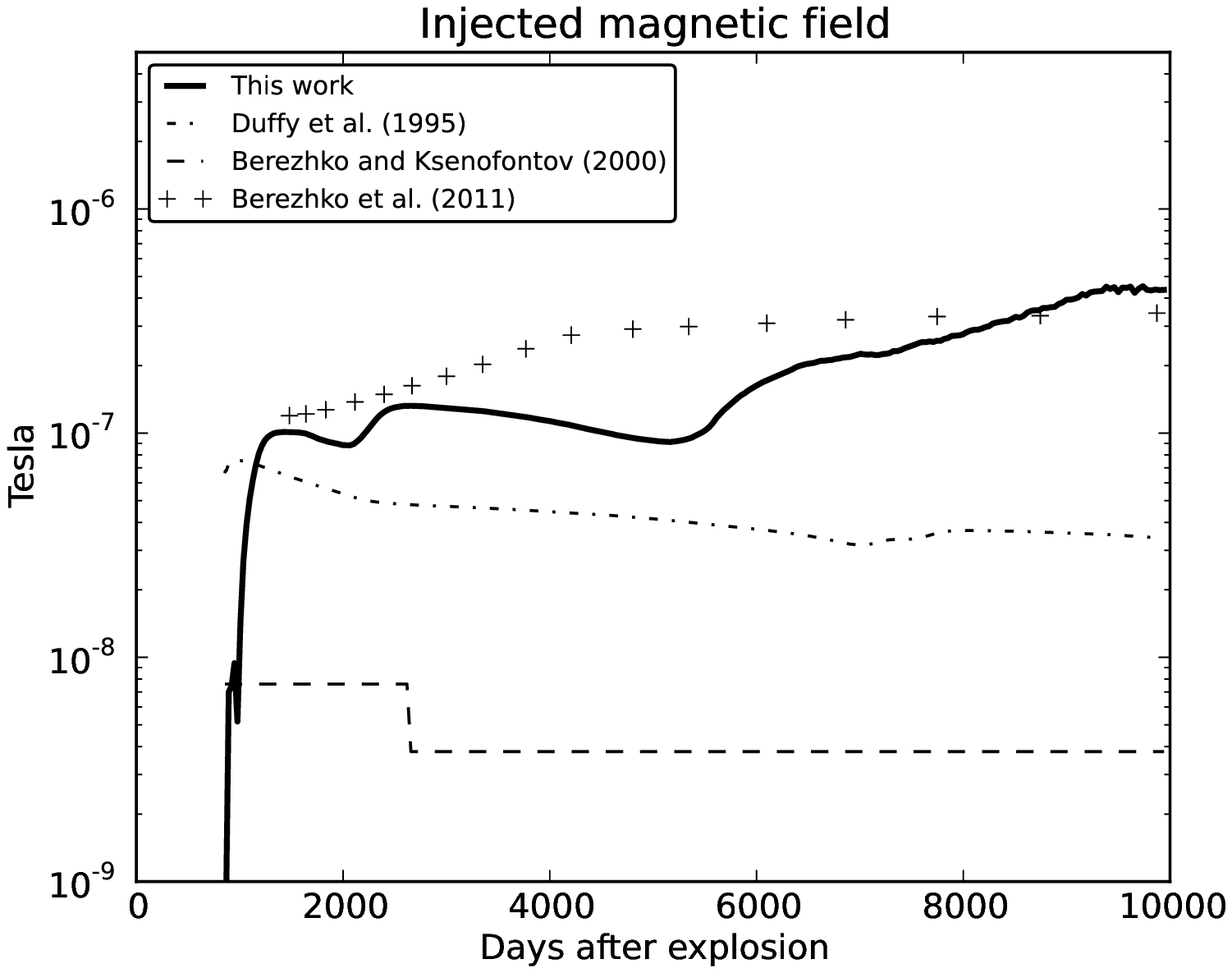,width=7.5cm,angle=0} &
\epsfig{file=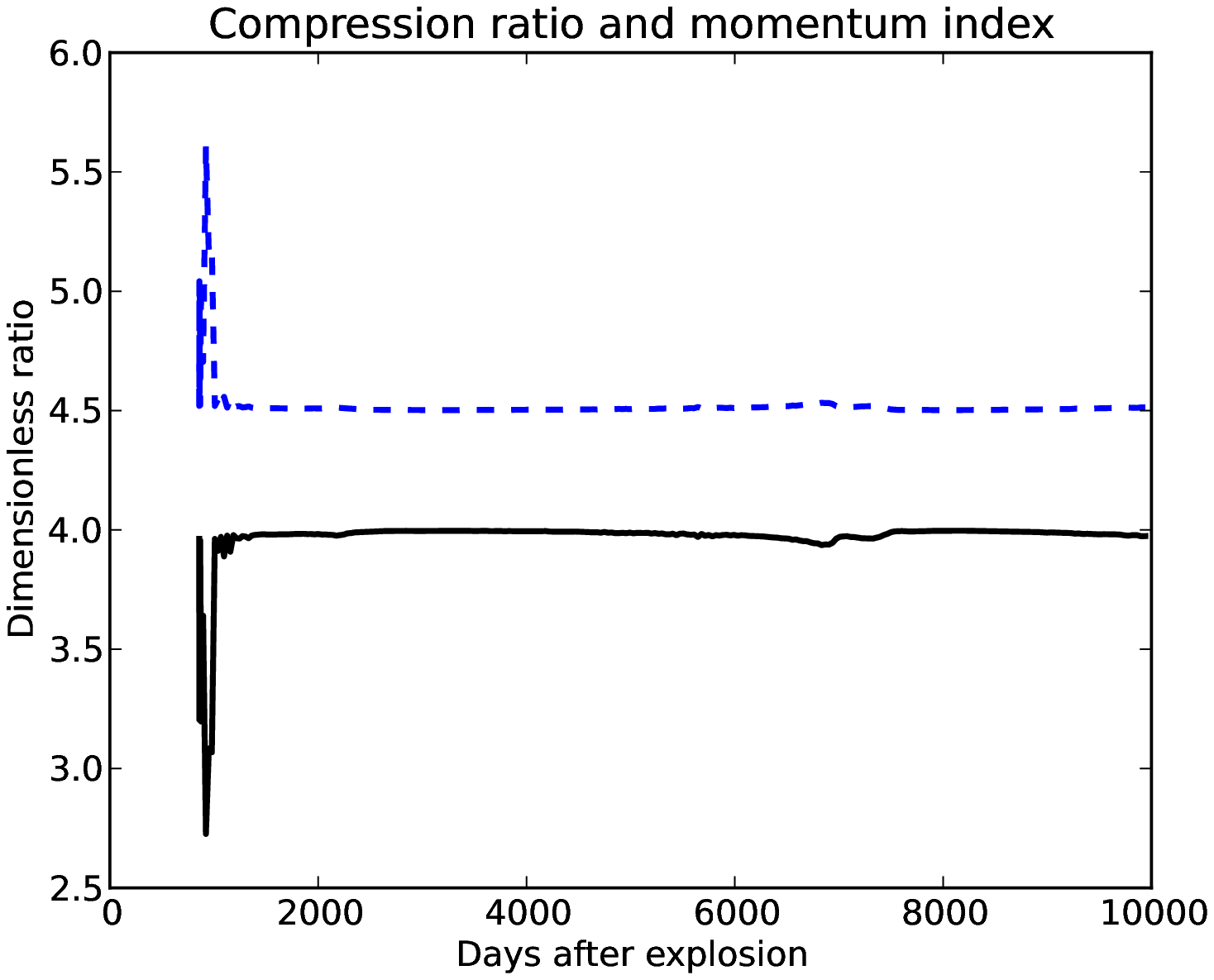,width=7.5cm,angle=0} \\
\end{array}
\end{math}
\end{center}
\caption{Injection parameters from the final model. In the top left is the injected density of energetic electrons $\chi_{el} n_e$; top right is the injected momentum $p_0$ ; bottom left is the injected magnetic field $B$ with the magnetic fields from \citet{Duffy:1995p18382}, \citet{Berezhko:2000p56} and \citet{Berezhko:2011p19115}. At bottom right is the compression ratio $\zeta$ and index $b$ on the isotropic momentum distribution. } \label{injection_parameters_plots}
\end{figure*}

As an independent consistency check to the semi-analytic injection physics of Section \ref{model_radio_thermal} we obtained the injection parameters set for newly shocked cells in the simulation.  The injection parameters at each timestep were obtained from a radio-weighted average of parameters from cells shocked during the previous three timesteps. We used the evolving radio emissivity at 843 MHz as the weight for the average at each timestep. In Figure \ref{injection_parameters_plots} is the result. 
At top left is the average number density of injected electrons, scaled by the fraction of injected electrons required to reproduce the observations. It is interesting to note two main jumps in the injected electron density. The first is from the shock encounter with the H\textsc{ii} region at day 2000 and second is the encounter with the equatorial ring around day 5500. The injected electron density as the shock crosses the H\textsc{ii} region is around $10^6$ m$^{-3}$. This is consistent with the pre-supernova electron density of the H\textsc{ii} region, after scaling by the injection efficiency and compression ratio. The higher electron injection density of $10^7$ m$^{-3}$ obtained at late times is consistent with the maximum scaled electron density of the ring. This suggests most of the radio emission at these times is arising from comparatively dense regions of the equatorial ring. 

At top right of Figure \ref{injection_parameters_plots} is the average injection momentum, in units of $m_e c$. As seen in Equation \ref{injection_momentum} we derived the injection momentum from the gas temperature at the downstream point. The maximum normalised injection momentum of $1.8$ set during the shock encounter with the H\textsc{ii} region is equivalent to a  shock temperature of $4 \times 10^{9}$ K. During propagation through the H\textsc{ii} region the injection momentum of 0.8 is equivalent to a temperature of $1 \times 10^{9}$ K. During the shock crossing of the ring, the injection momentum drops to 0.4, or a temperature of $4 \times 10^8$ K.

The injected magnetic field in the lower left panel of Figure \ref{injection_parameters_plots} shows that the amplified magnetic field is in the range $8 \times 10^{-8}- 5\times10^{-7}$ T. This is within an order of magnitude of the amplified magnetic field estimates in \citet{Duffy:1995p18382} and \citet{Berezhko:2011p19115}, but is an order of magnitude larger than the estimate in \citet{Berezhko:2000p56}.


The magnetic fields form these other works are also included in the figure for comparison. We believe the dip of the injected magnetic field around day 2000 is due to the lowering of shock velocity as the shocks crashed into the H\textsc{ii} region. This is probably an effect of the low resolution or an overestimated distance for the H\textsc{II} region. As the magnetic field is the only injection parameter to experience a dip at day 2000, we are confident this is responsible for the anomalous dip in seen around day 2000 in the flux density plots of Figure \ref{flux_vs_time}. 

For completeness, the compression ratio $\zeta$ and the index $b$ obtained at the shocks is plotted in the bottom right panel of Figure \ref{injection_parameters_plots}. Overall the compression ratio returned is fairly stable at the expected compression ratio of $\zeta=4$ for a strong shock, and an index of $b=4.5$ for sub-diffusive shock acceleration. This results in a spectral index of $\alpha=0.75$. A brief period of instability is observed at early times when the shock was established from the initial conditions of the simulation.

\subsection{Energy density at newly shocked cells}

\begin{figure}[h]
\begin{center}
\epsfig{file=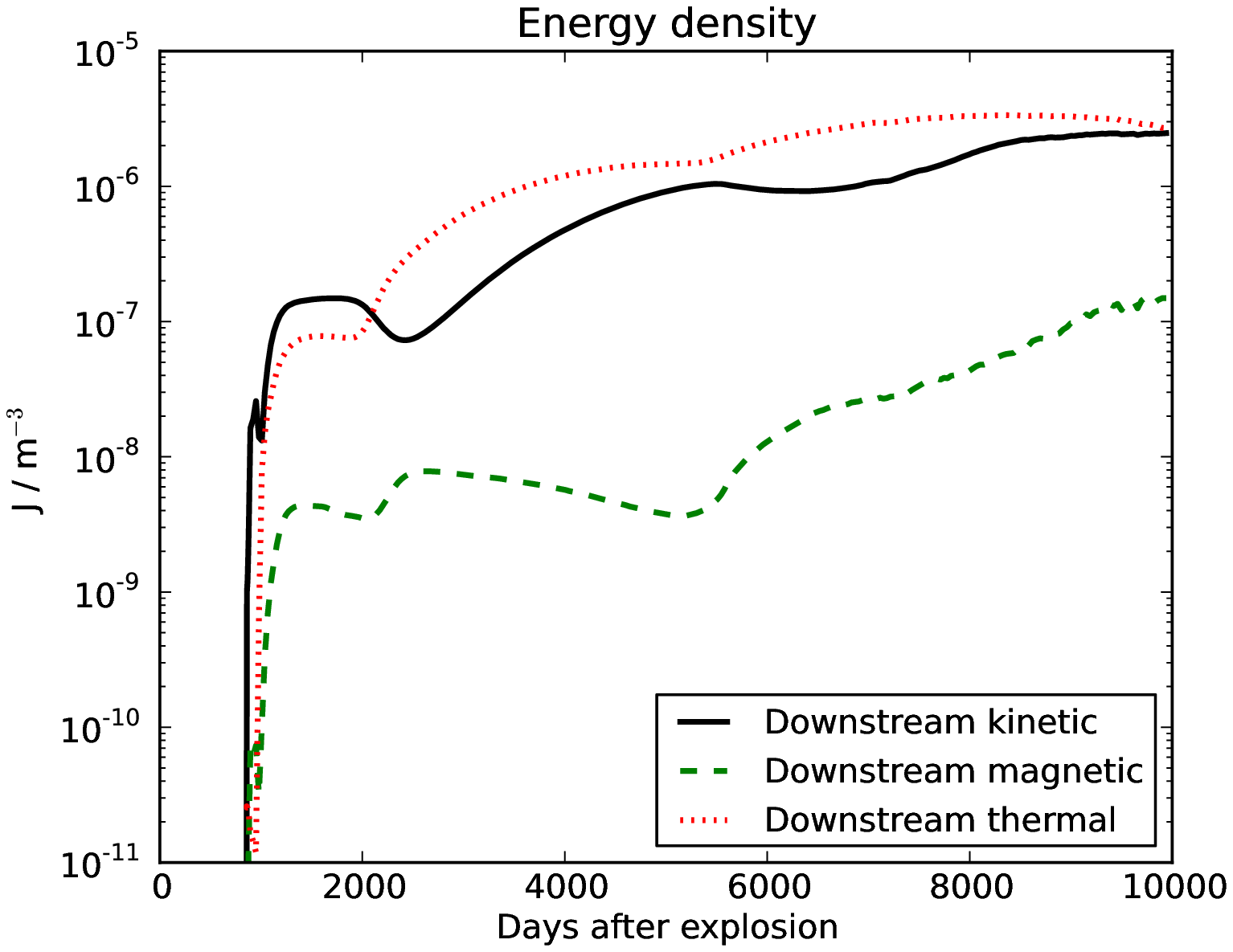,width=8.0cm,angle=0} 
\end{center}
\caption{Kinetic, thermal, electron, and magnetic field energy density at the shock. The kinetic energy density is dominant before the interaction of the shock with the H\textsc{ii} region. } \label{energy_densities}
\end{figure}

We also looked at the balance of energy density between kinetic, thermal, and magnetic processes at newly shocked points. Shown in Figure \ref{energy_densities} is the evolving energy density obtained by averaging in the same fashion as for the injection parameters. 
The magnetic energy density is nearly two orders of magnitude below the kinetic energy. This suggests that magnetic field amplification will have a negligible effect on shock evolution if the energy expended in amplifying a magnetic field is included in the energy budget. At early times prior to the encounter of the shock with the H\textsc{ii} region, the kinetic energy is clearly dominant. However this picture reverses soon after the encounter with the H\textsc{ii} region and thermal energy at the shock front becomes dominant for the rest of the simulation.

\section{Conclusions}

In this work we have sought to: (1) Test magnetic field amplification and an asymmetric explosion as the cause for the long term asymmetry in the radio remnant; (2) Refine the structure of the pre-supernova environment, (3) Obtain an estimate of the injection efficiency at the supernova shock, and (4) Provide a model that predicts future behaviour of the expanding radio remnant. We have addressed these questions by using a hydrodynamical simulation and a semi-analytic method incorporating Diffusive Shock Acceleration and magnetic field amplification to estimate power law distributions of electron momenta and the magnetic field in the downstream region of the shock. The distributions and magnetic field were evolved with the downstream flow. Morphological comparisons of the simulated radio emission with real observations shows that magnetic field amplification combined with an asymmetric explosion is able to reproduce the persistent asymmetry seen in radio observations of SN 1987A. The asymmetry in radio emission is primarily the result of non-linear dependence of the amplified magnetic field on the shock velocity. The evolving radio emission from the simulation was compared to a number of time-varying observations from SN 1987A, such as: radius,  flux density, morphology, opening angle, and spectral index. These comparisons formed the objective functions for an inverse problem, and allowed us to refine the model of the initial supernova environment. 

Essential features of the model are an asymmetric explosion, a blue supergiant wind, an H\textsc{ii} region and an equatorial ring at the waist of an hourglass. We fixed the energy and mass of the explosion at $1.5\times10^{44}$ J and 10 solar masses. From the radius and flux density comparisons we find that a termination shock distance of $(3.2-3.8)\times 10^{15}$ m $(0\farcs43-0\farcs51)$ provides a good fit for the turn on of radio emission around day 1200. An H\textsc{ii} region with an innermost radius of $(4.71\pm0.07) \times 10^{15}$ m $(0\farcs63 \pm 0\farcs01)$ and a maximum gas number density of $(7.11\pm1.78) \times 10^7 $ m$^{-3}$ provides a good fit to the shock deceleration around day 2000 and subsequent radius and flux density evolution to day 5500. The addition of clouds within the ring with a radius of $2.25 \times 10^{14}$ m, a peak number density of $3.1\times10^{10}$ m$^{-3}$ and a total mass $3.5\times10^{-2} M_{\odot}$ results in an abrupt increase in the flux density around day 5500, given a constant injection efficiency. It also results in a rapid reduction in opening angle and beading in the radio morphology after day 7000. 

Three dimensional renderings of the computational domain show that the period of apparent deceleration in shock velocity between days 7000 and 8000 may be the result of the forward shock leaving the equatorial ring. The forward shock emerged from the eastern lobe of the ring first around day 7000. It then emerged from the western lobe around day 8000. Following day 7000, the exit of the forward shock from the eastern lobe of equatorial ring leaves strong radio-emitting components in the western lobe.

The shock radii returned by truncated shell model fits to the simulation appear to be substantially larger than the expectation of radius from the simulation, and appears to follow the forward shock of the simulation. As we do not know how the radio emission of the real remnant is distributed between forward and reverse shocks we are unable to determine if truncated shell modelling is also similarly biased toward the forward shock of the real remnant.

Comparisons between simulated and observed flux density show that during the supernova shock traversal of the HII region, the flux densities of simulation and observation are in agreement if the fraction of electrons injected into the shock is around $4\%$. We arrive at this figure by making the somewhat speculative assumption that the electrons are in thermal equilibrium with the ions and are injected into the shock from the downstream region at a momentum consistent with their thermal velocity. There is a discrepancy between simulated and observed flux density around day 2000 due to a reduction in the amplified magnetic field caused by a stalled shock velocity as the shock encountered the H\textsc{ii} region. This problem might be rectified with higher resolution simulations. It may also mean that the radial distance of of the $H\textsc{ii}$ region has been overestimated. The flux density is also not in agreement with observations from day 5500 onwards as the shock encounters the thickest parts of the ring. This may be due to the reasonably coarse resolution of the simulation or lower $\chi_{el}$ arising from yet to be understood microphysics at the shock as it collides with the ring. 

As a result of the absence of cosmic ray feedback, the compression ratio, and hence the index on the inverse power law for the electron distribution remains constant. This produces a spectral index for radio emission which is inconsistent with the large dip seen in spectral index of from the real remnant \citep{Zanardo:2010p17425}. We expect that future models of the remnant that incorporate non-linear feedback \citep{lee:2014, ferrand:2014} or magnetic field topology \citep{bell:2011} will be able to address this discrepancy. 

By capturing the injection parameters at the shock and performing a radio emission weighted average, we also obtained estimates of the number density,  momentum, magnetic field, compression ratio, and energy density of the supernova shock. This permitted a consistency test of the semi-analytic method in use. The density of electrons injected into the shock is consistent with the upstream density (scaled by $\chi_{\mathrm{el}}$) of the medium into which the shock propagates. The injection momentum is consistent with a shock that has a temperature in the range $1-4\times10^{9}$ K. The injected magnetic field is in the range $8\times 10^{-8} - 5 \times 10^{-7}$ T, which is broadly consistent with \citet{Duffy:1995p18382} and \citet{Berezhko:2011p19115} but an order of magnitude higher than the estimate in \citet{Berezhko:2000p56}.


The ratio of energy densities at the shock clearly shows that kinetic and thermal energy are approximately two orders of magnitude stronger than magnetic energy density. It is interesting to note that the downstream thermal energy occupies the largest fraction of the available shock energy after the shock encounters the H\textsc{ii} region around day 2000.

 In terms of future predictions, the model indicates that the asymmetry in radio morphology may temporarily reverse in coming years as radio emitting spots in the western lobe of the ring decrease in brightness more slowly than their eastern counterparts. This is because the shock leaves the eastern ring more quickly. Synthetic images of the future radio morphology indicate that radio emission is concentrated in hotspots centred on overdense blobs within the equatorial ring. \\
 
We look forward to how this amazing young supernova remnant evolves in years to come.

\acknowledgments{}

We are grateful to John Kirk for his insightful and extremely helpful input in all stages of this project. We appreciate the support of staff at iVEC and ICRAR for providing supercomputing resources for our use. In addition we thank the Centre for Petroleum Geoscience and CO2 Sequestration for providing visualisation infrastructure to produce volumetric renderings of our simulations. The software (FLASH) used in this work was in part developed by the DOE - supported ASC/Alliances Center for Astrophysical Thermonuclear Flashes at the University of Chicago.








\begin{appendices}

\appendix

\section{Progenitor}\label{progenitor_appendix}

From \citet{Truelove:1999p17810} if an explosion were to propagate into a vacuum, it would expand with the velocity $v_{ej}$ and have radius $R_{ej}=v_{ej}t$. The velocity as a function of $r$ and $t$ is given by

\begin{eqnarray}
v(r,t)=\left \lbrace \begin{array}{c} \frac{r}{t}, \ \ r< R_{\mathrm{ej}} \\  0, \ \ r> R_{\mathrm{ej}}. \end{array}  \right.
\end{eqnarray}

The density profile of the supernova exploding into a power law environment with density $\rho(r)=\rho_s r^{-s}$ is given in terms of a structure function $f(v/v_{\mathrm{ej}})=f(w,n)$, and the ejecta mass $M_{\mathrm{env}}$. We introduce asymmetry in the progenitor by multiplying the density in \citet{Truelove:1999p17810} by the asymmetry function $(1+k \sin{\theta} \cos{\phi})$. 

\beq
\rho(r,t)=\left \lbrace \begin{array}{cc} \frac{M_{\mathrm{env}}}{(v_{ej} t)^3 } f \left ( \frac{v}{v_{\mathrm{ej}}} \right ) (1+k \sin{\theta} \cos{\phi}) , & r < R_{\mathrm{ej}} \\ 
\rho_{s}r^{-s}, & r > R_{\mathrm{ej}}. \end{array}  \right. 
\eeq 

Where the constant $k$ controls the degree of asymmetry in the progenitor. We chose $s=2$ for the environment surrounding the progenitor as we assume a constant velocity wind. The density scaling $\rho_{s}$ is determined by the wind velocity $v_{\mathrm{wind}}$ and progenitor mass loss rate $\dot{M}$

\beq
\rho_{s}=\frac{\dot{M}}{4 \pi v_{\mathrm{wind}}}.
\eeq

The structure function $f(w,n)$ specifies the shape of the solution given the exponent n

\beq
f(w,n)=\left \lbrace \begin{array}{cc} f_{\mathrm{n}} w_{\mathrm{core}}^{-n}, \ \ 0 \leq w \leq w_{\mathrm{core}} \\ 
f_{\mathrm{n}} w^{-n}, \ \ w_{\mathrm{core}} \leq w \leq 1. \end{array}  \right. 
\eeq 

\citet{Chevalier:1995p4450} used $n=9$ as a best fit to the supernova. The purpose of a core is to avoid a singularity when n is large with $w_{\mathrm{core}}=\frac{v_{\mathrm{core}}}{v_{\mathrm{ej}}}$ as a free parameter. Truelove and McKee recommend small values ($w_{\mathrm{core}}$=0.001-0.1) thus setting a small core velocity. We have adopted $w_{\mathrm{core}}=0.001$. Requiring that the density profile integrate to $M_{\mathrm{env}}$ reveals $f_{n}$ as

\beq
f_{n}=\frac{3}{4 \pi} \left [ \frac{3-n}{3-n w_{\mathrm{core}}^{3-n}} \right ].
\eeq

The kinetic energy of the explosion is determined by integration:

\beq
E_{\mathrm{kin}}= \frac{1}{2} M_{\mathrm{ej}} v_{\mathrm{ej}}^2  \int_{0}^{\pi} d\theta \int_{0}^{2 \pi} d\phi \int_{0}^{1} dw w^4 f(w,n)  (1+k \sin{\theta} \cos{\phi}) \sin{\theta}
\eeq

If the ratio of kinetic energies in the Eastern hemisphere to the Western hemisphere is $\chi_{\mathrm{ke}}$ then the constant $k$ may be obtained by taking the ratio of kinetic energy across the two hemispheres

\beq
k=-2 \left ( \frac{\chi_{\mathrm{ke}}-1}{\chi_{\mathrm{ke}}+1} \right ).
\eeq

Since $E_{\mathrm{kin}}$ has been specified as a fraction $\chi$ of the total explosion energy, we can solve for  $v_{\mathrm{ej}}$ to obtain

\beq \label{vej}
v_{\mathrm{ej}}=\sqrt{\frac{2E_\mathrm{kin}}{M_{\mathrm{ej}}} \frac{5}{3} \left ( \frac{5-n}{3-n} \right ) \left ( \frac{3-nw_{\mathrm{core}}^{3-n}}{5-n w_{\mathrm{core}}^{5-n}} \right ) }.
\eeq

In order to incorporate pressure we assume it is related to  density via an adiabatic process $P=k_{P} \rho(r,t)^{\gamma}$. Internal energy is derived from pressure through the ideal gas equation of state: 

\beq
E_{\mathrm{int}}= \left ( \frac{M_{\mathrm{ej}}}{(v_{\mathrm{ej}}t)^3}\right )^{\gamma}  \frac{ (v_{\mathrm{ej}} t)^3 k_P}{(\gamma - 1)}  \int_{0}^{\pi} d\theta \int_{0}^{2 \pi} d\phi \int_{0}^{1} dw w^2 f(w,n)^{\gamma} (1+k \sin{\theta} \cos{\phi})^{\gamma} \sin{\theta}.
\eeq

Since the internal energy is $(1-\chi)E_{\mathrm{tot}}$, the constant $k_P$ is given by

\beq
k_P=\frac{E_{\mathrm{int}}(\gamma-1)}{ f_n^{\gamma} (v_{\mathrm{ej}}t)^3}  \left ( \frac{M_{\mathrm{ej}}}{t^3 v_{\mathrm{ej}}^3} \right )^{-\gamma} \frac{3(3-n\gamma)}{3-n\gamma w_{\mathrm{core}}^{3-n\gamma}} \frac{1}{\int_{0}^{\pi} d\theta \int_{0}^{2 \pi} d\phi (1+k_2 \sin{\theta} \cos{\phi})^{\gamma} \sin{\theta}}.
\eeq

The position of the forward shock $(w_b)$ as a function of time is calculated from the differential equation of the shock motion, 

\beq
\frac{dw_b}{dt}=\frac{-w_bt }{  1+\left ( \frac{M_{\mathrm{env}}}{v_{\mathrm{ej}}^{3-s} \rho_s} \right )^{1/2} \frac{1}{l_{\mathrm{ed}}} \left [ \frac{f(w_b/l_{\mathrm{ed}})}{\sigma_{\mathrm{ed}}} \right ]^{1/2} w_b^{s/2} t^{(s-3)/2}   } \label{dwdt}.
\eeq

The constants $\sigma_{\mathrm{ed}}=0.212$ and $l_{\mathrm{ed}}=1.19$ are adopted for $n=9$ from Table 6 of \citet{Truelove:1999p17810}.

Regarding the position of the blast wave as a function of time, Truelove and McKee adopt the following characteristic values for position, time and mass

\begin{eqnarray}
R_{\mathrm{ch}} & = & M_{\mathrm{env}}^{1/(3-s)}\rho_s^{-1/(3-s)} \\
t_{\mathrm{ch}} & = &E_{\mathrm{kin}}^{-1/2}M_{\mathrm{env}}^{[(5-s)/(2(3-s))]}\rho_s^{-1/(3-s)} \label{tch} \\
M_{\mathrm{ch}} & = &M_{\mathrm{env}}.
\end{eqnarray}

Assuming an initial condition $w_b(0)=l_{\mathrm{ed}}$, \citet{Truelove:1999p17810} integrate Equation \ref{dwdt} to find the normalised forward blast position $(R^{\star}_b)$ as a function of time

\beq
R^{\star}_b= \frac{R_b}{R_{\mathrm{ch}}} = \left (  \frac{(3-s)}{2} \left ( \frac{l_{\mathrm{ed}}}{\sigma_{\mathrm{ed}}} \right )^{\frac{1}{2}}\int_{w_b/l_{\mathrm{ed}}}^{1}[wf(w)]^{\frac{1}{2}}dw \right )^{\frac{s-3}{2}}.
\eeq

They then invert this solution for the normalised time taken to get to $R_b$

\beq \label{tstar}
t^{\star}=\left (\frac{E_{\mathrm{kin}}}{M_{\mathrm{env}} v_{\mathrm{ej}}^2} \right )^{\frac{1}{2}} \frac{R^{\star}_b}{l_{\mathrm{ed}}} \left [ 1- \left ( \frac{3-n}{3-s} \right ) \left ( \frac{\sigma_{\mathrm{ed}}}{l_{\mathrm{ed}} f_n} \right )^{\frac{1}{2}} R_b^{\star \frac{3-s}{2}} \right ]^{-\frac{2}{3-n}}.
\eeq

 
 

  
Therefore, given the energies $E_\mathrm{kin}$ and $E_\mathrm{int}$, a progenitor mass  $M_{\mathrm{env}}, n, w_{\mathrm{core}}$, and $s$, we can completely describe the early expansion of the supernova. 

\section{Localising a shock and determining upstream and downstream fluid variables}\label{advection_shock_localise}

The scheme FLASH 3.2 uses to locate voxels undergoing a shock is to look for compression as well as a significant pressure gradient. We look for compression by finding velocity divergence $(\nabla \cdot \textbf{v})$ using central differencing. A negative velocity divergence indicates compression. If the voxel is indeed in a shock then it should also have a pressure gradient $||\nabla P||$ defined over the width of a numerical shock $s_w$. If $P_1$ is the upstream pressure and $P_2$ is the downstream pressure then the pressure gradient $||\nabla P||$ is

\beq
||\nabla P||=\left ( P_1 \left  (\frac{P_2}{P_1}-1 \right )/(s_w) \right ).
\eeq

Supposing we are looking for shocks with a compression ratio of at least $r_{\mathrm{min}}$. We can derive a minimum pressure ratio $r_{p,min}$ from Equation \ref{comprat}

\beq
r_{p,min}=\frac{(\gamma+1) r_{\mathrm{min}} -(\gamma-1)}{(\gamma+1)-(\gamma-1) r_{\mathrm{min}}}.
\eeq

Further supposing that the pressure $P$ in a voxel is at least the upstream pressure of a shock, $P_1$, then the pressure gradient in a voxel should be larger than

\beq \label{findshock}
||\nabla P|| > P \left  (r_{p,min}-1 \right )/(s_w).
\eeq

\begin{figure}[h]
\centering
\begin{tabular}{cc}
\epsfig{file=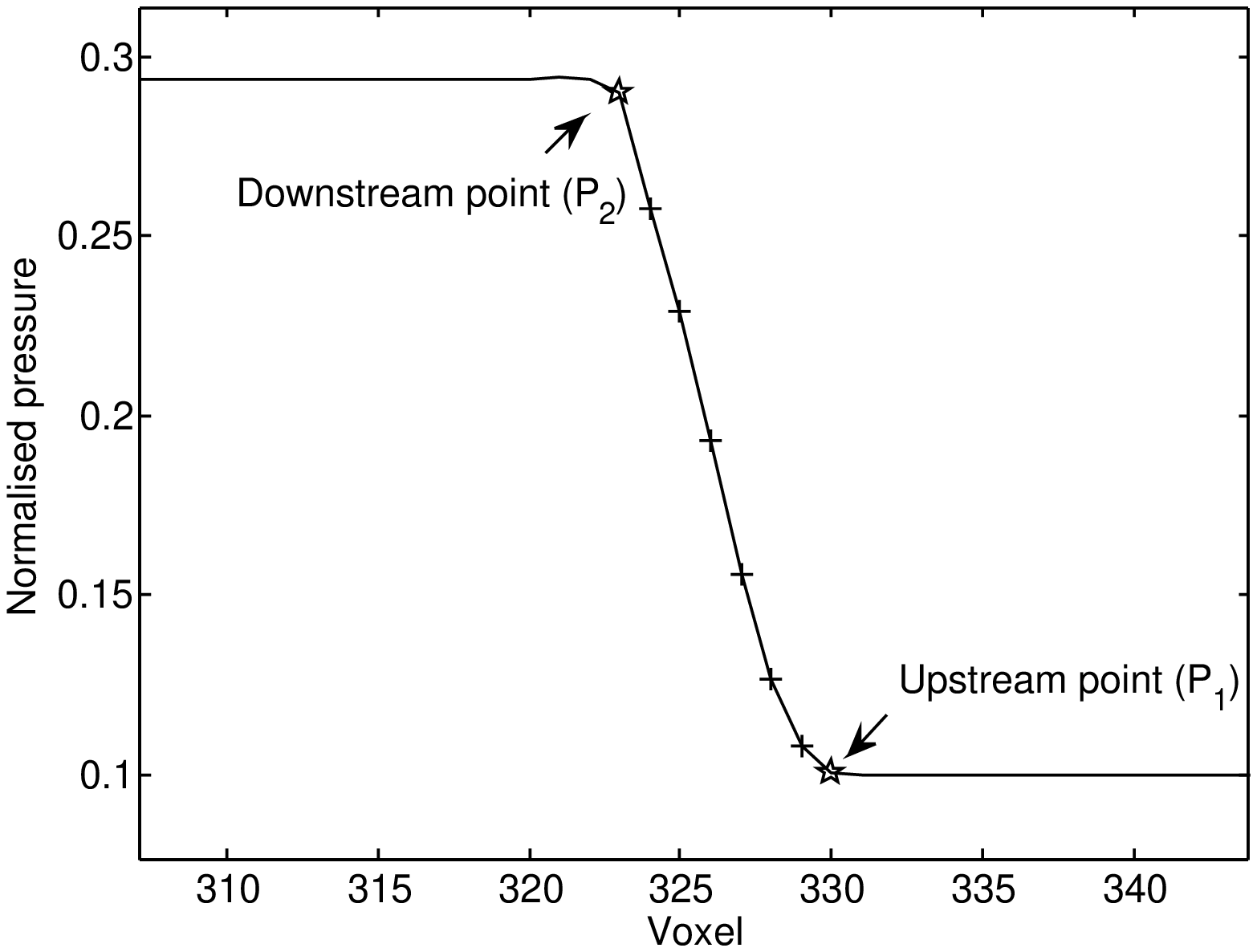,width=7.5cm} 
\epsfig{file=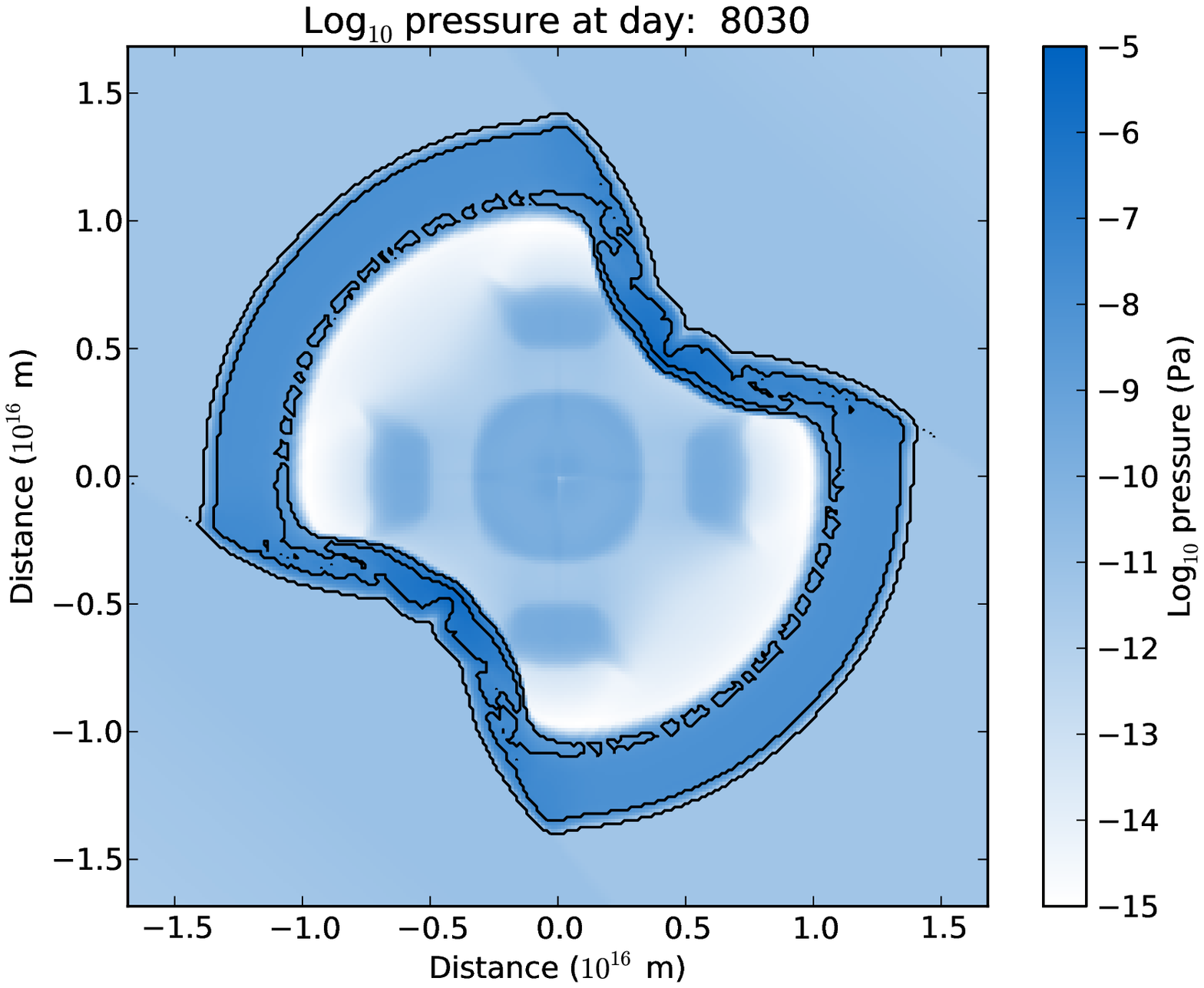,width=7.0cm} \label{find_shock_plot_overlay}
\end{tabular}
\caption{(Left) Points localised at the shock of a Sod shock tube problem. Crosses represent the localised points and the stars represent the upstream and downstream points $P_1$ and $P_2$ located by following the pressure gradient. (Right) The shock location algorithm applied to the supernova simulation at day 8030. Shown in blue is a slice across the computational domain in the log of pressure variable. Overlaid is a contour plot (with 1 level) of voxels that were determined to be within a shock. Outwards from the centre, the inner and outer contours mark the reverse and forward shock.} \label{find_shock_plot}
\end{figure}

A shock is crossing a voxel if \ref{findshock} is true and $(\nabla \cdot \textbf{v})<0$. Using $s_w=10$ cells at maximum mesh refinement and $r_{\mathrm{min}}=2.0$ is a good compromise on sensitivity. This technique does well to locate forward and reverse shocks, however it does not adequately locate shock boundaries. In Figure \ref{find_shock_plot} is a pressure profile of the forward shock from a Sod shock tube problem. The shock locator has found six points (shown as crosses) in the middle of the shock. The algorithm located points $P_1$ and $P_2$ by following the pressure gradient in the upstream and downstream directions until the slope fails to fulfil equation \ref{findshock} or the pressure gradient turns back on itself.

\section{Advecting a scalar variable}\label{advection_appendix}

Given the advection equation for a scalar variable $\Psi$, a velocity field $\textbf{u}$, and constant $\kappa$

\beq\label{full_psi_eq}
\frac{\partial \Psi}{\partial t}+\nabla \cdot (\textbf{u} \Psi) = (\nabla \cdot \textbf{u})\Psi \kappa.
\eeq

The method of solution is similar to that  \citep{toro:2009}, pages 533-535. We first solve the associated homogeneous equation 

\beq \label{homogeneous_psi_eq}
\frac{\partial \Psi}{\partial t}+\nabla \cdot (\textbf{u} \Psi) =0.
\eeq

Solving this equation gives a temporary solution $\Psi^*$ which is the solution to $\frac{d \Psi}{dt}=0$. The full solution of equation \ref{full_psi_eq} is completed by solving the following ODE in way following the prescription of \citep{toro:2009}, pages 533-535

\beq \label{dpsidt}
\frac{d \Psi}{dt}=(\nabla \cdot \textbf{u})\Psi^{*} \kappa.
\eeq 

In order to solve equation \ref{homogeneous_psi_eq} we integrate over the spatial-temporal cell dimensions to obtain the exact solution in terms of fluxes entering each interface. Given at timestep $\Delta t$ and grid spacing $\Delta x$ the exact solution is

\beq
\Psi^{* (n+1)}=\Psi^n-\frac{\Delta t}{\Delta x} \left (\mbox{F}^{n+1/2}_{x+1/2}-\mbox{F}_{x-1/2}^{n+1/2}+\mbox{F}^{n+1/2}_{y+1/2}-\mbox{F}_{y-1/2}^{n+1/2}+\mbox{F}^{n+1/2}_{z+1/2}-\mbox{F}_{z-1/2}^{n+1/2} \right ).
\eeq

To determine the fluxes we use the framework from page 457-461 of \citep{toro:2009}. Using the velocity field $\textbf{v}=[v^n_x,v^n_y,v^n_z]$ from the hydro simulation, the flux entering the cell from the left x direction is given by

\beq \label{update_fp}
\mbox{F}^{n+1/2}_{x-1/2}=\frac{1}{2}(1+\sigma_{x-1/2})v^{n}_{x,x-1} \ \Psi^{n}_{x-1}+\frac{1}{2}(1-\sigma_{x-1/2})v^{n}_{x,x} \ \Psi^{n}_{x},
\eeq

where $\sigma_{x-1/2}$ is a flux limiter function. We use the simple up-wind flux limiter. Given the average velocity across the cell interface $v^{n+1/2}_{x,x-1/2}=\frac{1}{2}(v^{n}_{x,x-1}+v^{n}_{x,x})$ the flux limiter is defined as 

\beq
\begin{array}{ccc}
\sigma_{x-1/2} & = & \left \{
\begin{array}{ccc}  1 &  \mbox{if} &  v^{n+1/2}_{x,x-1/2}>0 \\ 
 -1 & \mbox{if} & v^{n+1/2}_{x,x-1/2}<=0. \\
\end{array} \right. \\
\end{array}
\eeq

Once the solution to \ref{homogeneous_psi_eq} has been approximated the full solution is obtained through the analytic solution to equation \ref{dpsidt}
 
\beq
\Psi^{n+1}=\exp{\left ( ( \nabla \cdot \textbf{u})\Psi^{* (n+1)} \kappa \Delta t \right )}.
\eeq
 
\section{Pre-supernova environment formation simulation}\label{remnant_formation}

The beautiful hourglass structure of SN 1987A is thought to arise as a blue supergiant wind interacts with material from past evolutionary phases of the progenitor. Simulations of remnant formation have more or less been able to replicate the beautiful hourglass surrounding SN 1987A by placing a spherically-symmetric blue supergiant wind inside an asymmetric environment \citep{Blondin:1993p14977, Soker:1999p17968, Tanaka:2002p19, Podsiadlowski:2007p18006}. 

Previous simulations of remnant formation of SN 1987A \citet{Blondin:1993p14977} have shown that a supersonic blue supergiant (BSG) wind extends radially outwards from the progenitor. The density profile of the wind scales with radius as $r^{-2}$ since the flow is essentially a free-flowing wind. The free wind ends in a termination shock around $(2.99-3.74) \times 10^{15}$ m $(0\farcs4-0\farcs5)$ from the progenitor. Material downstream from the termination shock is hot due to adiabatic compression and forms the bubble responsible for inflating the hourglass. 

In order to obtain the density and temperature profiles of material in the free-wind and shocked-wind regions prior to the explosion, we simulated the formation of the pre-supernova environment in three dimensions. As with the supernova simulation we used the standard hydrodynamics solver in FLASH with radiative cooling as discussed in Section \ref{radiative_cooling}. The computational domain was constructed as a rectangular grid of dimensions $256\times256\times640$ at the finest level of mesh refinement. This corresponds to a box of dimensions $(2.4 \times 2.4 \times 6.0) \times 10^{16}$  m or $(3\farcs2\times3\farcs2\times8\farcs0)$ at a distance of 50 kpc. The same Cartesian geometry was employed, as in section \ref{orientation}, however we did not incline the environment in this instance.

The general idea of the formation simulation is to have a BSG wind interact with an asymmetric RSG wind. For a star with mass loss rate $\dot{M}$ and radial wind velocity $v_w$, the radial density profile of the free-wind is
\beq
\rho(r)=\frac{\dot{M}}{4\pi v_w r^2}\label{bsg_dens_profile}.
\eeq
Under the assumption of adiabatic flow, pressure is expected to scale with density as $P(r)\propto \rho(r)^{\gamma}$. In the centre of the grid and at the origin we fixed a "star" - a spherical region of radius $r_s=9.4$ cells ($r_s=8.77 \times 10^{14} $ m) at the highest level of refinement. Within the star we set constant boundary conditions using equation \ref{bsg_dens_profile} and $ \dot{M}=7.5 \times 10^{-8}$ $M_{\sun}$  $\mathrm{yr}^{-1}$, $450 \ \mathrm{km} \ \mathrm{s}^{-1}$ from \citet{Chevalier:1995p4450}. Everywhere within the star we set a constant radial velocity of $v_w=450 \ \mathrm{km} \ \mathrm{s}^{-1}$.

 The pressure profile within the ``star" was generated assuming adiabatic flow and a wind temperature of $16,000$ K at the stellar surface where $r=3.0\times10^{10}\mathrm{m}$ \citep{Woosley:1988p18039}. 

For the the initial environment of the asymmetric RSG wind we used the wind profile from \citet{Blondin:1993p14977}. If $\theta=\sin^{-1}( z^{\prime}/r)$ is the angle from the $z^{\prime}$ axis (see section \ref{orientation} where these axes are defined),  $\dot{M}_{\mbox{\tiny{RSG}}}$ is the mass loss rate of the red supergiant (RSG), $v_{w,\mbox{\tiny{RSG}}}$ is the RSG wind velocity, the density of the environment is described in terms of the asymmetry parameter A:

\beq
\rho(r,\theta)=\frac{3 \dot{M}_{\mbox{\tiny{RSG}}}}{4 v_{w,\mbox{\tiny{RSG}}} \pi r^2(3-A)}(1-A \cos^2 \theta).
\eeq

We used $\dot{M}_{\mbox{\tiny{RSG}}}=2.0 \times 10^{-5}$ $M_{\sun}$  $\mathrm{yr}^{-1}$, $v_{w,\mbox{\tiny{RSG}}}=5 \ \mathrm{km} \ \mathrm{s}^{-1}$ and $A=0.95$ from the best-fit model of \citet{Blondin:1993p14977}. For $A=0.95$, half of the RSG mass was lost  within a half-opening angle of $21^{\circ}$ from the equatorial plane and the equatorial-to-polar density ratio is 20:1. The pressure profile of the relic RSG wind was set by keeping the wind temperature constant at 500 K. 

\begin{figure}[h]
\begin{center}
\includegraphics[width=18.0cm, angle=0]{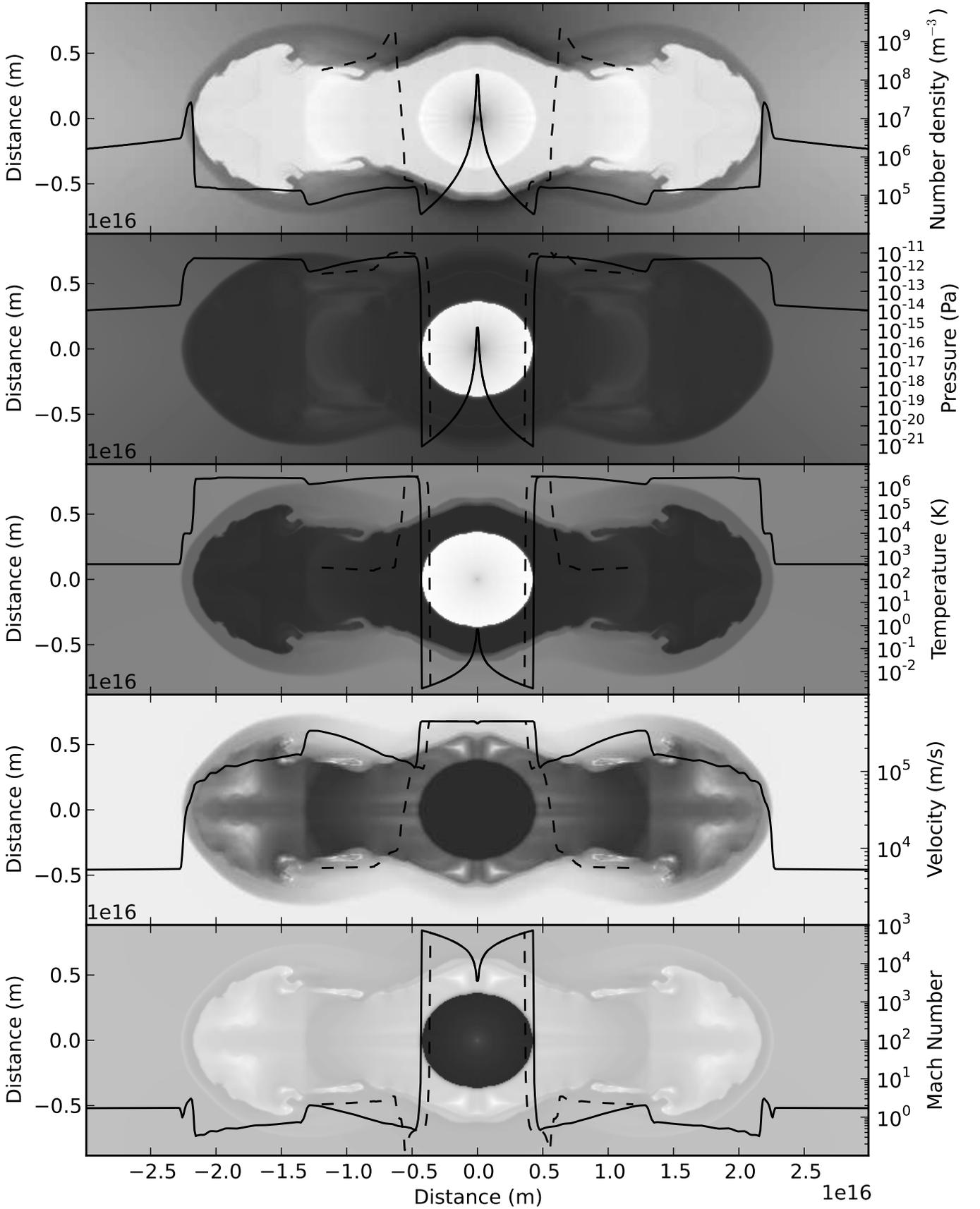} 
\end{center}
\caption{Slices of the formation simulation around 19,725 years after the initial conditions when the distance to waist of the hourglass approximates the radius of the observed equatorial ring. The slices are formed by cuts halfway along the $x$ axis. Shown are log-scaled images in different hydrodynamical variables overlaid by one-dimensional profiles. The horizontal (polar) and vertical (equatorial) profiles are represented by solid and dashed lines. Note the highly supersonic BSG wind bubble in the centre, surrounded by a hot bubble of shocked BSG wind at a temperature of around $10^6$ K.} \label{formation_sequence}
\end{figure}

The simulation was evolved until the waist of the bipolar inflated bubble matched the radius of the equatorial ring from the observations. For $A=0.95$ this occurred around 19,725 simulated years from the initial conditions. This is consistent with other estimates of around 20,000 years for the time taken for the BSG to inflate the hourglass \citep{Podsiadlowski:2007p18006}. Figure \ref{formation_sequence} shows slices of the computational domain formed by cuts halfway along the $x$ axis. Shown are the slices in different hydrodynamical variables overlaid by one-dimensional profiles. The horizontal and vertical profiles are represented by solid and dashed lines. 

Outwards from the star, a rarefied and fast, blue supergiant (BSG) wind extends to a termination shock located at a radial distance of  $(3.62-4.12) \times 10^{15}$ m ($0\farcs48-0\farcs55$) along the polar axis, and $(4.27 - 4.77 ) \times 10^{15}$ m ($0\farcs57-0\farcs64$) along the equatorial axis. The BSG wind within this region is rarefied and highly supersonic with little variation from the specified velocity of $450$ km s$^{-1}$. Given the density and pressures of the environment this corresponds Mach numbers ranging from $19$ at the BSG surface, $23,000$ at the inner boundary conditions of the ``star" and $72,000$ just inside the polar termination shock, where the Mach number crosses unity. 

Exterior to the termination shock is a hot bubble comprising shocked BSG wind. Overall, the shocked BSG wind has approximately constant gas properties with a particle density of $1.33\times10^{5}$ m$^{-3}$ and a temperature of $2.4 \times 10^6$ K. Due to the bipolar nature of the outflow, another shock known as a Mach disk forms at a distance $(1.29-1.35) \times 10^{16}$ m  ($1\farcs72-1\farcs81$) along the polar axes. 

At at the expanding edge of the bubble, the hot BSG wind interacts with the relic RSG wind in two places. The inner edge of the bubble is the interface between BSG and RSG winds. From the inner edge a forward disturbance propagates outwards to become the outer edge of the expanding bubble. The ``shocked" RSG material between the inner and outer edges of the bubble is associated with the H\textsc{ii} region from \citep{Chevalier:1995p4450}. The H\textsc{ii} region in this simulation has a particle density in the range $10^{7}- 10^{8}$ m$^{-3}$ and a temperature around $10^4$ K. 

In order to derive profiles for use in the supernova simulations polynomials were fitted to the density, pressure and velocity profiles within the expanding bubble. If $r_0$ is the radius of the BSG at $3.0\times10^{10}$ m and $r^{\prime}=\frac{r}{r_0}$ then the polynomial to be fitted is $y=10^{a (\log_{10}{r^{\prime}})^2+b (\log_{10}{r^{\prime}})+c}$ with the resulting units are in SI (units of density are in kg m$^{-3}$). The coefficients of the fit are listed in Table \ref{efstable}.

\begin{table}
\caption{Parameters of the fit to hydrodynamical variables in the environment formation simulation} 
\centering
\begin{tabular}{cccccc}
\hline \hline
 &  Radial Distance &  &  &  &  \\
Region & (m) & Parameter  & $a$ & $b$ & $c$ \\ \hline
\multirow{3}{*}{BSG wind} & \multirow{3}{*}{$7\times 10^{13} - 4.27\times 10^{15}$}  &  Density  & $-0.1574$ & $-0.5961$ & $-15.1352$ \\
 &  &  Pressure & $-0.2156$ & $-1.4396$ & $-7.8483$ \\
 &  &  Velocity & $-0.0122$ & $0.1132 $ & $5.3911$ \\ \hline
\multirow{3}{*}{Shocked BSG wind to Mach disk} & \multirow{3}{*}{$(4.77-12.87) \times 10^{15}$}  &  Density  & $0.0$ & $-1.0848$ & $-15.9787$ \\
 &  &  Pressure & $0.0$ & $-1.8930$ & $-1.1412$ \\
 &  &  Velocity & $0.0$ & $1.2061 $ & $-1.2405$ \\ \hline
 \multirow{3}{*}{Mach disk to edge of bubble} & \multirow{3}{*}{$(13.48-21.57) \times 10^{15}$}  &  Density  & $0.0$ & $0.0$ & $-21.7899$ \\
 &  &  Pressure & $0.0$ & $0.0$ & $-11.2969$ \\
 &  &  Velocity & $0.0$ & $0.0 $ & $5.0924 $ \\ \hline
\end{tabular} \label{efstable}
\end{table} 

We compared these fits of the BSG wind profiles to theoretical estimates of the density, pressure and velocity profiles on the assumption of ballistic flow. We find that the properties of the free BSG wind is in agreement with the adiabatic approximation. The average deviation of the fits to the theoretical profiles are $6\%$ in density, $9\%$ in pressure and $0.4\%$ in the velocity profiles.
 



\end{appendices}

\end{document}